\documentclass{aa} 
\usepackage{subfigure}
\usepackage{txfonts}
\usepackage[authoryear]{natbib}
\usepackage{lscape}                                
\usepackage{color}
\usepackage{graphicx}
\usepackage{longtable}
\newcommand{\iraf} {{\sc iraf}}

\newcommand{\elecd}{$n_{\rm e}$}
\newcommand{\te}{$T_{\rm e}$}
\newcommand{\hb}{H$\beta$}

\newcommand{\fci}{[C~{\sc i}]}

\newcommand{\foiii}{[O~{\sc iii}]}

\newcommand{\foi}{[O~{\sc i}]}
\newcommand{\foii}{[O~{\sc ii}]}
\newcommand{\fsi}{[S~{\sc i}]}
\newcommand{\fsii}{[S~{\sc ii}]}
\newcommand{\fsiii}{[S~{\sc iii}]}
\newcommand{\fnitroi}{[N~{\sc i}]}
\newcommand{\fnii}{[N~{\sc ii}]}
\newcommand{\sfmgi}{Mg~{\sc i}]}
\newcommand{\mgii}{Mg~{\sc ii}}
\newcommand{\fariii}{[Ar~{\sc iii}]}
\newcommand{\fariv}{[Ar~{\sc iv}]}
\newcommand{\farv}{[Ar~{\sc v}]}
\newcommand{\fcaii}{[Ca~{\sc ii}]}
\newcommand{\fclii}{[Cl~{\sc ii}]}
\newcommand{\fcliii}{[Cl~{\sc iii}]}
\newcommand{\fcliv}{[Cl~{\sc iv}]}
\newcommand{\fcrii}{[Cr~{\sc ii}]}
\newcommand{\fcriii}{[Cr~{\sc iii}]}
\newcommand{\fcriv}{[Cr~{\sc iv}]}
\newcommand{\fneiii}{[Ne~{\sc iii}]}
\newcommand{\fneiv}{[Ne~{\sc iv}]}
\newcommand{\fnev}{[Ne~{\sc v}]}
\newcommand{\fkriii}{[Kr~{\sc iii}]}
\newcommand{\fkriv}{[Kr~{\sc iv}]}
\newcommand{\fxeiii}{[Xe~{\sc iii}]}

\newcommand{\fniqii}{[Ni~{\sc ii}]}
\newcommand{\fniqiii}{[Ni~{\sc iii}]}
\newcommand{\fniqiv}{[Ni~{\sc iv}]}
\newcommand{\fkiv}{[K~{\sc iv}]}

\newcommand{\fmniv}{[Mn~{\sc iv}]}
\newcommand{\fmnv}{[Mn~{\sc v}]}
\newcommand{\ffeii}{[Fe~{\sc ii}]}
\newcommand{\ffeiii}{[Fe~{\sc iii}]}
\newcommand{\ffeiv}{[Fe~{\sc iv}]}

\newcommand{\fpii}{[P~{\sc ii}]}
\newcommand{\fseii}{[Se~{\sc ii}]}
\newcommand{\oiii}{O~{\sc iii}}
\newcommand{\nitroi}{N~{\sc i}}
\newcommand{\nii}{N~{\sc ii}}
\newcommand{\niii}{N~{\sc iii}}
\newcommand{\sili}{Si~{\sc i}}
\newcommand{\silii}{Si~{\sc ii}}
\newcommand{\oi}{O~{\sc i}}
\newcommand{\oii}{O~{\sc ii}}
\newcommand{\ci}{C~{\sc i}}
\newcommand{\cii}{C~{\sc ii}}
\newcommand{\ciii}{C~{\sc iii}}
\newcommand{\civ}{C~{\sc iv}}
\newcommand{\nei}{Ne~{\sc i}}
\newcommand{\neii}{Ne~{\sc ii}}

\newcommand{\sii}{S~{\sc ii}}
\newcommand{\siii}{S~{\sc iii}}

\newcommand{\cli}{Cl~{\sc i}}

\newcommand{\hi}{H\,{\sc i}}
\newcommand{\hii}{H~{\sc ii}}

\newcommand{\hei}{He~{\sc i}}
\newcommand{\heii}{He~{\sc ii}}

\newcommand{\mc}{\multicolumn}
\newcommand{\nodata}{---}

\def\msun{\mbox{{\rm M}$_\odot$}}

\begin{document}

\title{Analysis of chemical abundances in planetary nebulae with [WC] central stars. I. Line intensities and physical conditions
\thanks{Based on data obtained at Las Campanas Observatory, Carnegie Institution.} 
} 
\author{
Jorge Garc\'ia-Rojas\inst{1,2}, 
 Miriam Pe\~na\inst{3}, Christophe Morisset\inst{1,3}, Adal Mesa-Delgado\inst{1,2}, and Mar\'ia Teresa Ruiz\inst{4}} 
\offprints{J. Garc\'ia-Rojas} 
\institute{
1 Instituto de Astrof\'isica de Canarias, E-38200 La Laguna, Tenerife, Spain\\ 
2 Departamento de Astrof\'{\i}sica. Universidad de La Laguna, E38205 La Laguna, Tenerife, Spain \\
3 Instituto de Astronom\'ia, Universidad Nacional Aut\'onoma de M\'exico,
Apdo. Postal 70264, M\'ex. D. F., 04510 M\'exico\\
4 Departamento de Astronom\'ia, Universidad de Chile, Casilla 36 D, Las Condes, Santiago, Chile.\\
\email{jogarcia@iac.es; miriam@astro.unam.mx; chris.morisset@gmail.com; amd@iac.es; mtruiz@das.uchile.cl}  }
\date{Received xxxxxxxxxxx; accepted xxxxxxxxxxxx} 

\titlerunning{PNe with a [WC] central stars. I.}

\authorrunning{Garc{\'\i}a-Rojas, Pe\~na, Morisset, Mesa-Delgado \& Ruiz} 


\abstract 
{Planetary nebulae (PNe) around Wolf-Rayet [WR] central stars ([WR]PNe) constitute a particular photoionized nebula class 
that represents about 10\% of the PNe with classified central stars.} 
{We analyse deep high-resolution spectrophotometric data of 12 [WR] PNe. This sample of [WR]PNe represents the most extensive 
analysed so far, at such high spectral resolution. We aim to select the optimal physical conditions in the nebulae to be used 
in ionic abundance calculations that will be presented in a forthcoming paper.} 
{We acquired spectra at Las Campanas Observatory with the 6.5-m telescope and the Magellan Inamori Kyocera (MIKE) spectrograph, 
covering a wavelength range from 3350 \AA~to
9400 \AA. The spectra were exposed deep enough to detect, with signal-to-noise ratio higher than three, the weak optical 
recombination lines (ORLs) of {\oii}, {\cii}, and other species. We detect and identify about 2980 emission lines, 
which, to date, is the most complete set of spectrophotometric data published for this type of objects. 
From our deep data, numerous diagnostic line ratios for {\te} and {\elecd} are determined from collisionally excited lines (CELs), 
ORLs, and continuum measurements ({\hi} Paschen continuum in particular).} 
{Densities are closely described by the average of all determined values for objects with {\elecd}$<$10$^4$ cm$^{-3}$, and 
by {\elecd}({\fcliii}) for the densest objects. For some objects, {\elecd}({\fariv}) is adopted as the characteristic density of the 
high ionization zone. For {\te}, we adopt a three-zone ionization scheme, where the low ionization zone is characterised by 
{\te}({\fnii}), the medium ionization zone by {\te}({\foiii}), and the highest ionization one by {\te}({\fariv}) when available. 
We compute {\te} from the {\hi} Paschen discontinuity and from {\hei} lines. For each object, {\te}({\hi}) is, in general, 
consistent with {\te} derived from CELs, although it has a very large error. Values of {\te}({\hei}) are systematically lower than the 
{\te} derived from CELs. 
When comparing {\te}({\hi}) and {\te}({\hei}) it is unclear whether the behaviour of both temperatures agrees with the predictions of the 
temperature fluctuations paradigm, owing to the large errors in {\te}({\hi}). 
We do not find any evidence of low-temperature, high-density clumps in our [WR]PNe from the analysis of 
faint {\oii} and {\nii} plasma diagnostics, although uncertainties dominate the observed line ratios in most objects. 
The behaviour of {\te}({\foiii})/{\te}({\fnii}), which is smaller for high ionization degrees, can be reproduced by a set of 
combined matter-bounded and radiation-bounded models, although, for the smallest temperature ratios, a too high metallicity seem to be required. }
{}

\maketitle 


\section{Introduction}

Planetary nebulae (PNe) constitute the evolutionary end point of most stars in the Universe, and they play a major role in the 
chemical enrichment history of  the interstellar medium (ISM) of  galaxies. 
Their chemical compositions allow us to determine  the abundances of
some chemical elements present in the ISM when their progenitor stars
were born and to analyse the contribution of these stars to galactic chemical enrichment by elements processed 
in the nucleus and ejected into the ISM. Planetary nebulae are produced by stars with initial masses of between 
$\sim$  1 \msun \ and $\sim$  8 \msun \ and a large
age spread \citep[from 0.1 to 9 Gyr, ][]{allenetal98}. The central stars (CS) are evolved stars (post-AGB) in the 
pre-white dwarf stage. Among them, there is an interesting group 
that has both very intense stellar winds with mass-loss rates in the range of 10$^{-7}- 10^{-5}$ 
\msun \ per year \citep[][and references therein]{koesterke01} and terminal velocities from several hundreds to several 
thousands km s$^{-1}$ and a hydrogen deficient chemical composition. All these stars have been catalogued as Wolf-Rayet of the C-sequence 
\citep[for a quantitative classification scheme of these stars, see][and references therein]{ackerneiner03}. 
Until recently, the [WR] CSPNe were thought to represent about 5-7\% of all CSPNe \citep{tylendaetal93, ackerneiner03}, 
but that value should be a lower limit owing to several selection effects, and the real percentage could be larger \citep{depewetal11}.
Among these CSPNe, about half of them, 
the so-called [WC]-early stars, have very high surface temperatures from 80 kK to about 150 kK (spectral classes are [WC4-5], [WO1-3]) 
while the others, called [WC]-late stars (spectral classes [WC6] to [WC11]) have temperatures between 
80 kK and about 30 kK \citep[][]{koesterke01}.

Many papers have been devoted to the analysis of these PNe and their CS because they 
can provide several clues about the processes of the chemical enrichment of both the shells and the interstellar medium with 
freshly made products, the processes of the shell ejection and its additional interaction with the stellar winds, etc. 
Several mechanisms have been proposed to explain the presence of the strong winds and the peculiar H-deficient 
chemical composition of [WC] stars, among them a final thermal pulse at the AGB phase (AFTP), a late thermal pulse (LTP), 
and a very late 
thermal pulse occurring in the white dwarf cooling track, \citep[e.g.,][and references therein]{wernerherwig06}.

Recently, \citet{depewetal11} reported the discovery of several more central stars of the [WR] type, which 
together with the sample reported by \citet{gornyetal04} have increased the number of known [WR]PNe central stars, 
to over a hundred. 

Some years ago, we initiated a project to obtain deep high-resolution spectroscopy of PNe around [WC] stars ([WR]PNe), 
in order to analyse the chemical behaviour of the photoionized plasma as based on collisionally excited lines 
(CELs) or on optical recombination lines (ORLs). It is known that a discrepancy in the abundances of a factor 
of two or more is commonly found  in photoionized nebulae. This discrepancy is measured through the abundance 
discrepancy factor, ADF, defined as:
\centerline{ADF(X$^{i+}$) = (X$^{i+}$/H$^+)_{\rm ORLs}$ / (X$^{i+}$/H$^+)_{\rm CELs}$,} 
where X$^{i+}$ is the $i+$ ionic abundance of element X and H$^+$ is the abundance of ionized hydrogen.

The presence of H-deficient inclusions is one of the mechanisms suggested to explain ADFs larger than one 
\citep[see e.g., ][and references therein]{liuetal06}. Since [WR] PNe are ionized by H-deficient stars whose atmospheres are almost pure He and C
\citep[i.e.,][]{koesterke01} and since they lose mass at high
rates, it seems plausible that the presence of tiny H-deficient
knots in the ionized plasma could cause large ADFs.

In a previous paper \citep{garciarojasetal09}, three [WR]PNe were analysed based on echelle data obtained with the 
same instrument as in this work: PB\,8, a young nebula around an 
uncommon [WN]/[WC] central star \citep{todtetal10}; NGC\,2867, an evolved nebula around a [WC4] star; and PB\,6, a highly 
excited [WR]PNe around a [WO2] star. Some evidences against the presence of H-deficient metal-rich knots in the PNe coming from a 
late thermal pulse event were presented, based on the low (C/O)$_{ORLs}$/(C/O)$_{CELs}$ observed in both objects with 
detected ORLs of C and O ions (PB\,8 and NGC\,2867) and the similarity between {\oii} and {\foiii} heliocentric velocities. 
These results seem to argue against a ``C-rich knots ejected in a late thermal pulse'' scenario as the origin of the observed ADFs.

In this work, we present additional high-resolution spectrophotometric data 
obtained for 12 objects, where the faint {\oii} and {\cii} ORLs have been 
detected. The main characteristics of these nebulae are presented in Table~\ref{log} where the log of 
observations also can be found. An extensive catalog of lines (hundreds of lines have been detected for each 
analysed object). For each emission line, we provide the line identification, radial velocity, and observed and dereddened line fluxes. 
In the following, we describe the observations 
and data reduction procedures (Sect. 2), the determinations of line intensities and reddening (Sect. 3), and the analysis 
of the plasma physical conditions as derived from the diagnostic line ratios (Sect. 4). In Sect. 5, we discuss the results. 
In a forthcoming paper (Garc\'{\i}a-Rojas et al., in preparation, Paper II), we discuss the abundance pattern and 
the ADFs found for these objects.

\section{Observations and data reduction}

High spectral resolution data were obtained at Las Campanas Observatory (Carnegie Institution) with the Clay 6.5-m telescope 
and the double echelle Magellan Inamori Kyocera spectrograph (MIKE) on September 2009 and June 2010. 
This spectrograph operates with two arms that allow us to obtain a blue and a red spectrum simultaneously (Berstein et al. 2003). 
The standard set of gratings were employed, providing a wavelength coverage from 3350 \AA~to 5050 \AA~in the blue and 
from 4950 \AA~to 9400 \AA~in the red. Long and short exposure time observations were carried out for each object, 
in order to ensure an appropriate signal-to-noise ratio in the faint lines and unsaturated data in the strongest lines. The log of 
observations is presented in Table 1, where we list the observing date, the exposure times 
for each object and the averaged airmass during each set of observations. For all the observed PNe, the 
slit dimensions were 1$''$ along the dispersion axis, and 5$''$ in the spatial direction.  A binning of 2$\times$2 
was used, obtaining a spacial scale of 0.2608 \arcsec/pix. As usual, series of bias, milky-flats and flats with 
the internal incandescent lamp were acquired for data reduction.  A Th-Ar lamp exposure was observed after each science exposure for 
wavelength calibrations. The spectral resolution obtained varied from 0.14 \AA~to 0.17 \AA~(about 10.8 
km s$^{-1}$) in the blue, and from 0.23 \AA~to 0.27 \AA~(about 12.8 km s$^{-1}$) in the red, as measured from the half-width half maximum 
(HWHM) of the lines of the Th-Ar comparison lamp.

As shown in Table~\ref{log}, all the objects were observed at zenith distances smaller than or about 30$^o$ (as 
recommended by the MIKE User Manual) covering airmasses between 1.004 and 1.25, thus the atmospheric refraction was 
not expected to affect the spectra.

In the September 2009 run, the seeing was about 1$''$-1.2$''$. In the June 2010 run, the seeing was better than 1$''$, being as 
good as 0.6-0.7$''$ for some hours.


\setcounter{table}{0}
\begin{table*}[t!] 
\begin{center}
\caption{Object characteristics and log of observations (slit dimensions were always 1$''$ $\times$5$''$) } 
\label{log}
\begin{tabular}{lccccrcrcc} 
\noalign{\smallskip} \noalign{\hrule} \noalign{\smallskip}
\noalign{\smallskip} \noalign{\hrule} \noalign{\smallskip}
PN G &Object & R.A & Dec& Spec T &T$_\star$ (kK)& obs. date & exp. time &Av. air mass & comments$^{\rm a}$ \\
 &  &  &  & & kK & yy-mm-dd& s,times x s &  & ref for T$_\star$ \\ 
\noalign{\smallskip} \noalign{\hrule} \noalign{\smallskip}
\hline
002.2$-$09.4	&  Cn\,1-5     & 18 29 11.7 & $-$31 29 59 & [WO4]pec & $<$57 & 09-09-09	& 30,2$\times$2700  		& 1.07 	& 1	\\
003.1+02.9    	&  Hb\,4       & 17 41 52.8 & $-$24 42 08 & [WO3]    & 86    & 10-06-05	& 60,3$\times$1500  		& 1.004	& 1	\\
300.7$-$02.0    &  He\,2-86    & 12 30 30.4 & $-$64 52 06 & [WC4]    &       & 10-06-06 & 60,3$\times$1200  		& 1.24  &	\\
004.9+04.9    	&  M\,1-25     & 17 38 30.3 & $-$22 08 39 & [WC5]    & 60    & 10-06-05	& 60,3$\times$1200  		& 1.033	& 2 	\\
355.9$-$04.2   	&  M\,1-30     & 17 52 58.9 & $-$34 38 23 & wels     &       & 10-06-06 & 60,3$\times$1200   		& 1.008 &	\\
011.9+04.2   	&  M\,1-32     & 17 56 20.0 & $-$16 29 04 & [WC4]pec & 66    & 10-06-06	& 60,3$\times$1500 (knot)  	& 1.11 	& 1	\\
019.4$-$05.3    &  M\,1-61     & 18 45 55.1 & $-$14 27 38 & wels     &       & 10-06-05 & 10,30,60,2$\times$120		& 1.12	&	\\
006.8+04.1  	&  M\,3-15     & 17 45 31.7 & $-$20 58 02 & [WC4]    & 55    & 09-09-09 & 60,600,1800,2400    		& 1.03 	& 3	\\
307.2$-$03.4   	&  NGC\,5189   & 13 33 33.0 & $-$65 58 27 & [WO1]    & 135   & 10-06-05 & 3$\times$1800,120 (knot)   	& 1.253 & 4     \\
002.4+05.8  	&  NGC\,6369   & 17 29 20.4 & $-$23 45 34 & [WO3]    & 150   & 10-06-06 & 60,3$\times$1800 (knot)  	& 1.04 	& 4     \\
336.3$-$06.9   	&  PC\,14      & 17 06 14.8 & $-$52 30 00 & [WO4]    &       & 10-06-05 & 60,3$\times$1200     		& 1.15 	&	\\
285.4+01.5  	&  Pe\,1-1     & 10 38 27.6 & $-$56 47 06 & [WO4]    &       & 10-06-05	& 60,900,2$\times$1200       	& 1.15  &	\\
\noalign{\smallskip} \noalign{\hrule} \noalign{\smallskip}
\end{tabular}
\end{center}
\begin{description}
\item[$^{\rm a}$] (1) \citet{tylendastasinska94}, (2) \citet{leuenhagenetal96}, (3) \citet{zhangkwok93}, (4) \citet{koesterkehamman97}.
\end{description}
\end{table*}

Standard data reduction procedures were performed. Two-dimensional (2D) echellograms were bias-subtracted and flat-fielded using 
{\iraf}\footnote{{\sc iraf} is distributed by the National Optical Astronomy Observatories, which is operated the 
Association of Universities for Research in Astronomy, Inc., under contract to the National Science Fundation.} 
$echelle$ reduction packages. 
We used the $apscatter$ task to substract the background-scattered light contribution in the 2D-echellograms.  
The spectra were extracted with an extraction window of 3$\farcs$72 to fit the width of the normalized flat-field windows, 
and they were wavelength-calibrated using the Th-Ar exposures. The fits to the wavelength function give an rms of about 
0.005, which translates 
into a precision of about 0.01 \AA\  in the wavelength-calibrated spectra. The data were flux-calibrated using the 
spectrophotometric standard stars Feige~110, LDS~749B, and NGC~7293 \citep{oke90}. For these objects, the slit width was 
increased to 2$''$ to maximize the stellar flux entering the slit. The estimated error in the absolute flux 
calibration was about 5\%. We estimated this error by self-calibrating the standard stars by themselves and 
comparing the computed flux with that obtained from the spectrophotometric data of these stars \citep{oke90}. 
The bluest ($\lambda$$<$3700 \AA\ ) flux calibration of the objects observed on June 6 (He\,2-86, M\,1-30, M\,1-32, NGC\,6369) 
was unreliable owing to problems deriving the stellar continuum associated with strong {\hi} absorption lines 
in the blue spectrum of the selected standard star.

\section{Line intensities and reddening correction}

\begin{figure}[!htb] 
\begin{center}
\includegraphics[width=\columnwidth]{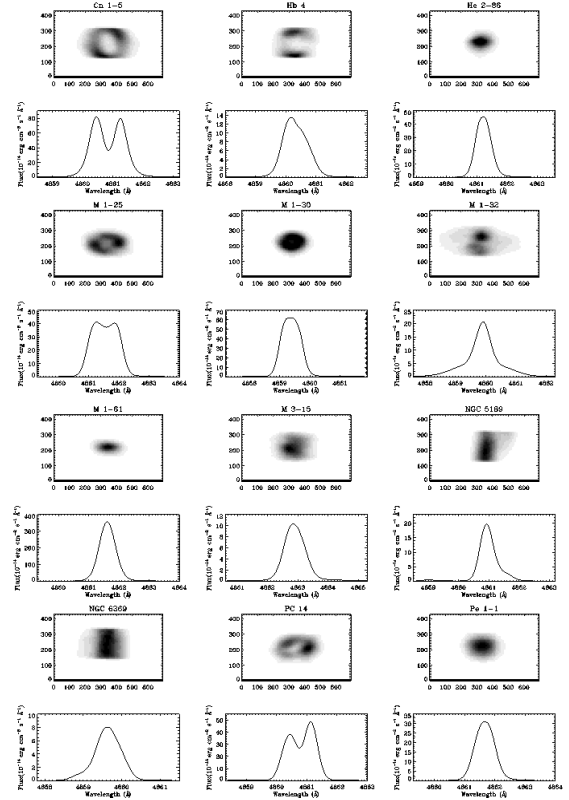}
\caption{Portion of the 2D echellograms and the extracted spectra showing the spatially resolved {\hb} 
line for all objects in our sample.}
\label{profiles}
\end{center}
\end{figure}

The {\sc splot} routine of the {\sc iraf} package was used to measure the line intensities. In general, as most of 
the lines have either a double-peak or a complex velocity structure, all the flux in the 
line was integrated between two given limits, over a local continuum estimated by eye. In the case of objects with a simple 
velocity structure showing very tight blends among nebular lines or with telluric emission lines, the analysis was performed 
via a multiple Gaussian fitting. In Fig.~\ref{profiles}, we present the {\hb} emission line as seen in the 2D-echellograms for all 
the objects in our sample, 
to show the spatially and spectroscopically resolved emission.  The extracted spectra are also shown. It can be 
seen that several 
objects have very complicated line profiles.

All the lines of a given spectrum were normalized to a particular bright emission line in the common range 
between both the blue and red spectrum. In the blue range we used {\hb}, and in the red range we used either the 
{\foiii} $\lambda$4959 line or the {\hei} $\lambda$5015 line, when the {\foiii} line was saturated in the 
long exposure. To produce a homogeneous data set of line flux 
ratios, all of them were rescaled to the {\hb} flux by using the {\foiii} $\lambda$4959/{\hb} or the {\hei} 
$\lambda$5015/{\hb} flux ratios measured in the 
blue range. Some lines that were saturated in the long exposures were measured in the short ones and rescaled 
to the {\hb} flux. 
Differences of up to 10\% between the integrated flux of the common lines, in the blue and red ranges, were measured. 
These differences are probably caused by common lines (namely {\hb}, {\foiii} 
$\lambda\lambda$4959,5007 and {\hei} $\lambda$5015) being at the red and blue extremes of the CCDs, where the 
flat-field correction might be less reliable. Nevertheless, we were always able to find a reasonable agreement between 
blue and red measured fluxes of {\foiii} $\lambda$4959 in the short exposure, with differences amounting  
to a maximum of 11\% in two objects (H\,b4 and NGC\,5189), 9\% in M\,3-15, 
and remaining below or of the order of the adopted flux calibration uncertainty (5\%) in the remaining objects. 
We do not expect the final results to be substancially affected because  only line ratios are used in our analysis. 

Owing to the small area covered by our slit and because our objects are extended, we were unable to extract a sky 
spectrum. However, taking into account the peculiar profile of the emission lines in each object, it was 
easy to distinguish telluric features from the nebular emission lines. The cases 
in which nebular emission lines were severely blended with sky telluric features are labelled in the table of 
line identifications (Table~\ref{lineid}). 


\newpage
\setcounter{table}{1}
\begin{table*}[t!] 
\caption{Extinction coefficients} 
\label{chb}
\begin{center}
\begin{tabular}{lccccc} 
\noalign{\smallskip} \noalign{\hrule} \noalign{\smallskip}
Object & c({\hb}) & {\hi} lines$^{\rm a}$ & {\te} (K), {\elecd} (cm$^{-3}$) & Literature & Ref.$^{\rm c}$ \\
\noalign{\smallskip} \noalign{\hrule} \noalign{\smallskip}
Cn\,1-5      & 0.56$\pm$0.05 & H16-H3, P25-P10 & 8700, 5000  & 0.42, 0.1 & 1,2 \\
Hb\,4	     & 1.81$\pm$0.14 & H16-H3, P25-P10 & 10000, 6700 & 1.76, 2.3 & 1,2 \\
He\,2-86     & 2.10$\pm$0.10 & H16-H3, P25-P10 & 9000, 12000 & 2.49	 & 1 \\
M\,1-25      & 1.41$\pm$0.09 & H15-H3, P25-P10  & 8000, 8000 & 1.46, 1.0, 1.42 & 1,2,3 \\
M\,1-30$^{\rm b}$& 1.00$\pm$0.08 & H25-H3, P25-P10  & 7000, 4900 & 1.01, 1.04	& 1,3 \\
M\,1-32$^{\rm b}$& 1.30$\pm$0.13 & H16-H3, P25-P10  & 9750, 9250 & 1.30, 1.9 & 1,2 \\
M\,1-61      & 1.24$\pm$0.10 & H16-H3, P25-P10  & 9000, 15000 &	1.71	& 1 \\
M\,3-15      & 2.09$\pm$0.13 & H6-H3, P25-P10  & 11000, 5000 & 2.08, 2.3, 2.12 & 1,2,3 \\
NGC\,5189    & 0.47$\pm$0.08 & H16-H3, P25-P10  & 10000, 500 & 0.44	& 1 \\
NGC\,6369$^{\rm b}$& 1.93$\pm$0.06 & H16-H3, P25-P10  & 8600, 1750 & 1.91, 1.9 & 1,2 \\
PC\,14       & 0.63$\pm$0.06 & H16-H3, P25-P10  & 10000,3000 & 0.65	& 1 \\
Pe\,1-1      & 1.80$\pm$0.09 & H16-H3, P25-P10  & 10000, 18000 & 2.16	& 1 \\
\noalign{\smallskip} \noalign{\hrule} \noalign{\smallskip}
\end{tabular}
\end{center}
\begin{description}
\item[$^{\rm a}$] H14, H10, and H8 (blended with other emission lines) and Paschen lines blended with telluric emission lines are excluded.
\item[$^{\rm b}$] H7 also excluded
\item[$^{\rm c}$] (1) \citet{girardetal07}, (2) \citet{penaetal01}, (3) \citet{gornyetal09}.
\end{description}
\end{table*}

\begin{figure}[!htb] 
\begin{center}
\includegraphics[width=\columnwidth]{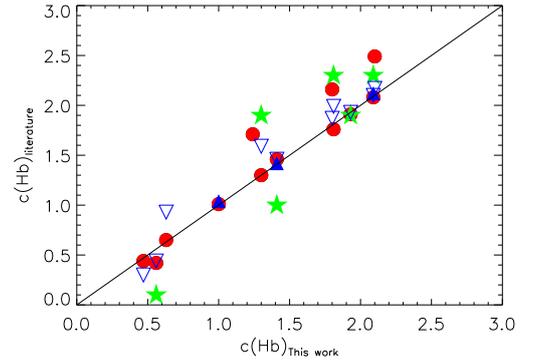}
\caption{Comparison between our derived c({\hb}) and those derived in the literature for common objects. Red filled circles: 
\citet{girardetal07}; green filled stars: \citet{penaetal01}; blue filled triangles: \citet{gornyetal09}; blue open inverse 
triangles: \citet{ackerneiner03}. It is clear that there is a good overall agreement, except for some data (see text).}
\label{figchb}
\end{center}
\end{figure}

For the reddening correction we assumed the standard extinction law for the Milky Way parametrized by \citet{seaton79},  
with $R_v$=3.1. The logarithmic redddening coefficient, c(H$\beta$), was derived in each case by fitting  the observed 
Balmer decrement, $F$({\hi} Balmer)/$F$({\hb}), and the observed $F$({\hi} Paschen)/ $F$({\hb}) to the theoretical 
values computed by \citet{storeyhummer95} for an electron  temperature, {\te}, and a density, {\elecd}, as given for 
each object by \citet{girardetal07}. 
In Table~\ref{chb}, we present the reddening coefficient, c({\hb}), the {\hi} lines used for deriving it, the physical conditions 
({\te} and {\elecd}) adopted for each case, and previous determinations found in the literature. In Fig.~\ref{figchb}, 
we can see that 
there is a good overall agreement between our derived c({\hb}) values and those reported in the literature, except 
for some objects of \citet{girardetal07} 
(red filled circles in the on-line version) and the data of \citet{penaetal01} (green stars in the on-line version) where we found 
significant differences in the derived 
c({\hb}) for most of the objects in common between both samples. 
These differences could be due to difficulties in separating the nebular from the stellar emission in the case of 
\citet{girardetal07} data, or to 
problems in the flux calibration in the zones of bluer {\hi} lines in \citet{penaetal01} data.  

Table~\ref{lineid} presents the observed and reddening corrected emission line intensities measured in all our PNe. 
A sample of Table~\ref{lineid} (available on-line only in machine readable format) is shown. 
The first column presents the adopted laboratory 
wavelength, $\lambda_0$. The second and third columns provide the ion and multiplet number or series for each line. Column 4 lists the 
observed wavelength, while column 5 shows the heliocentric radial velocity determined for the line. Columns 6, 7, and 8 present the 
observed and dereddened flux relative to H$\beta$  and the observational 1$\sigma$ error (in percentage) associated with the line 
flux. The uncertainties in the flux measurement, the flux calibration, and the error propagation in the reddening coefficient are included 
in the observational error. We provide notes in the last column.

The identification and adopted laboratory wavelength for the lines were  obtained from several previous identifications in 
the literature \citep[][and references therein]{sharpeeetal04, zhangetal05, garciarojasetal04, estebanetal04, mesadelgadoetal09, fangliu11}.

\setcounter{table}{2}
\begin{table*}[]
\begin{center}
\caption{Observed and reddening corrected line ratios (F(H$\beta$) = 100) and line identifications. (An example. The full version of 
this table is on-line) }
\label{lineid}
\begin{tabular}{lcccccccl}
\noalign{\smallskip} \noalign{\hrule} \noalign{\smallskip}
\noalign{\smallskip} \noalign{\hrule} \noalign{\smallskip}
 & & & \multicolumn{4}{c}{Cn\,1-5} \\
$\lambda_o (\AA)$  & Ion & Mult. & $\lambda_{\rm obs} (\AA)$ & V$_{\rm rad}$ (km s$^{-1}$) & F($\lambda$)/F(H$\beta$ & I($\lambda$)/I(H$\beta$ & err (\%)& notes      \\
\noalign{\smallskip} \noalign{\hrule} \noalign{\smallskip}
 3447.59 &           {\hei} &            7 & 3447.24 & -30.45 &   0.286 &   0.441 &   16 &     \\
 3478.97 &           {\hei} &           43 & 3478.76 & -18.09 &   0.169 &   0.257 &   23 &     \\
 3487.73 &           {\hei} &           42 & 3487.44 & -24.93 &   0.106 &   0.161 &   33 &     \\
 3512.51 &           {\hei} &           38 & 3512.17 & -29.03 &   0.225 &   0.338 &   18 &     \\
 3530.50 &           {\hei} &           36 & 3530.15 & -29.73 &   0.202 &   0.302 &   20 &     \\
 3554.42 &           {\hei} &           34 & 3554.05 & -31.20 &   0.232 &   0.343 &   18 &     \\
 3587.28 &           {\hei} &           32 & 3586.89 & -32.61 &   0.350 &   0.512 &   13 &     \\
 3613.64 &           {\hei} &            6 & 3613.26 & -31.52 &   0.363 &   0.525 &   13 &     \\
 3634.25 &           {\hei} &           28 & 3633.85 & -32.99 &   0.484 &   0.696 &   11 &     \\
 3669.47 &            {\hi} &          H25 & 3669.07 & -32.68 &   0.352 &   0.500 &   13 &     \\
 3671.48 &            {\hi} &          H24 & 3671.07 & -33.47 &   0.399 &   0.567 &   12 &     \\
 3673.76 &            {\hi} &          H23 & 3673.34 & -34.27 &   0.452 &   0.641 &   11 &     \\
 3676.37 &            {\hi} &          H22 & 3675.97 & -32.63 &   0.518 &   0.733 &   10 &     \\
 3679.36 &            {\hi} &          H21 & 3678.94 & -34.24 &   0.492 &   0.696 &   11 &     \\
 3682.81 &            {\hi} &          H20 & 3682.48 & -26.87 &   0.636 &   0.898 &    9 &     \\
 3686.83 &            {\hi} &          H19 & 3686.46 & -30.10 &   0.586 &   0.827 &   10 &     \\
 3691.56 &            {\hi} &          H18 & 3691.17 & -31.69 &   0.606 &   0.854 &   10 &     \\
 3694.22 &          {\neii} &            1 & 3693.80 & -34.08 &   0.124 &   0.175 &   29 &     \\
 3697.15 &            {\hi} &          H17 & 3696.75 & -32.43 &   0.745 &   1.047 &    9 &     \\
 3703.86 &            {\hi} &          H16 & 3703.47 & -31.58 &   0.914 &   1.279 &    8 &     \\
 3705.04 &           {\hei} &           25 & 3704.63 & -33.19 &   0.740 &   1.036 &    9 &     \\
 3711.97 &            {\hi} &          H15 & 3711.60 & -29.88 &   1.063 &   1.485 &    7 &     \\
 3721.83 &         {\fsiii} &           2F & 3721.42 & -33.04 &   2.386 &   3.321 &    6 &     \\
 3721.93 &            {\hi} &          H14 &       * &       * &       * &       * &    * &    \\
 3726.03 &          {\foii} &           1F & 3725.62 & -32.98 &  51.615 &  71.752 &    6 &     \\
 3728.82 &          {\foii} &           1F & 3728.37 & -36.18 &  25.956 &  36.053 &    6 &     \\
 3734.37 &            {\hi} &          H13 & 3733.97 & -32.13 &   1.610 &   2.233 &    7 &     \\
 3750.15 &            {\hi} &          H12 & 3749.72 & -34.37 &   2.195 &   3.029 &    6 &     \\
 3770.63 &            {\hi} &          H11 & 3770.22 & -32.59 &   2.675 &   3.669 &    6 &     \\
 3797.63 &         {\fsiii} &           2F & 3797.48 & -11.83 &   3.834 &   5.217 &    6 &     \\
 3797.90 &            {\hi} &          H10 &       * &       * &       * &       * &    * &    \\
 3819.61 &           {\hei} &           22 & 3819.24 & -29.05 &   1.317 &   1.781 &    7 &     \\
 3835.39 &            {\hi} &           H9 & 3834.97 & -32.83 &   5.738 &   7.725 &    6 &     \\
 3862.59 &         {\silii} &            1 & 3861.97 & -48.14 &   0.061 &   0.082 &    : &     \\
 3868.75 &        {\fneiii} &           1F & 3868.33 & -32.54 &  71.411 &  95.248 &    6 &     \\
 3871.82 &           {\hei} &           60 & 3871.35 & -36.39 &   0.086 &   0.115 &   39 &     \\
 3888.65 &           {\hei} &            2 & 3888.42 & -17.73 &  17.195 &  22.809 &    6 &     \\
 3889.05 &            {\hi} &           H8 &       * &       * &       * &       * &    * &    \\
\noalign{\smallskip} \noalign{\hrule} \noalign{\smallskip}
\end{tabular}
\end{center}
\end{table*}

We detected hundreds of lines in each object of our sample. The final set of detected and measured lines amounts to more 
than 3500 lines. Owing to the high resolution of our spectra, we were able to detect and deblend several faint lines that 
were partially blended 
with other features. In some cases, we included faint ORLs belonging to a multiplet with other clear identifications in our spectra. 
If the lines are brighter than 10\% of the principal detected component, we then included them 
in our list and labelled them as blended (e.g., lines of multiplet 2 of {\oii} blended with other features). 

Most of the lines detected in our spectra are permitted lines of {\hi}, {\hei}, and {\heii} but there also are many heavy 
element permitted lines, such as  {\ci}, {\cii}, {\ciii}, {\nitroi}, 
{\nii}, {\niii}, {\oi}, {\oii}, {\oiii}, {\nei}, {\neii}, 
{\sili}, {\silii}, {\sii}, {\siii}, {\cli}, and {\mgii}. Several of these lines are excited mainly by recombination and could 
be useful for abundance determinations (e.g., some multiplets of {\oi}, {\oii}, {\cii}, {\ciii}, {\nii}, {\neii}, and {\mgii}). 
The analysis and discussion about these lines will be presented in Paper II.

We also detected several forbidden and semi-forbidden lines from ions such as 
{\fnitroi}, {\fnii}, {\foi}, {\foii}, {\foiii}, 
{\fneiii}, {\fneiv}, {\fnev}, {\sfmgi}, {\fpii}, {\fsi}, {\fsii}, {\fsiii}, {\fclii}, {\fcliii}, {\fcliv}, {\fariii}, {\fariv}, {\farv}, 
{\fkiv}, {\fcaii}, {\fcrii}, {\fcriii}, {\fcriv}, {\fmniv}, {\fmnv}, {\ffeii}, {\ffeiii}, {\ffeiv}, 
{\fniqii}, {\fniqiii}, {\fniqiv}, {\fseii}, {\fkriii}, {\fkriv}, and {\fxeiii}. 
The abundance analysis of several of these ions will be presented in Paper II. 

In Fig.~\ref{spectra} (available only on-line), we show the complete spectra of our objects, from 3350 \AA~to 9400 \AA. 
The flux scale is such that $I$({\hb})=100.0. 

\onlfig{3}{
\begin{figure*} 
\begin{center}
\subfigure{
\includegraphics[width=18cm]{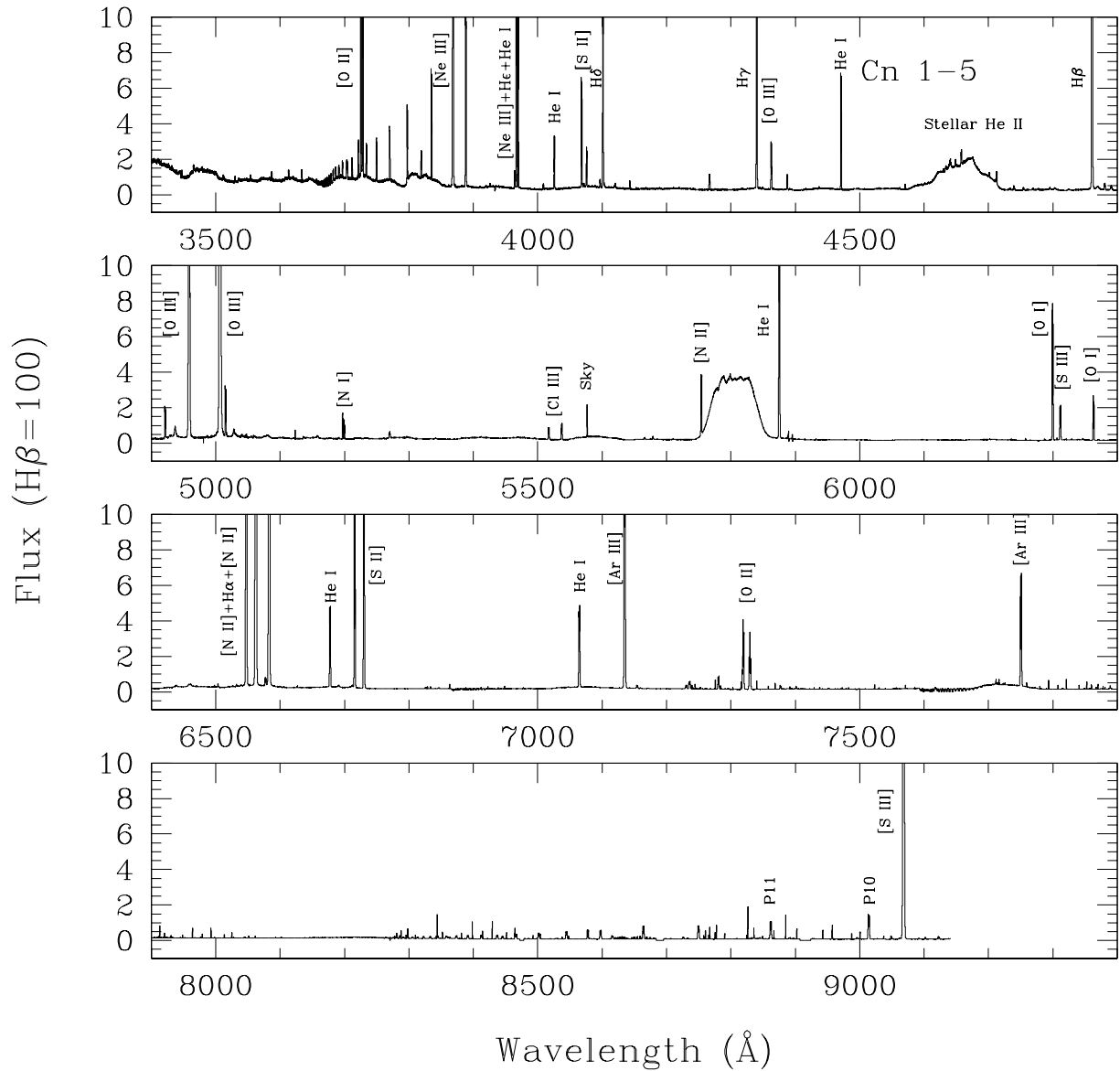}}
\caption{Complete MIKE-echelle spectra of Cn\,1-5. The most characteristic nebular emission lines have been labelled. Wide stellar WR features, 
generally stellar helium lines, are clearly visible in some spectral regions.}
\label{spectra}
\end{center}
\end{figure*}

\addtocounter{figure}{-1}
\begin{figure*}
\addtocounter{subfigure}{-1}
\begin{center}
\subfigure{
\includegraphics[width=18cm]{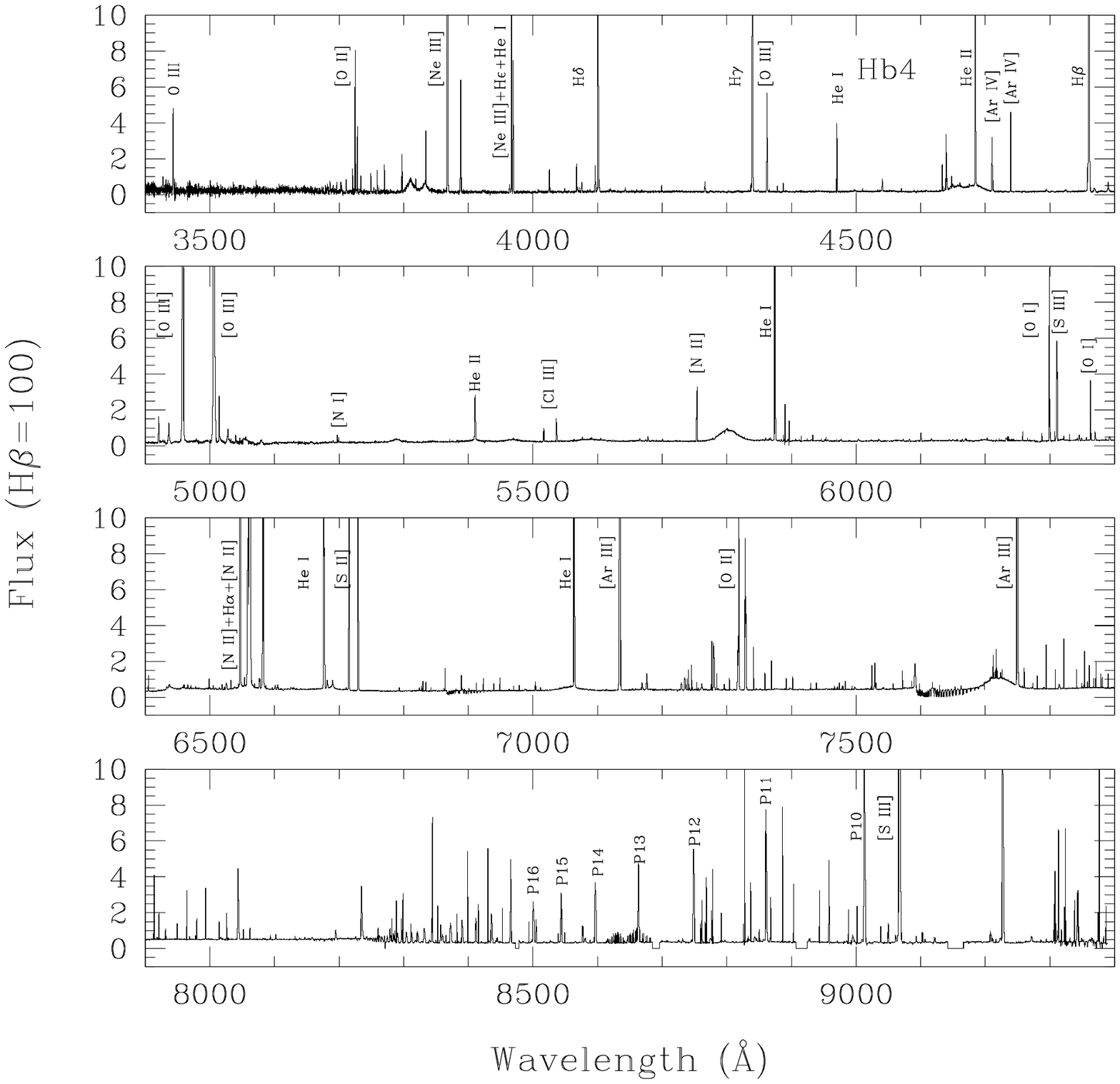}}
\caption{Complete MIKE-echelle spectra of H\,b4.}
\label{spectra}
\end{center}
\end{figure*}

\addtocounter{figure}{-1}
\begin{figure*}
\addtocounter{subfigure}{-1}
\begin{center}
\subfigure{
\includegraphics[width=18cm]{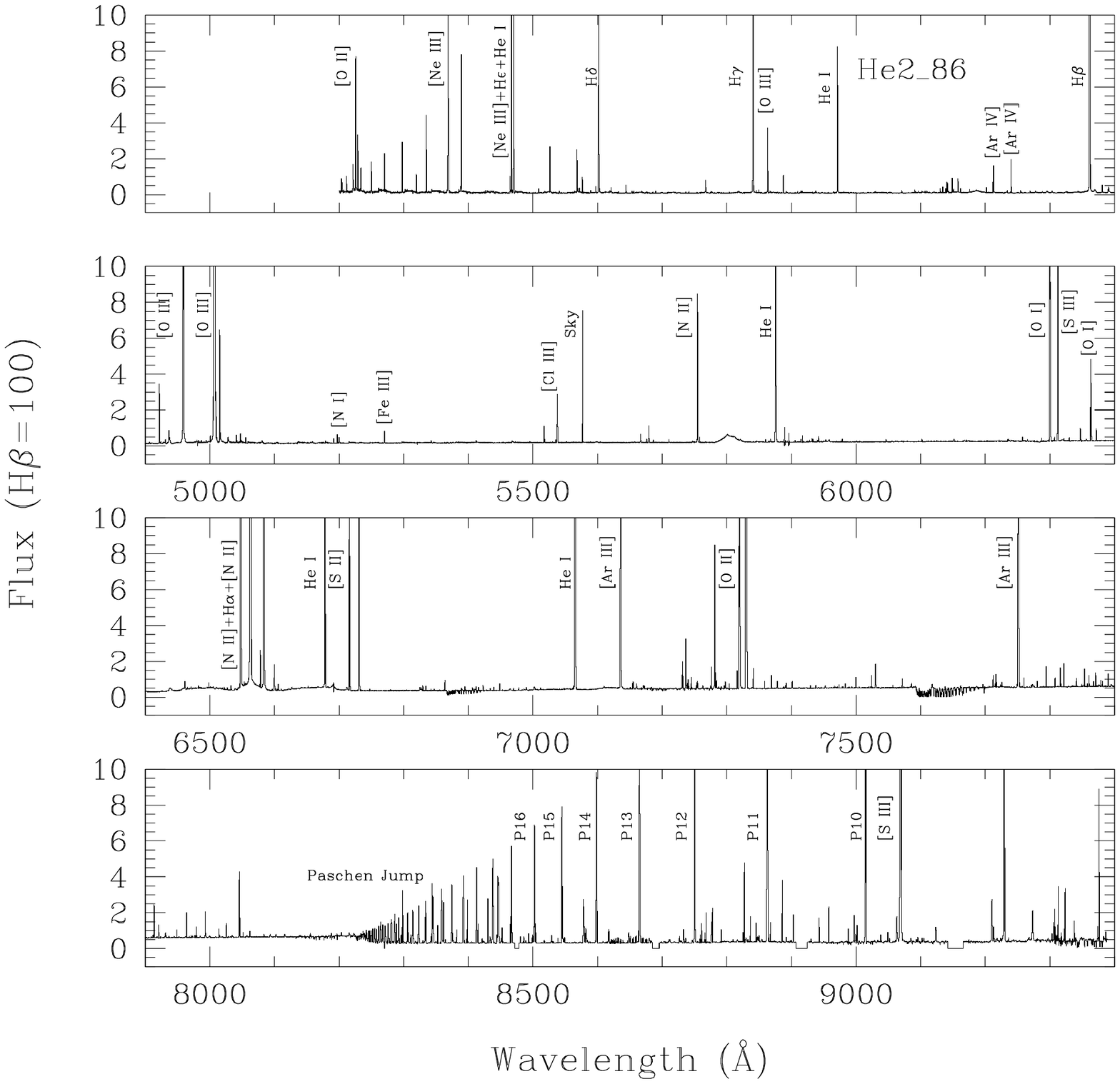}}
\caption{Complete MIKE-echelle spectra of He\,2-86.}
\label{spectra}
\end{center}
\end{figure*}

\addtocounter{figure}{-1}
\begin{figure*}
\addtocounter{subfigure}{-1}
\begin{center}
\subfigure{
\includegraphics[width=18cm]{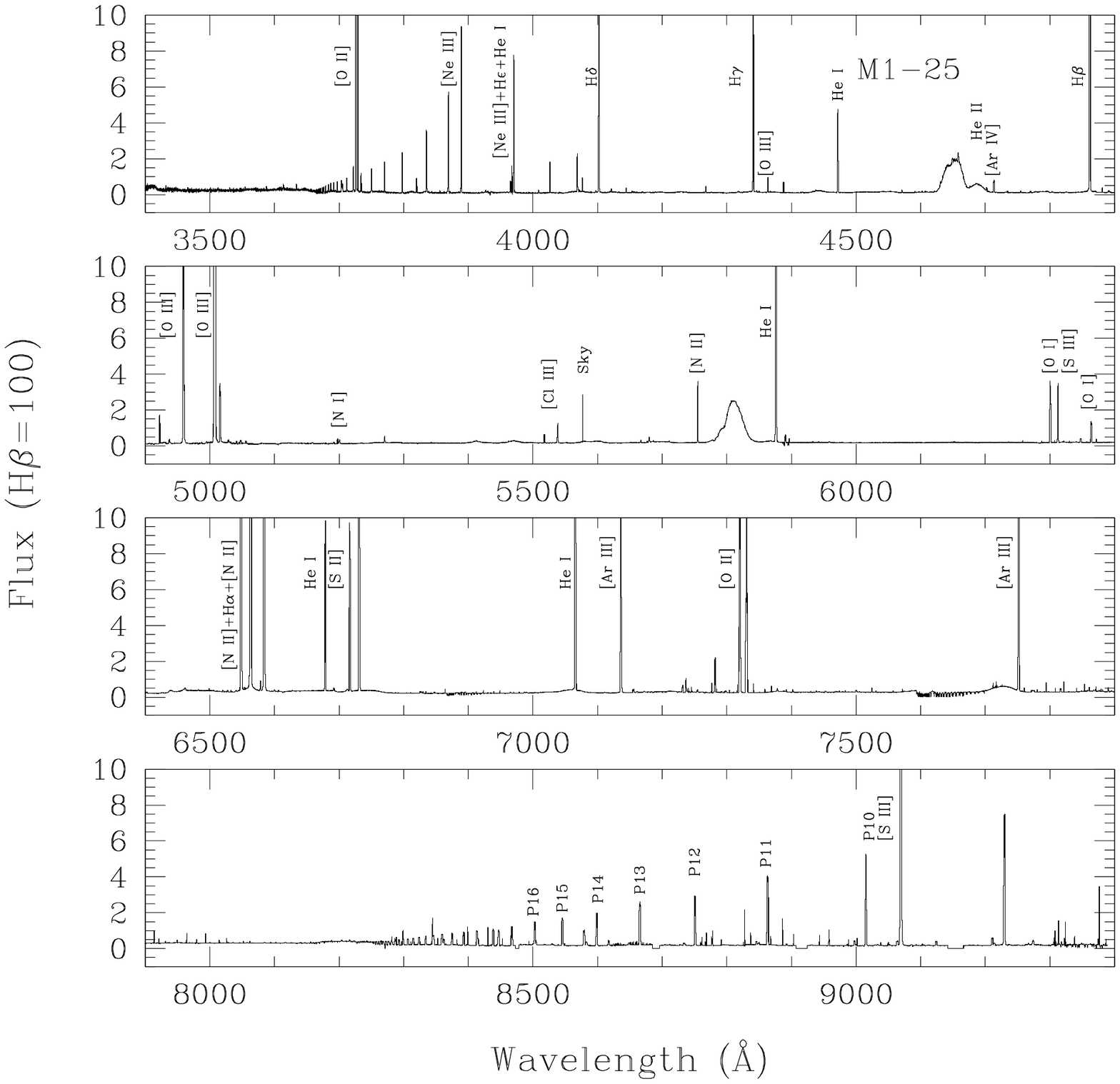}}
\caption{Complete MIKE-echelle spectra of M\,1-25.}
\label{spectra}
\end{center}
\end{figure*}

\addtocounter{figure}{-1}
\begin{figure*}
\addtocounter{subfigure}{-1}
\begin{center}
\subfigure{
\includegraphics[width=18cm]{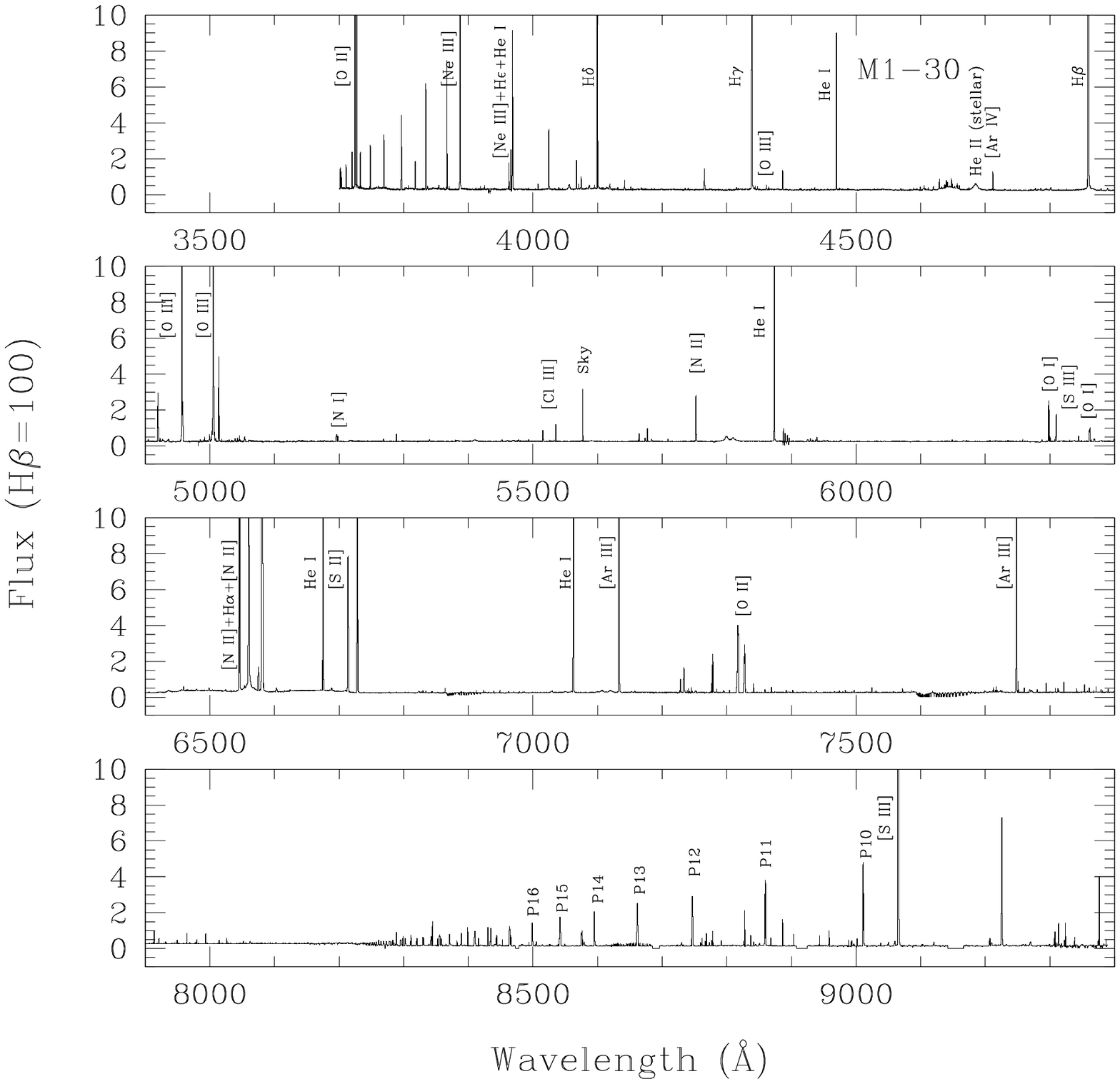}}
\caption{Complete MIKE-echelle spectra of M\,1-30.}
\label{spectra}
\end{center}
\end{figure*}

\addtocounter{figure}{-1}
\begin{figure*}
\addtocounter{subfigure}{-1}
\begin{center}
\subfigure{
\includegraphics[width=18cm]{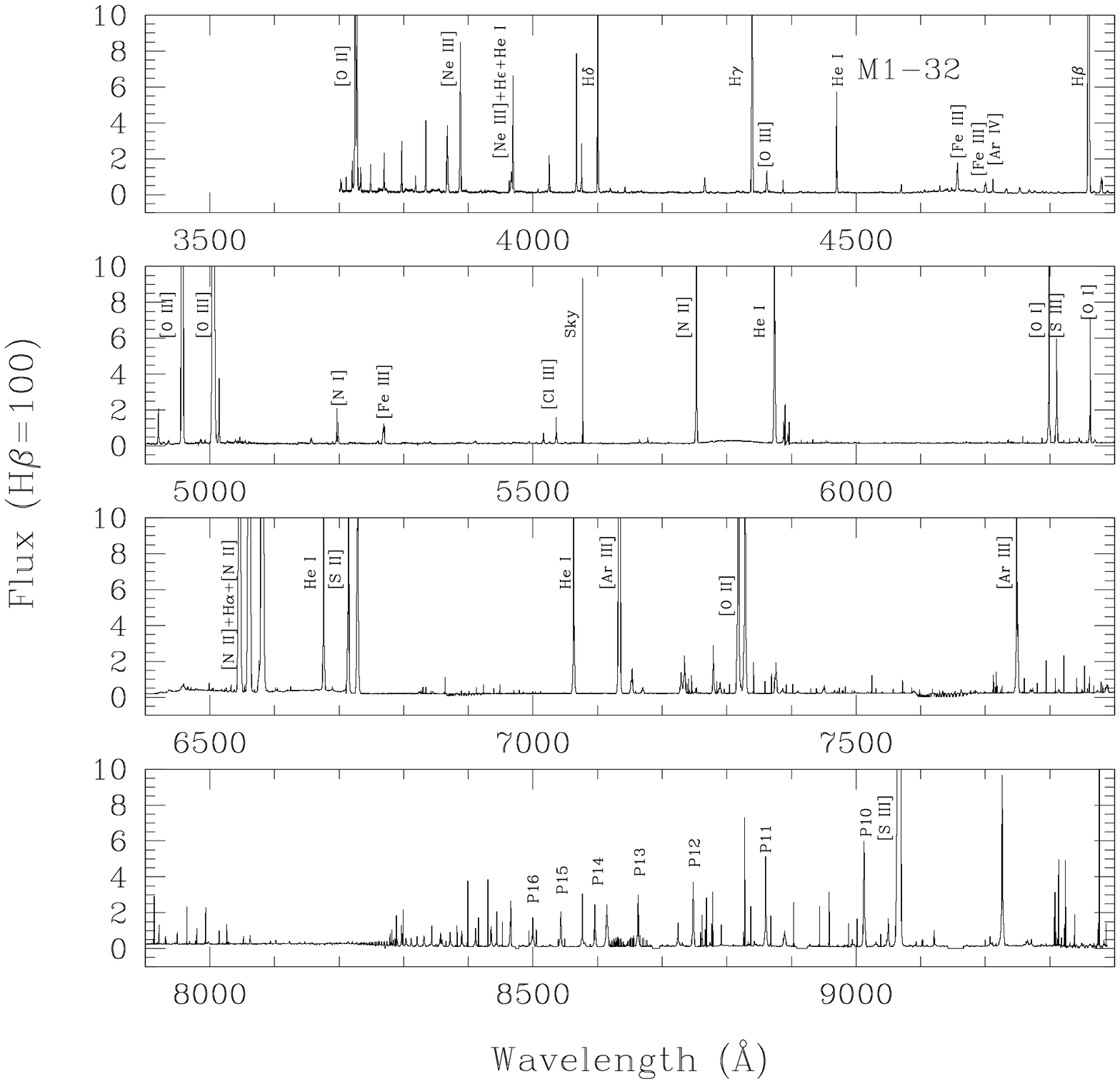}}
\caption{Complete MIKE-echelle spectra of M\,1-32.}
\label{spectra}
\end{center}
\end{figure*}

\addtocounter{figure}{-1}
\begin{figure*}
\addtocounter{subfigure}{-1}
\begin{center}
\subfigure{
\includegraphics[width=18cm]{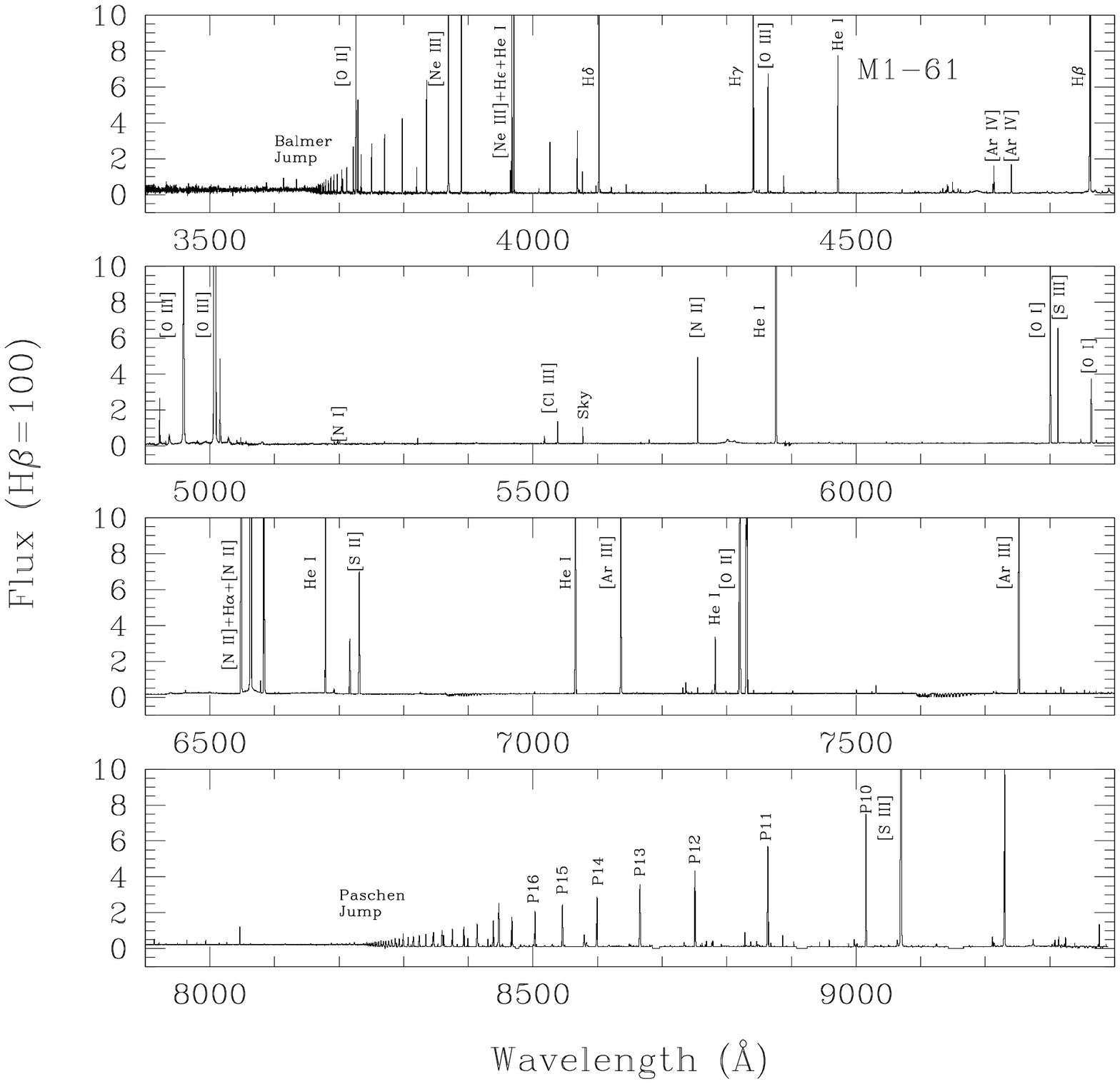}}
\caption{Complete MIKE-echelle spectra of M\,1-61.}
\label{spectra}
\end{center}
\end{figure*}

\addtocounter{figure}{-1}
\begin{figure*}
\addtocounter{subfigure}{-1}
\begin{center}
\subfigure{
\includegraphics[width=18cm]{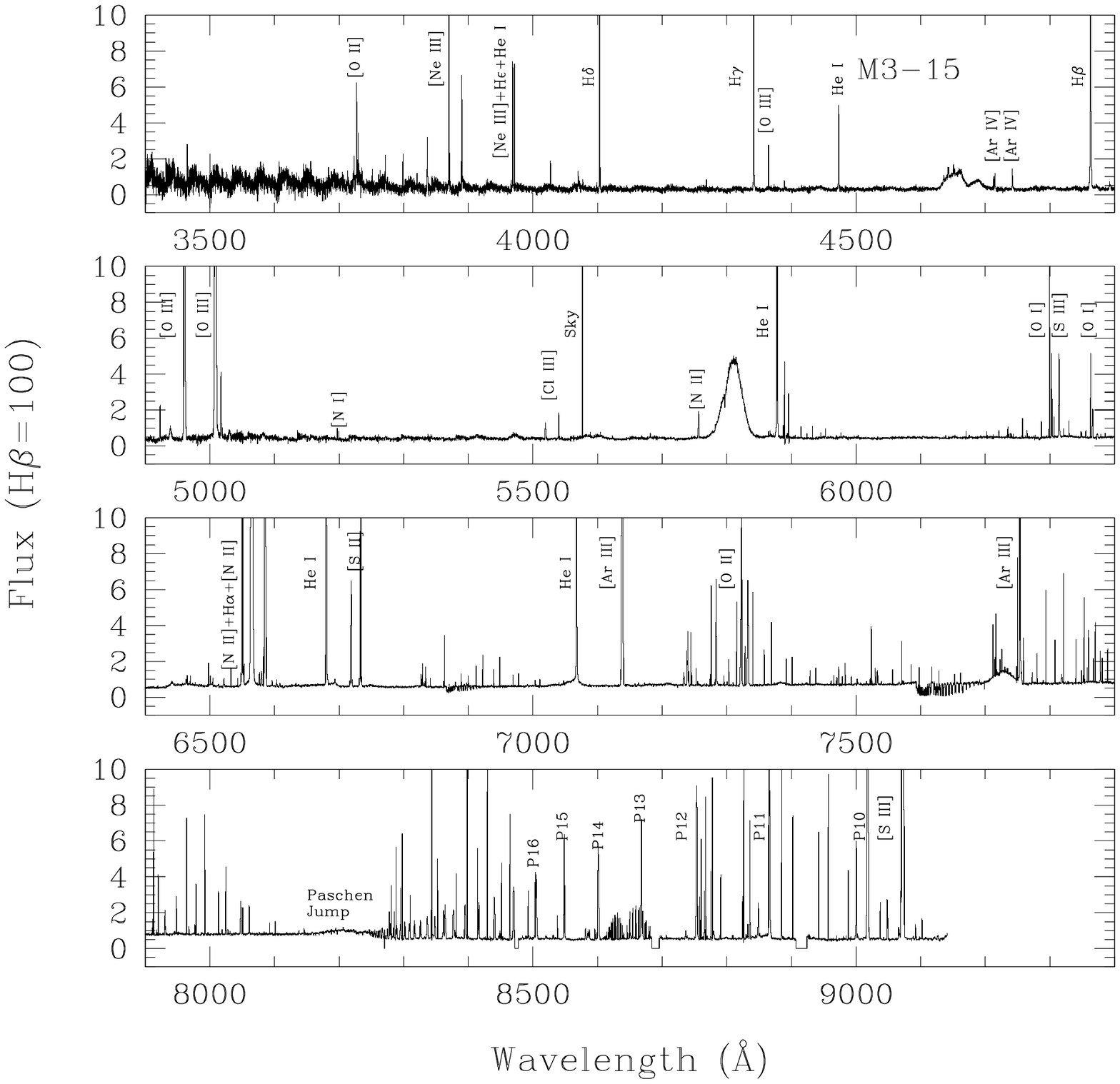}}
\caption{Complete MIKE-echelle spectra of M\,3-15.}
\label{spectra}
\end{center}
\end{figure*}

\addtocounter{figure}{-1}
\begin{figure*}
\addtocounter{subfigure}{-1}
\begin{center}
\subfigure{
\includegraphics[width=18cm]{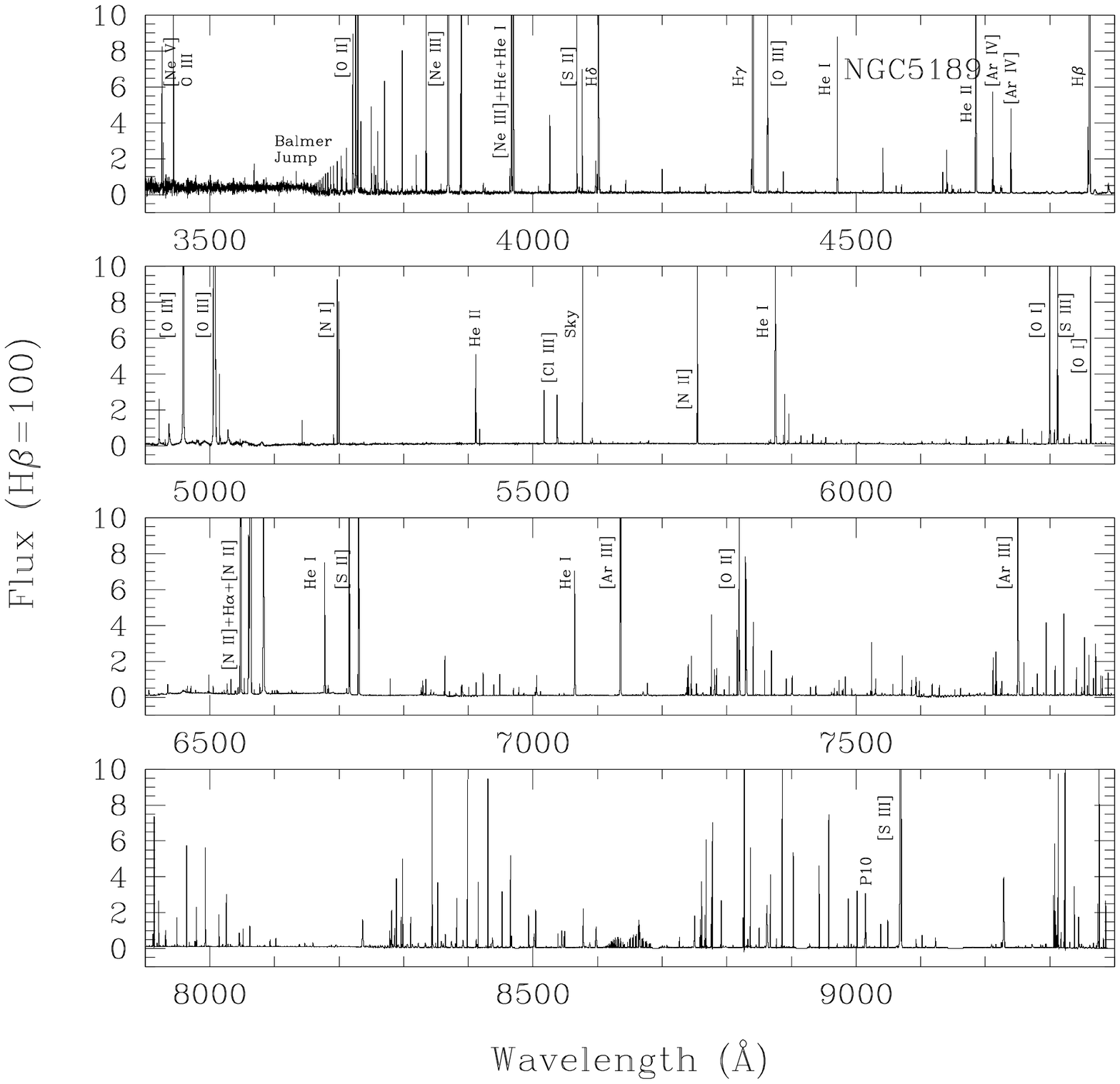}}
\caption{Complete MIKE-echelle spectra of NGC\,5189.}
\label{spectra}
\end{center}
\end{figure*}

\addtocounter{figure}{-1}
\begin{figure*}
\addtocounter{subfigure}{-1}
\begin{center}
\subfigure{
\includegraphics[width=18cm]{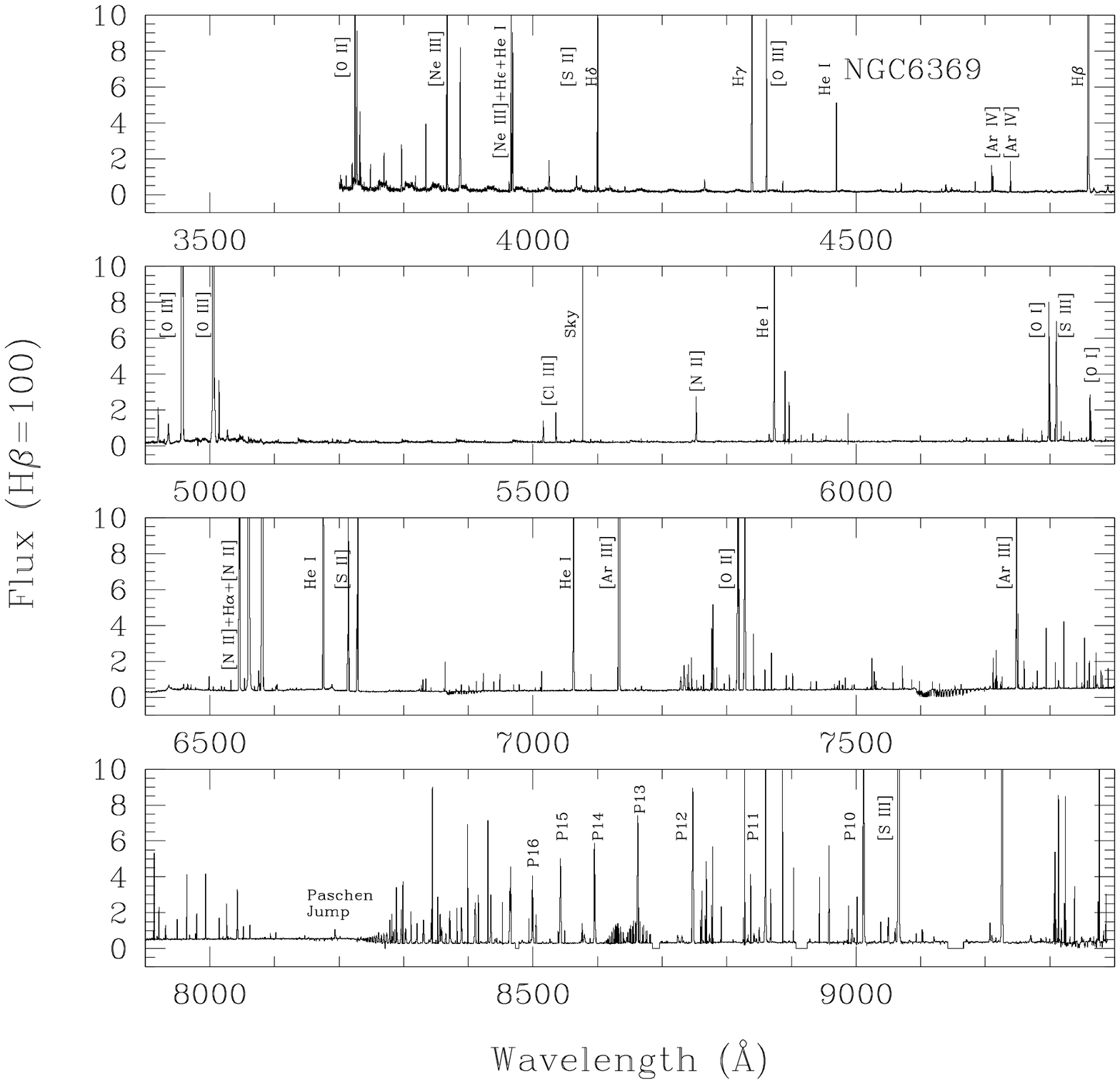}}
\caption{Complete MIKE-echelle spectra of NGC\,6369.}
\label{spectra}
\end{center}
\end{figure*}

\addtocounter{figure}{-1}
\begin{figure*}
\addtocounter{subfigure}{-1}
\begin{center}
\subfigure{
\includegraphics[width=18cm]{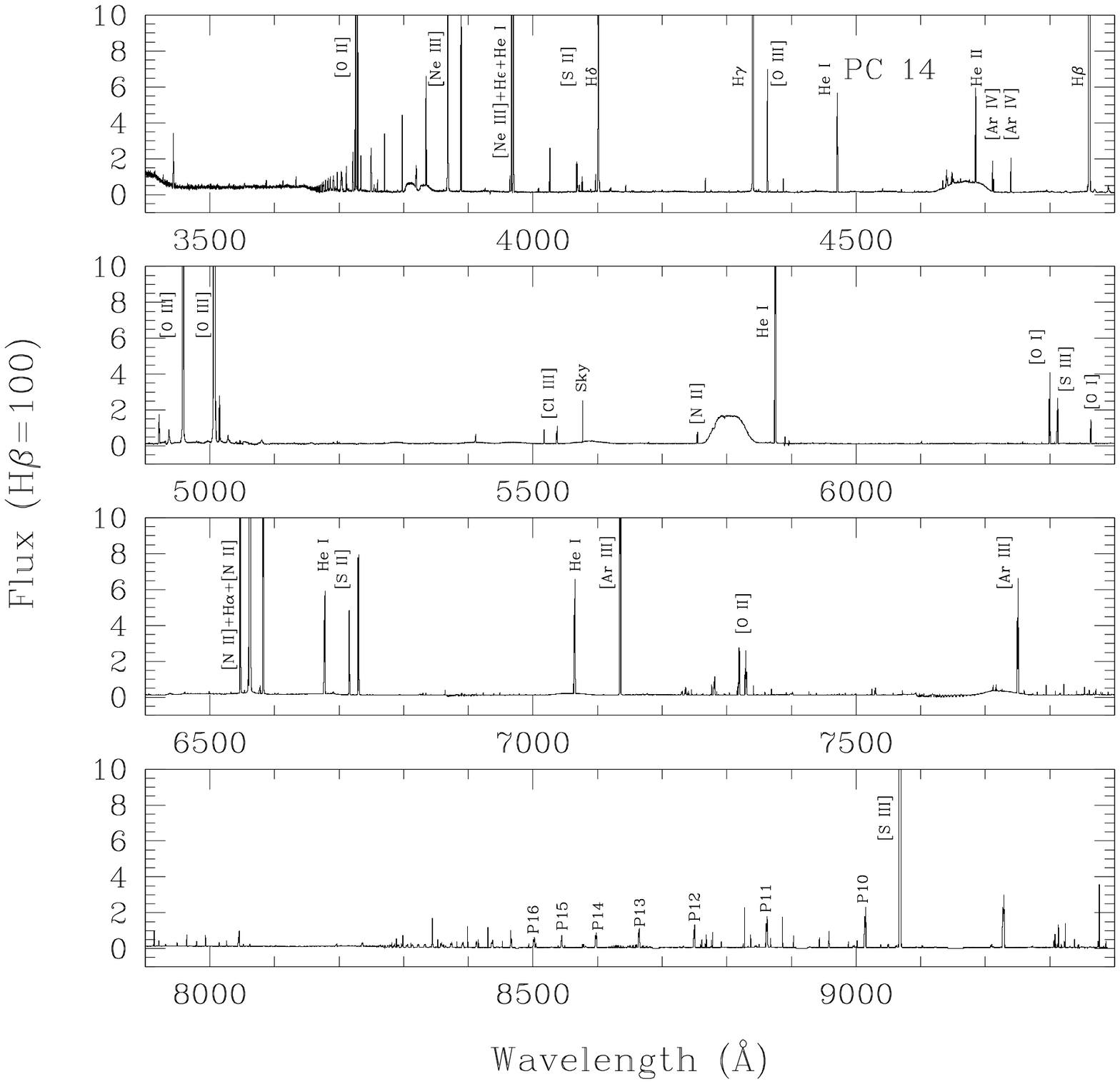}}
\caption{Complete MIKE-echelle spectra of PC14.}
\label{spectra}
\end{center}
\end{figure*}

\addtocounter{figure}{-1}
\begin{figure*}
\addtocounter{subfigure}{-1}
\begin{center}
\subfigure{
\includegraphics[width=18cm]{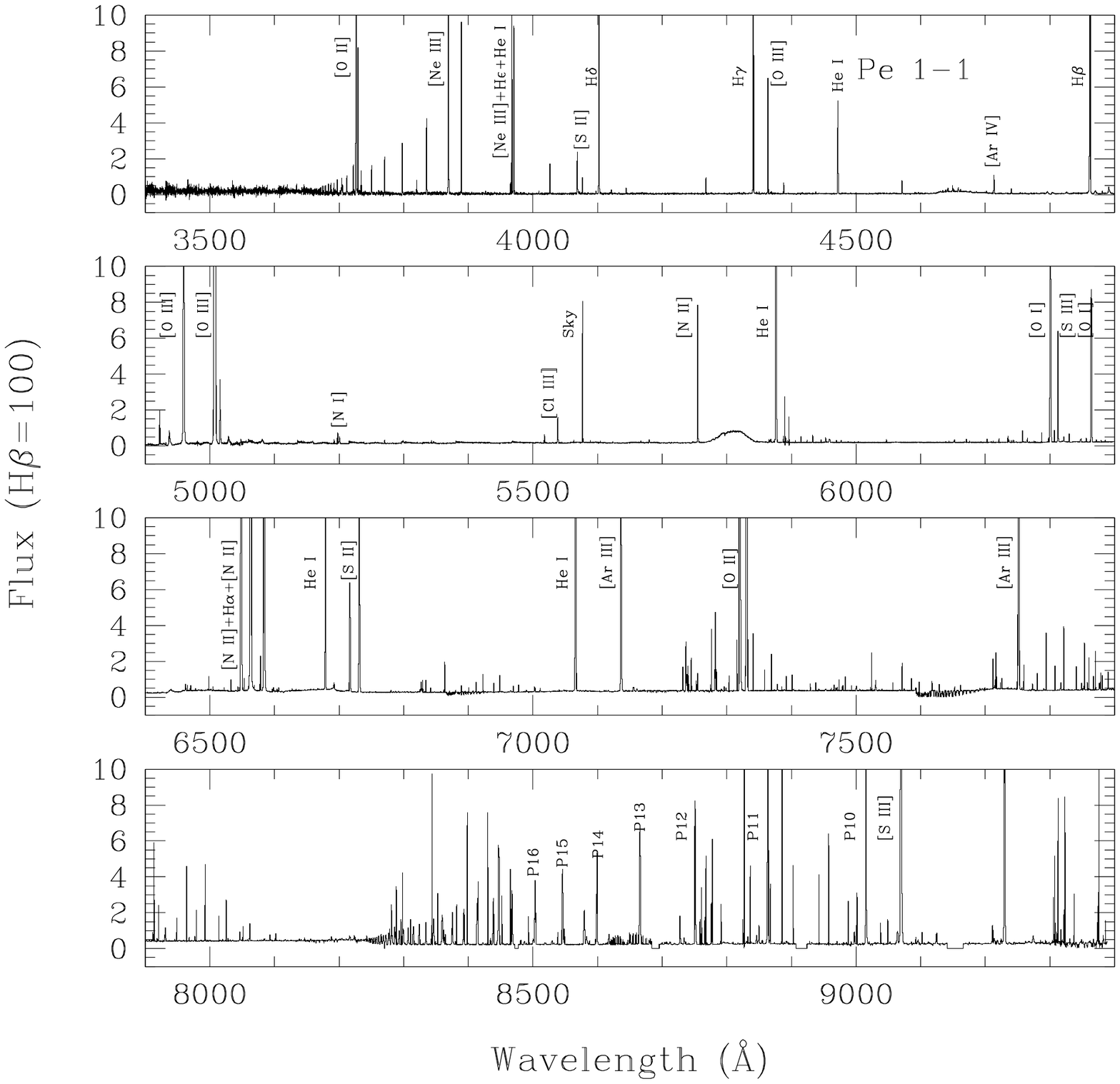}}
\caption{Complete MIKE-echelle spectra of Pe\,1-1.}
\label{spectra}
\end{center}
\end{figure*}
}

\section{Plasma diagnostic: temperatures and densities}

The large wavelength range covered by our spectra allows us to obtain a large number of diagnostic line ratios to determine the
physical conditions (electron temperature, {\te}, and density, {\elecd} ) in the plasma. The task {\sc temden} of the package {\sc nebular} 
\citep{shawdufour95}  in {\sc iraf} was used to determine T$_e$ and n$_e$. In our version of {\sc temden}, the atomic parameters 
(transition probabilities and collisional strengths) were updated. The atomic data were taken from Table 6 of  
\citet{garciarojasetal09} with the exception of the collisional strengths of {\fsii} lines that were updated to those computed by 
\citet{tayalzatsarinny10} . 

In this section we describe in detail the procedures and line ratios used to determine the plasma physical properties. 
In the following analysis, we include the data for PB~8 and NGC~2867 by \citet{garciarojasetal09} for completeness.

\subsection{Plasma diagnostics from CELs}

Electron densities were inferred 
from the {\fsii} $\lambda$6730/$\lambda$6717, {\foii} $\lambda$3726/$\lambda$3729, {\fcliii} $\lambda$5517/$\lambda$5537, 
and {\fariv} $\lambda$4740/$\lambda$4711 line ratios and, for some objects, from a set of {\ffeiii} lines (see discussion about 
this diagnostic in Sect.\ref{densities}).

In general, we considered a three-ionization-zone scheme. 
First, we assumed a {\te}=10000 K and computed the densities. These densities were then used to derive {\te} from different diagnostic ratios.
We adopted the averaged density, weighted by the errors, for the objects with a good agreement among the different density diagnostics,
within the uncertainties; in the cases with large differences among the {\fsii}, 
{\foii}, {\fcliii}, and {\fariv} nebular density diagnostics, we adopted {\elecd}({\fcliii}) as representative of 
the low-medium ionization zones (see Sect.~\ref{discuss}). 
For three objects (Cn\,1-5, He\,2-86, and M\,1-61), we adopted {\elecd}({\fariv}) as representative of the highest ionization zone.

For the low ionization zone of each 
object, the representative {\te} was computed from the line ratios 
{\fsii} ($\lambda\lambda$6716+31)/($\lambda\lambda$4069+76), {\foii} ($\lambda\lambda$3726+29)/($\lambda\lambda$7320+30), and 
{\fnii} ($\lambda\lambda$6548+83)/($\lambda$5755). For the higher ionization zone, we computed {\te} from   
{\foiii} ($\lambda$4959+$\lambda$5007)/($\lambda$4363), {\fsiii} ($\lambda\lambda$9069+530)/($\lambda$6312), and 
{\fariii} ($\lambda\lambda$7136+751)/($\lambda$5192) line ratios.  
The {\fariv} ($\lambda\lambda$4711+40)/($\lambda\lambda$7170+263) line ratio was used to compute {\te}({\fariv}), which is 
representative of the highest ionization zones in the nebulae. 
These temperatures were used to compute densities and we then iterated the process until convergence. 

\subsubsection{ The effect of recombination on some diagnostic line ratios}

To obtain {\te}({\foii}), it is necessary to subtract the contribution to {\foii} $\lambda$$\lambda$7320+7330 by 
recombination. \citet{liuetal00} found that this contribution can be fitted in the range 0.5$\le${\te}/10$^4$$\le$1.0 by

\begin{equation}
\frac{I_R(7320+7330)}{I({\rm H\beta})}
= 9.36\times(T_4)^{0.44} \times \frac{{\rm{O}}^{++}}{{\rm{H}}^+},
\end{equation}
where $T_4$=$T_e$/10$^4$. The importance of this contribution is relative. For instance, for PB~8,  
by assuming the values derived for O$^{++}$/H$^+$ from ORLs and using this equation,  
\citet{garciarojasetal09} estimated a contribution to the observed line intensities of approximately 59\% by recombination.
This yielded a {\te}({\foii})=7050 K, which was 4350 K lower than those derived without taking into account this 
contribution. 
For the objects in this work, we computed the recombination contribution to the {\foii} lines by adopting the abundance 
derived from multiplet 1 {\oii} ORLs. For all of these objects,   
the correction of {\te}({\foii}) was lower than 2000 K with a median of about 500 K.

\citet{liuetal00} found that the contribution to the intensity of the $\lambda$5755 
{\fnii} line by recombination can also be estimated from
\begin{equation}
\frac{I_R(5755)}{I({\rm H\beta})}
= 3.19\times(T_4)^{0.30} \times \frac{{\rm N}^{++}}{{\rm H}^+},
\end{equation}
in the range 0.5$\le$ $T_4$$\le$2.0. We used this expression to calculate the recombination contribution to $\lambda$5755. 
We derived the N$^{++}$/H$^+$ abundance, to a first approximation,  by assuming that 
N$^{++}$/H$^+$=O$^{++}$/H$^+$ $\times$ N$^{+}$/O$^{+}$ where 
O$^{++}$/H$^+$ is the value derived from recombination lines and N$^+$/O$^+$ is the fraction derived from CELs. 
In all the objects of our sample, the correction for {\te}({\fnii}) was smaller than 1500 K, with a median of about 550 K.
The objects with the largest corrections are Hb\,4 and M\,3-15, whose temperature decreased by 1300 K and 1500 K, respectively. On the other 
hand, M\,1-25, M\,1-30, and M\,1-32 are almost unaffected and are corrected by only about 50$-$150 K.

In several objects of our sample, we detected bright {\heii} lines. Taking into account the similarity between the ionization potentials 
of He$^{++}$ and O$^{3+}$, we expected to find a significant amount of oxygen in the form of O$^{3+}$, 
so we had to take into account the contribution of recombination  to the auroral {\foiii} $\lambda$4363 line, which we 
estimated using equation 3 of \citet{liuetal00}

\begin{equation}
\frac{I_R(4363)}{I({\rm H\beta})}
= 12.4\times(T_4)^{0.79} \times \frac{{\rm O}^{3+}}{{\rm H}^+}.
\end{equation}
We were unable to compute directly the O$^{3+}$ abundance from either CELs or ORLs, 
but we can estimate it from helium ionic abundances using ${\rm O}^{3+}/{ \rm H}^+ = [({\rm He}/{\rm He}^+)^{2/3} -1] \times 
({\rm O}^+/{\rm H}^+ +{\rm O}^{++}/{\rm H}^+ )$ \citep{kingsburghbarlow94}.  
With this expression, we found that the contribution of recombination to the {\foiii} 
$\lambda$4363 line is smaller than $\sim$3\% for the most ionized objects in our sample, which has negligible effects on 
the determination of {\te}({\foiii}).

The temperatures and densities derived from the various diagnostic line ratios  and their respective 1$\sigma$ errors, for each object 
in our sample, are presented in Table~\ref{plasma}. 


\setcounter{table}{3}
\begin{table*}
\tiny
\caption{Plasma diagnostics.}
\label{plasma}
\begin{tabular}{llccccccc}
\noalign{\smallskip} \noalign{\smallskip} \noalign{\hrule} \noalign{\smallskip}
           Parameter &                                  Line ratio &            Cn1-5 &              Hb4 &           He2-86 &            M1-25 &            M1-30 &            M1-32 &            M1-61\\
\noalign{\smallskip} \noalign{\hrule} \noalign{\smallskip}
{\elecd} (cm$^{-3}$) &                                    {\foii}  &  3450$\pm$   930 &  4900$\pm$  1770 &  7430$\pm$  3420 &  6650$\pm$  2940 &  2950$\pm$   800 &  5370$\pm$  2180 & 16040:          \\
                     &                                    {\fsii}  &  3780$\pm$  1400 &  5760:		 & 15450: 	    &  7740:		&  5180$\pm$  2770 &  8350: 	     & 20810:	       \\
		     &                   {\foii}$_{na}$$^{\rm c}$  &     \nodata      & 12300:$^{\rm d}$ & 25100$\pm$  2600 & 13150$\pm$  1320 &  7150$\pm$   700 & 19780$\pm$  2600 & 28050$\pm$  3750\\
		     &                   {\fsii}$_{na}$$^{\rm e}$  &  3775$\pm$   350 &  7700$\pm$  1100 & 18000$\pm$  1750 &  8900$\pm$   850 &  7250$\pm$   275 &  9180$\pm$  1120 & 20400$\pm$  2000\\
                     &                                  {\fcliii}  &  4320$\pm$   900 &  7360$\pm$  1480 & 23280$\pm$  4360 & 15100$\pm$  2600 &  8100$\pm$  1400 & 14800$\pm$  3300 & 22200$\pm$  4500\\
		     &                                  {\ffeiii}  & 13100$\pm$  6150 &      \nodata     & 30950$\pm$  6500 & 16500$\pm$ 11900 &     \nodata      & 16700$\pm$  3350 & 94100$\pm$  8000\\
                     &                                   {\fariv}  & 10050$\pm$  5700 &  7400$\pm$  1300 & 35810$\pm$  5990 &	  \nodata      &     \nodata      &	\nodata      & 33590$\pm$  8070\\
                     &                                             &		      & 		 &		    &		       &		  &		     &  	       \\
     $T_{\rm e}$ (K) &                          {\fnii}$^{\rm b}$  &  8650$\pm$   230 &  8600$\pm$   400 &  9300$\pm$	400 &  7720$\pm$   250 &  6560$\pm$   160 &  8350$\pm$   360 & 11800$\pm$   600\\
                     &       			{\foii}$^{\rm c}$  &     \nodata      & 12950:$^{\rm d}$ &  9700$\pm$   600 &  7300$\pm$   300 &  6300$\pm$   200 &  9700$\pm$   730 & 14350$\pm$  1880\\
                     &                          {\fsii}$^{\rm e}$  &  8380$\pm$   460 & 10000$\pm$  1200 &  7630$\pm$   500 &  5750$\pm$   250 &  6230$\pm$   250 &  6300$\pm$   370 & 10730$\pm$  1160\\
                     &                                             &		      & 		 &		    &		       &		  &		     &  	       \\
                     &                                   {\foiii}  &  8780$\pm$   160 &  9950$\pm$   240 &  8420$\pm$   150 &  7800$\pm$   150 &  6600$\pm$   150 &  9430$\pm$   220 &  9170$\pm$   180\\
                     &                                   {\fsiii}  &  8900$\pm$   310 &  9350$\pm$   750 &  9000$\pm$   500 &  7900$\pm$   360 &  6270$\pm$   200 &  8270$\pm$   540 &  9900$\pm$   620\\
                     &                                  {\fariii}  &  7830$\pm$   350 &  8240$\pm$   530 &  8950$\pm$   300 &     \nodata      &  6310$\pm$   240 &  7960$\pm$   640 &  8690$\pm$   360\\
                     &                                             &		      & 		 &		    &		       &		  &		     &  	       \\
                     &                                   {\fariv}  &	 \nodata      & 13200$\pm$  1050 &  7700$\pm$	400 &	  \nodata      &     \nodata      &	\nodata      &  8110$\pm$   690\\
                     &                                             &                  &                  &                  &                  &                  &                  &                 \\
                     &                           {\hei}$^{\rm f}$  &  7800$\pm$   140 &  7850$\pm$   220 &  7660$\pm$   140 &  7140$\pm$   140 &  5830$\pm$   110 &  7230$\pm$   250 &  8530$\pm$   170\\
                     &                           {\hei}$^{\rm g}$  &  5800$\pm$   550 &  7250$\pm$  1750 &  6850$\pm$   900 &  6400$\pm$   800 &  5700$\pm$   650 &  7300$\pm$  1050 &  7750$\pm$  1050\\
                     &                          Paschen decrement  &        \nodata   &  12450$^{+5700}_{-3700}$&  7560$^{+3050}_{-2060}$&7750$^{+3100}_{-2050}$&  5800$^{+2150}_{-1300}$ &  8000$^{+3200}_{-2100}$&  9800$^{+4150}_{-2700}$\\
\noalign{\smallskip} \noalign{\smallskip} \noalign{\hrule} \noalign{\smallskip}
           Parameter &                                 Line ratio &            M3-15 &          NGC5189 &          NGC6369 &             PC14 &            Pe1-1 &    PB8$^{\rm a}$ & NGC2867$^{\rm a}$\\
\noalign{\smallskip} \noalign{\hrule} \noalign{\smallskip}
{\elecd} (cm$^{-3}$) &                                    {\foii} &  7470:           &   900$\pm$   180 &  3020$\pm$   720 &  3050$\pm$   760 & 13160$\pm$ 10880 &  2650$\pm$   750 &  2675$\pm$   680\\
                     &                                    {\fsii} &  5660:  	     &   950$\pm$   240 &  3550$\pm$  1130 &  3080$\pm$   980 & 14000: 	 	 &  2450$\pm$  1000 &  3000$\pm$  1000\\
		     & 		        {\foii}$_{na}$$^{\rm c}$  &	  \nodata    &  1950$\pm$   260 &  5680$\pm$   520 &  4050$\pm$   400 & 29000$\pm$  2800 &  4950$\pm$   300 &     \nodata     \\
		     & 			{\fsii}$_{na}$$^{\rm e}$  &  7000$\pm$  1850 &  1550$\pm$   200 &     \nodata      &  3550$\pm$   350 & 17900$\pm$  1700 &     \nodata      &     \nodata     \\
                     &                                  {\fcliii} & 10250$\pm$  1730 &  1320$\pm$   520 &  4640$\pm$  1240 &  3850$\pm$   870 & 31360$\pm$  7800 &  2400$\pm$  1800 &  4750$\pm$  1150\\
                     &                                  {\ffeiii} &     \nodata      &     \nodata      &     \nodata      &  8700:           &    \nodata       &     \nodata      &     \nodata     \\      
                     &                                   {\fariv} &  7680$\pm$  4350 &  1290$\pm$   680 &  5000$\pm$  1240 &  4700$\pm$  1060 & 40950:           &     $<$6850      &  3900$\pm$ 1000\\
                     &                                            &                  &                  &                  &                  &                  &                  &                 \\
     $T_{\rm e}$ (K) &                         {\fnii}$^{\rm b}$  &  9500$\pm$   450 &  9050$\pm$   300 & 13380$\pm$   630 &  9800$\pm$   330 & 10100$\pm$   450 &  8900$\pm$   500 & 11750$\pm$   400\\
                     &                         {\foii}$^{\rm c}$  &     \nodata      & 11350$\pm$   600 & 18500:$^{\rm d}$ & 10400$\pm$   470 &  9640$\pm$   550 &  7050$\pm$   400 &     \nodata     \\
                     &                         {\fsii}$^{\rm e}$  &  8100$\pm$  1400 & 11100$\pm$   800 &     \nodata      &  9800$\pm$   650 &	 6700$\pm$350    &     \nodata      &     \nodata     \\
                     &                                            &  	   	     &		        &		   & 		      &		         &		    &		      \\
                     &                                  {\foiii}  &  8350$\pm$   230 & 11600$\pm$   280 & 10650$\pm$   230 &  9300$\pm$   180 &  9980$\pm$  220  &  6900$\pm$   150 & 11725$\pm$   300\\
                     &                                  {\fsiii}  &  8630$\pm$   600 & 10720$\pm$   600 & 10370$\pm$   480 &  9000$\pm$   350 &  9620$\pm$  540  &     \nodata      &     \nodata     \\
                     &                                 {\fariii}  &     \nodata      & 10100$\pm$   370 &  9100$\pm$   750 &  8750$\pm$   530 &  9170$\pm$  430  &     \nodata      & 11100$\pm$   700\\
                     &                                            &  		     &		        &		   & 		      &		         &	            &		      \\
                     &                                  {\fariv}  &     \nodata      & 17000$\pm$  1300 & 13560$\pm$  1440 & 12700$\pm$  1700 &   \nodata        &     \nodata      &     \nodata     \\
                     &                                            &                  &                  &                  &                  &                  &                  &                 \\
                     &                           {\hei}$^{\rm f}$ &  7320$\pm$   224 &  9200$\pm$   200 &  9880$\pm$   230 &  8420$\pm$   170 &  9000$\pm$  200  &  6250$\pm$   150 & 10600$\pm$   400\\
                     &                           {\hei}$^{\rm g}$ &  7400$\pm$  1300 &  8450$\pm$  1700 &  8300$\pm$   950 &  6600$\pm$   950 &  8550$\pm$ 1050  &  5850$\pm$   750 &  7450$\pm$  1000\\
                     &                         Paschen decrement  &  9800$^{+4150}_{-2800}$   & 9200$^{+3600}_{-2300}$ &  11350$^{+5000}_{-3200}$&  8500$^{+3250}_{-2050}$&10300$^{+4500}_{-2950}$&  5100$^{+1300}_{-900}$ &  8950$^{+2900}_{-1900}$\\
\noalign{\smallskip} \noalign{\hrule} \noalign{\smallskip}
\end{tabular}
\begin{description}
\item[$^{\rm a}$] Data from \citet{garciarojasetal09}.
\item[$^{\rm b}$] Corrected for recombination contribution to {\fnii} $\lambda$5755 line (see text).
\item[$^{\rm c}$] Corrected for recombination contribution to {\foii} $\lambda$$\lambda$7320+30 lines (see text).
\item[$^{\rm d}$] {\foii} $\lambda$$\lambda$7320+30 lines affected by telluric emission lines.
\item[$^{\rm e}$] {\fsii} $\lambda$$\Lambda$4068.60, 4076.35 lines corrected for the contributions of 
{\oii} $\lambda\lambda$4069.62+4069.86 lines and/or {\ciii} $\lambda$4067.94 line and/or {\oii} $\lambda$4075.86, respectively.
\item[$^{\rm f}$] $T_e$({\hei}) derived using the method developed by \citet{apeimbertetal02}.
\item[$^{\rm g}$] $T_e$({\hei}) derived using the method of \citet{zhangetal05}.
\end{description}
\end{table*}

\subsection{Plasma diagnostics from ORLs and continua}

\subsubsection{Electron temperatures from Paschen discontinuity}

\begin{figure}[] 
\begin{center}
\includegraphics[width=8cm]{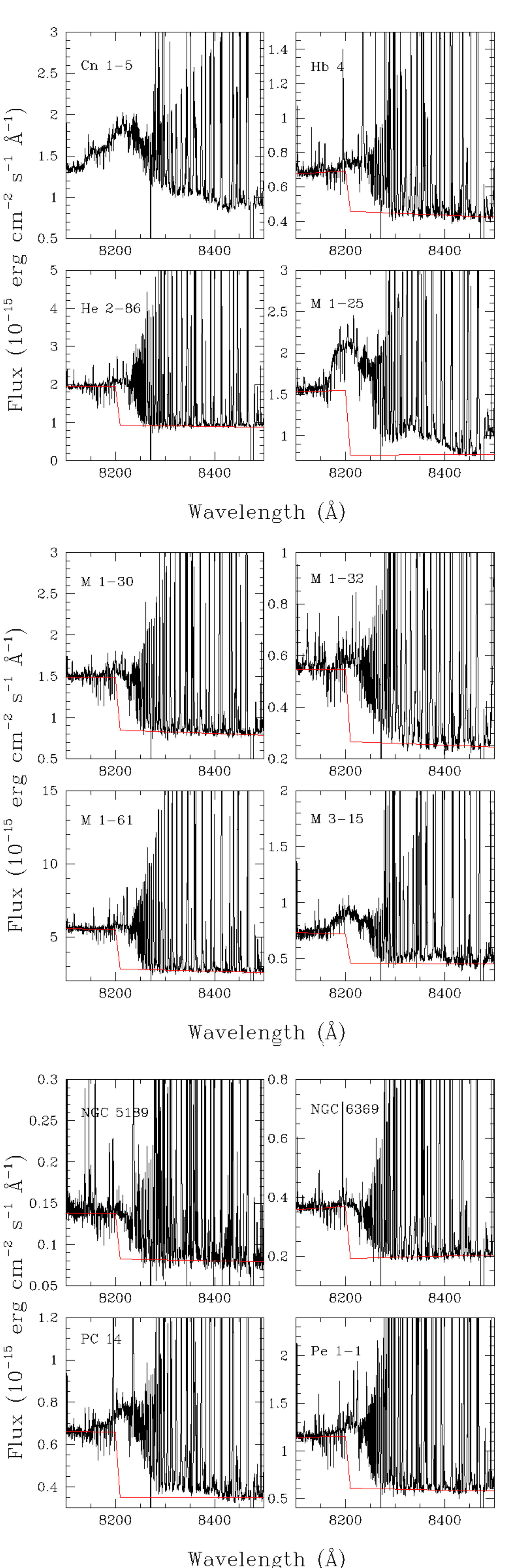}
\caption{Section of the echelle spectra showing the Paschen discontinuity in all the objects of our sample. In several 
objects, we can clearly see the wide stellar emission line that contaminates the spectrum in the blue part of 
the discontinuity }
\label{paschen}
\end{center}
\end{figure}

Electron temperature can be derived from the ratio of the Balmer or Paschen discontinuities to the {\hi} lines belonging 
to the Balmer or Paschen series, respectively. The high resolution of our spectra allows us to measure the 
continuum flux in zones very close to 
the discontinuities. In principle, this technique would allow us to establish whether other continuum contributions 
apart from pure nebular processes 
affect the observed spectrum. However, we have to handle continuum measurements in the {\hi} discontinuities 
with care, because in most of 
our spectra the central star  was included in the slit, and [WC] CS show in their spectra prominently wide emission 
lines that can affect the continuum. Balmer discontinuities are indeed strongly affected by intense  wide stellar {\hei} 
emission lines, hence 
are unsuitable for computing temperatures. In the objects where this effect was negligible, the signal-to-noise ratio 
was too low to derive  {\te} with any precision. 
In the Paschen discontinuity, the effect is smaller, but non-negligible. In Fig.~\ref{paschen}, we show the spectral region near 
the Paschen limit for all our objets. The presence of strong {\hei} $\lambda$8155.5 and $\lambda$8203.9 stellar emission 
lines, very close to the Paschen discontinuity, is quite clear in all of them. We attempted to 
compute {\te} from the Paschen discontinuity paying special 
attention to the zones where continuum flux determinations were reliable and we then made a linear fit to the blue 
and red continua to compute the Paschen jump ($I_c$($Pac$)) 
at $\lambda$8204. This fit is  overplotted in Fig.~\ref{paschen}. Only for Cn\,1-5 was it impossible to fit the red continuum owing 
to the presence of some quite strong wide stellar features.  

The Paschen continuum temperature was derived by fitting the relation between $I_c$($Pac$)/$I$(P$n$) and {\te} 
for $10\le n\le 16$. The emissivities as a function of the electron temperature for the nebular 
continuum and the {\hi} Paschen lines were taken from \citet{brownmathews70} and 
\citet{storeyhummer95}, respectively. The finally adopted value of {\te}($Pac$) is the average of those obtained using 
the different {\hi} lines, 
neglecting those affected by atmospheric features. 
Errors were computed considering the {\hi} line uncertainties and an error of $\sim$15\% in the Paschen discontinuity 
determination, and propagating the error by means of an 
extinction correction;  finally, we added quadratically the dispersion in the values obtained from different continuum-to-line ratios. 
Table~\ref{plasma} shows the {\te} values derived from the {\hi} Paschen decrement.
Although {\te}(Paschen decrement) are in relatively good agreement with {\te} from other diagnostics, the errors are quite large.

\subsubsection{Electron temperatures from helium lines}

The {\hei} recombination lines can be used as a diagnostic of electron temperature. Two approaches have been proposed 
for deriving {\te} from these lines. 
In the first one, \citet{apeimbertetal02} claimed that in the 
presence of temperature fluctuations in the ionized gas, the temperature in the zone where {\hei} is present  
is a function of the temperatures in the O$^{+}$ and O$^{++}$ zones and of the temperature fluctuation parameter, $t^2$. 
These authors proposed a maximum likelihood method to obtain simultaneously the $n_e$({\hei}), $T_e$({\hei}), He$^+$/H$^+$ 
ratio, and the optical depth of the {\hei} $\lambda$3889 line. 
In a different approach, \citet{zhangetal05} used the analytic expressions of the emissivity of {\hei} lines given by 
\citet{benjaminetal99} to compute the temperature-dependent {\hei} singlet $\lambda$7281/$\lambda $6678 line ratio.
A matter of concern about this method is that there might be 
a departure from case B to case A for the {\hei} singlet lines by means of the destruction of {\hei} Lyman photons 
by the photoionization of neutral hydrogen and/or the absorption by dust grains \citep{liuetal01, fangliu11}. Unfortunately, 
only tailored photoionization modelling constrained by multiwavelength data could quantitatively compute this effect.  

\begin{figure}[] 
\begin{center}
\includegraphics[width=8cm]{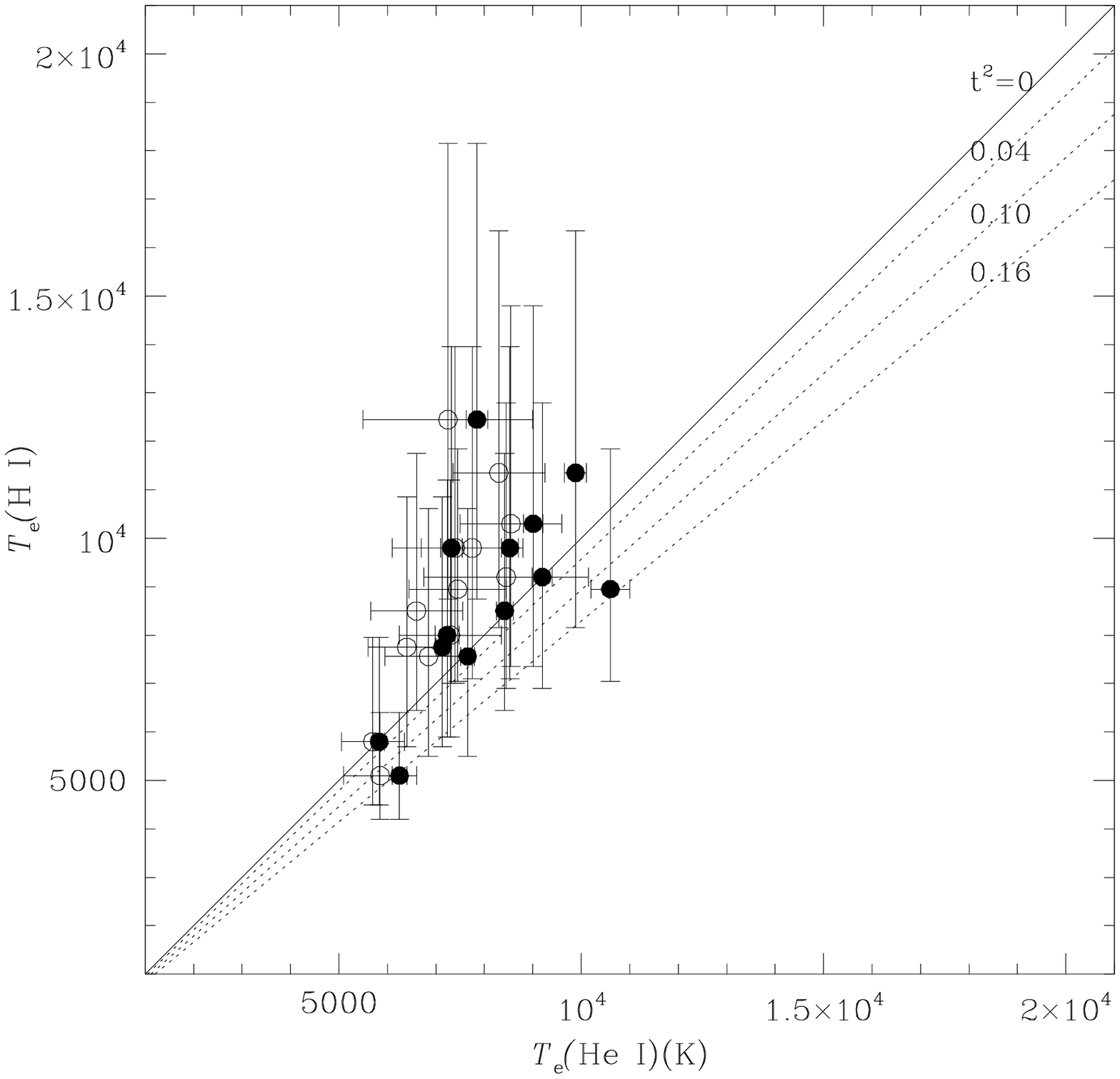}
\caption{$T_e$({\hei}) $vs.$ $T_e$({\hi}) diagram. Two methods for deriving $T_e$({\hei}) have been used.
One  follows the method of \citet{zhangetal05}, based on the line ratio $\lambda$7281/$\lambda $6678 (open circles).
The other follows the \citet{apeimbertetal02} procedure and computes $T_e$({\hei}) based on several {\hei}lines (black circles).
Solid line represents equality (t$^2$=0), and dotted lines show the variation in $T_e$({\hi}) as a function of $T_e$({\hei})
considering different $t^2$ values.}
\label{thi_thei}
\end{center}
\end{figure}

In Table~\ref{plasma}, we show {\te}({\hei}) for our objects, 
derived from the two methods. For the {\te}({\hei}) computed using the \citet{apeimbertetal02} method, we assumed a 
temperature fluctuations parameter, $t^2$, given by the ADF(O$^{++}$) (see Paper II for more details about $t^2$ determinations) 
and a minimum of 11 {\hei} lines were used. In Fig.~\ref{thi_thei}, we show the comparison between $T_e$({\hi}) and $T_e$({\hei}) 
derived using both methods: the 
open circles correspond to values using the \citet{zhangetal05} method, 
and the black ones are the values derived with the \citet{apeimbertetal02} procedure. In this figure, we have overplotted lines 
representing equality ($t^2$ = 0, solid line) and representing different values of $t^2$ (dashed lines).
It is observed that, for our objects, the $T_e$({\hei}) derived from many lines (Peimbert et al. method) are in general
higher (by up to 2-3 thousand degrees) than the values derived from Zhang et al. method. We decided to adopt the measurements obtained using 
the \citet{apeimbertetal02} method, which, being based on the observation of several lines, more tightly constrains the
temperature. It is not clear in this figure that the  behaviour of our [WC]PNe 
corresponds to objects with a moderate $t^2$. Most of the objects, in particular the hotter ones,
have a $T_e$({\hi}) that is higher than $T_e$({\hei}), in some cases by up to several thousand degrees. Similar results were obtained
by \citet{zhangetal05} for their PN sample, which they interpreted as evidence of H-deficient knots in the plasma.
However, our large uncertainties, mainly in the derived $T_e$({\hi}) for the hotter objects, cannot rule out an overall agreement 
with the predictions of the temperature fluctuations paradigm. We note that in our work, it is particularly difficult 
to properly determine $T_e$ from {\hi}
discontinuities, owing to the wide emission stellar features that are clearly present in the continuum, 
thus causing large uncertainties.

\subsubsection{Electron temperatures and densities from {\oii} and {\nii} recombination lines}

The intensities of the ORLs, originating from states with different orbital angular momentum, have different 
dependences on {\te}. 
Thus, by comparing the intensity of an {\oii}  $3d-4f$ transition to that of an {\oii} $3s-3p$ transition, it is possible to 
compute the electron temperature \citep{tsamisetal04, wessonetal03}. 
However, this method has its difficulties: first, the dependence of the intensity ratio on the temperature 
is very weak, so extremely high quality spectra are required to measure accurately these faint lines; second, the relative 
intensities of the {\oii} lines can be affected by departures from the local thermodynamic 
equilibrium (LTE) in the fundamental recombination level of the O$^{++}$ ion, $^3P$. \citet{tsamisetal04} argued 
that the intensity ratio of the $\lambda$4089.29 {\oii} $3d-4f$ transition  and the $\lambda$4649.14 {\oii} $3p-3s$ transition 
(of multiplet 1) was adequate to derive {\te} 
because these lines originate from states of high total 
angular momentum, $^3$P$_2$, and therefore must be affected in a similar way by this effect. Moreover, 
high densities (which is the case for most of our objects) also minimize this effect, which is negligible for {\elecd} $>$ 10$^4$ cm$^{-3}$. 

The temperature-sensitive ratio $I$($\lambda$4089.29)/$I$(4649.14) was measured for the first time by 
\citet{wessonetal03} for Abell 30, to determine the {\te} of the plasma in which 
{\oii} recombination lines arise. These authors found very low {\te} in two H-deficient knots in this PN. 
This method was later applied  by several authors to other PNe, with similar results 
\citep[e.g.,][]{tsamisetal04,yliuetal04a,wessonetal05}. 
These results are consistent with a scenario in which there is a cold H-deficient component of the gas emitting the bulk 
of the {\oii} lines and a ``normal'' gas component that emits the bulk of the CELs. Independently of the origin of this 
cold H-deficient component which, to this day, has not been resolved, this scenario provides an explanation to the 
abundance discrepancy. However, \citet{garciarojasesteban07} measured this ratio in three {\hii} regions with 
moderate ADFs, and did not find any evidence of this cold component. 
Our sample of [WC]PNe also have moderate values of the ADF, similar to the values found in 
{\hii} regions (see Paper II). 
We checked the $I$($\lambda$4089.29)/$I$(4649.14) ratio for all our objects where the very faint 
{\oii} $\lambda$4089.29 line was detected. 
In Table~\ref{te_orls}, we present the values of the {\te} obtained from {\oii} ORLs for our objects. 
These results are discussed later. 

Recently, \citet{fangetal11} computed new effective recombination coefficients for the {\nii} 
recombination spectrum, including radiative and dielectronic recombination. These authors found a set of recombination 
coefficients that allows us to construct density and temperature diagnostics based on the different 
dependences of the emissivities of different transitions on {\te} and n$_e$. On the basis of their method, we  
computed electron temperature and density simultaneously using the {\it loci} of the recombination line 
ratios $\lambda$5679/$\lambda$5666 versus ($vs.$)  $\lambda$5679/$\lambda$4041  of {\nii}.  
The results are presented in Table~\ref{te_orls}. However, prior to discussing them,
we note that several authors \citep{grandi76, escalantemorisset05} have claimed that 
resonance fluorescence coming from both the absorption of continuum or line photons, 
is necessary to reproduce the observed intensities of multiplet 3 {\nii} lines. 
In particular, \citet{grandi76} found that the resonance fluorescence of the $4s$ $^3P^0$ term by 
the recombination line {\hei} $1s^2$ $^1S-^1s8p$ $^1P^0$  was the dominant excitation mechanism, over starlight and 
recombination, for several multiplets of {\nii} 
(including {\nii} multiplet 3, i.e., $\lambda$5666 and $\lambda$5679 lines). \citet{escalantemorisset05} found that, 
in the Orion nebula, recombination contributed a minor part of the observed intensities of lines from $3p$ and $3d$ 
levels connected to the ground state (e.g. lines of multiplet 3), but that it was the dominant mechanism producing the 
intensity of lines coming from $4f$ levels (e.g. {\nii} $\lambda$4041 line). 
These results suggest that we have to be careful when dealing with temperatures and abundances from {\nii} lines, especially in low 
ionization PNe, where the potential effects of fluorescence on these lines would be stronger\footnote{The recombination 
contribution to {\nii} lines comes from the capture of a free electron  to an excited level of N$^{++}$ 
followed by a radiative transition, and dominates when the ionization degree is high, while the fluorescence contribution 
comes from the excitation of N$^+$ by fluorescence 
photons followed by a radiative transition, and could be very important in relatively low ionization PNe.}. 

\begin{figure}[] 
\begin{center}
\includegraphics[width=8cm]{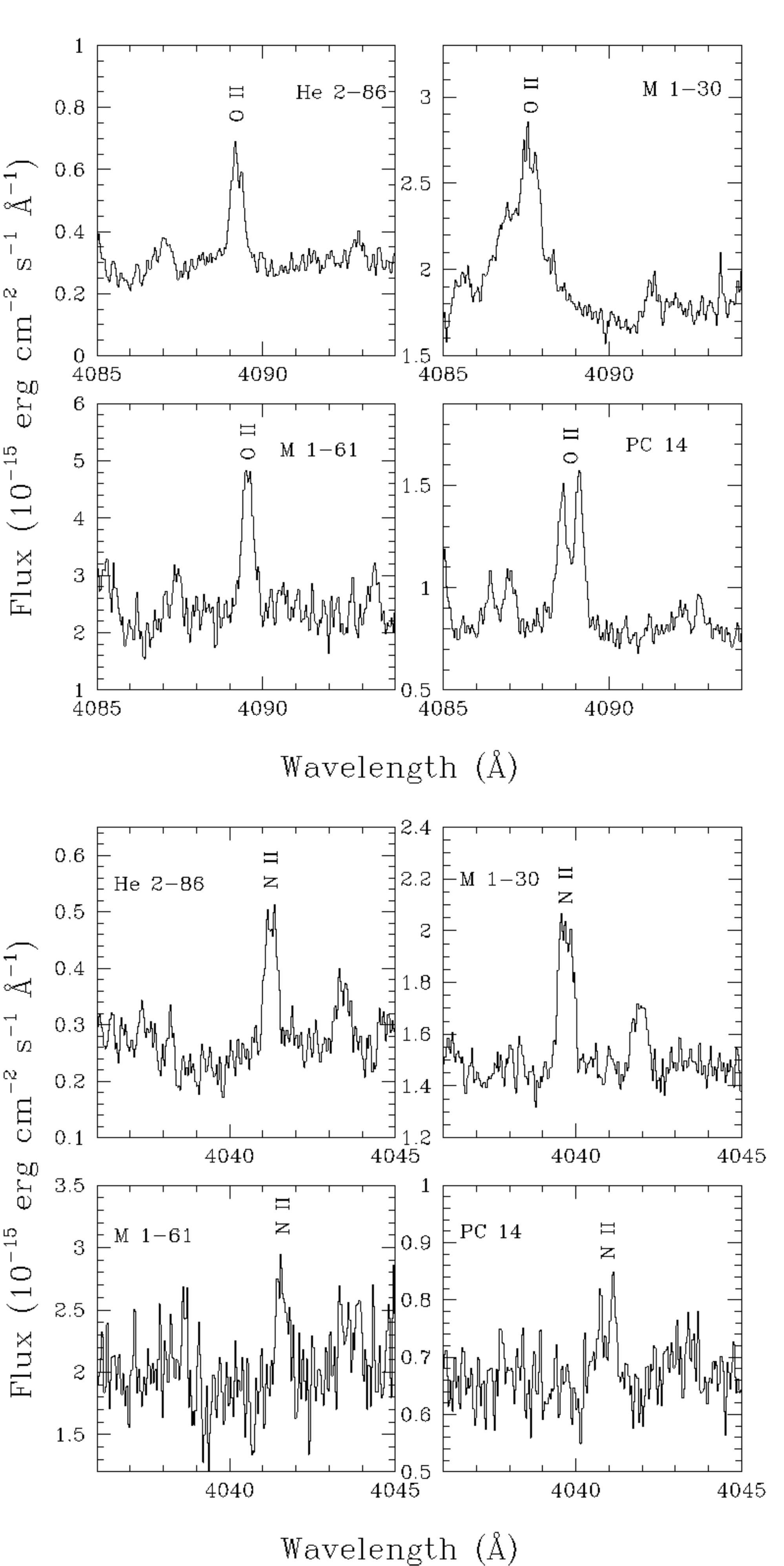}
\caption{Part of the spectrum of four objects in which the faint {\oii} $\lambda$4089 (upper four panels) and {\nii} 
$\lambda$4041 (lower four panels) lines were most reliably measured.}
\label{oii_nii_lines}
\end{center}
\end{figure}

In Table~\ref{te_orls}, we show the physical conditions obtained from
{\oii} and {\nii} permitted lines for the objects
of our sample. In this table, objects are ordered from higher to lower
ionization degree, i.e., from low to
high potential influence of fluorescence in the {\nii} lines. Owing to
the large uncertainties in the faint {\oii} $\lambda$4089 and {\nii}
$\lambda$4041 line fluxes of most objects, it makes no sense to
calculate the single {\te} and {\elecd} values obtained from the
measured line ratios, because they are several times beyond the validity range
of the computed diagnostics.
Instead, we show the upper or lower limits given by the error boxes
in the diagrams and, additionally, we show in brackets the values
obtained without considering the uncertainties. However, for some
objects, we can compute {\te}({\oii}) with the corresponding uncertainties.
The values obtained from {\oii} and {\nii} diagnostics seem to show that the uncertainties 
in the measurement of the faint {\oii} $\lambda$4089 and {\nii}
$\lambda$4041 lines dominate over any other effect, real or not, so
it is really hard to conclude anything about the origin of the
emission of {\oii} and {\nii} permitted lines. In addition, we did not find any systematic trend 
to very low temperatures similar to that found in other objects \citep{wessonetal03, wessonetal05}. 
To illustrate the quality of these diagnostics, in Figure~\ref{oii_nii_lines}
we present {\oii} $\lambda$4089 and {\nii} $\lambda$4041 lines in the
four PNe with the highest signal-to-noise ratios.

\setcounter{table}{4}
\begin{table}[] 
\begin{center}
\caption{Physical conditions derived from {\oii} and {\nii} recombination lines$^{\rm a}$. } 
\label{te_orls}
\begin{tabular}{lcccc} 
\noalign{\smallskip} \noalign{\hrule} \noalign{\smallskip}
\noalign{\smallskip} \noalign{\hrule} \noalign{\smallskip}
Object & {\te}({\oii}) (K) & {\te}({\nii}) (K) & {\elecd}({\nii}) (cm$^{-3}$) & log(O$^{++}$/O$^{+}$) \\
\noalign{\smallskip} \noalign{\hrule} \noalign{\smallskip}
NGC\,6369    & \nodata			& \nodata			& \nodata 		& 1.90	\\
M\,3-15      & \nodata			& \nodata			& \nodata 		& 1.56	\\
PC\,14       & 2900$^{+7000}_{-2000}$	& $<6300$ [2500]		& $>$100 [1300]   	& 1.30	\\
M\,1-61      & $>$5000 [$>10000$]	& $>3200$$^{\rm b}$		& $>$600$^{\rm b}$	& 1.30	\\
He\,2-86     & 6200$^{+4000}_{-2700}$	& $>7900$ [11200]		& $>$2000 [6300]	& 1.12	\\
Hb\,4	     & \nodata 			& $>4000$$^{\rm b}$		& $>$2000$^{\rm b}$	& 1.05	\\
Cn\,1-5      & 8500$^{+>2000}_{-6700}$ 	& $>1500$ [12000]		& $<$6500 [1600]  	& 0.60	\\
Pe\,1-1      & 10000$^{+>2000}_{-8000}$	& \nodata			& \nodata  		& 0.57	\\
M\,1-25      & $>$5500 [$>10000$]	& $>$5000$^{\rm b}$		& $>$600$^{\rm b}$	& 0.16	\\
NGC\,5189    & \nodata			& $<$15800 [7100]		& $<$4000 [650]   	& 0.13	\\
M\,1-30	     & $>$6000 [$>10000$]	& $>$15800$^{\rm b}$		& $>$1000$^{\rm b}$	& $-$0.12\\
M\,1-32      & \nodata			& \nodata			& \nodata 		& $-$0.27\\
\noalign{\smallskip} \noalign{\hrule} \noalign{\smallskip}
\end{tabular}
\begin{description}
\item[$^{\rm a}$]  The {\te} and {\elecd} values shown in the table are upper or lower limits given by the error boxes in the diagrams. Additionally, we show 
in brackets the values obtained without considering the uncertainties.
\item[$^{\rm b}$] Line ratios (without errors) are out of range
\end{description}
\end{center}
\end{table}

\section{Discussion
\label{discuss}}

The correct knowledge of chemical abundances in PNe is essential to constrain the different models of post-AGB evolution. 
The most accurate evaluation of these chemical abundances has several steps in which we have to make the right decisions 
to avoid errors that would lead us to incorrect conclusions. Adopting a correct set of plasma conditions for each ion is 
one of those fundamental steps that we discuss. In a forthcoming paper (Paper II), 
we will focus on whether ORLs or CELs are more reliable options for computing abundances and on the issue of 
ionization correction factors. 

\begin{figure*}[] 
\begin{center}
\includegraphics[width=18cm]{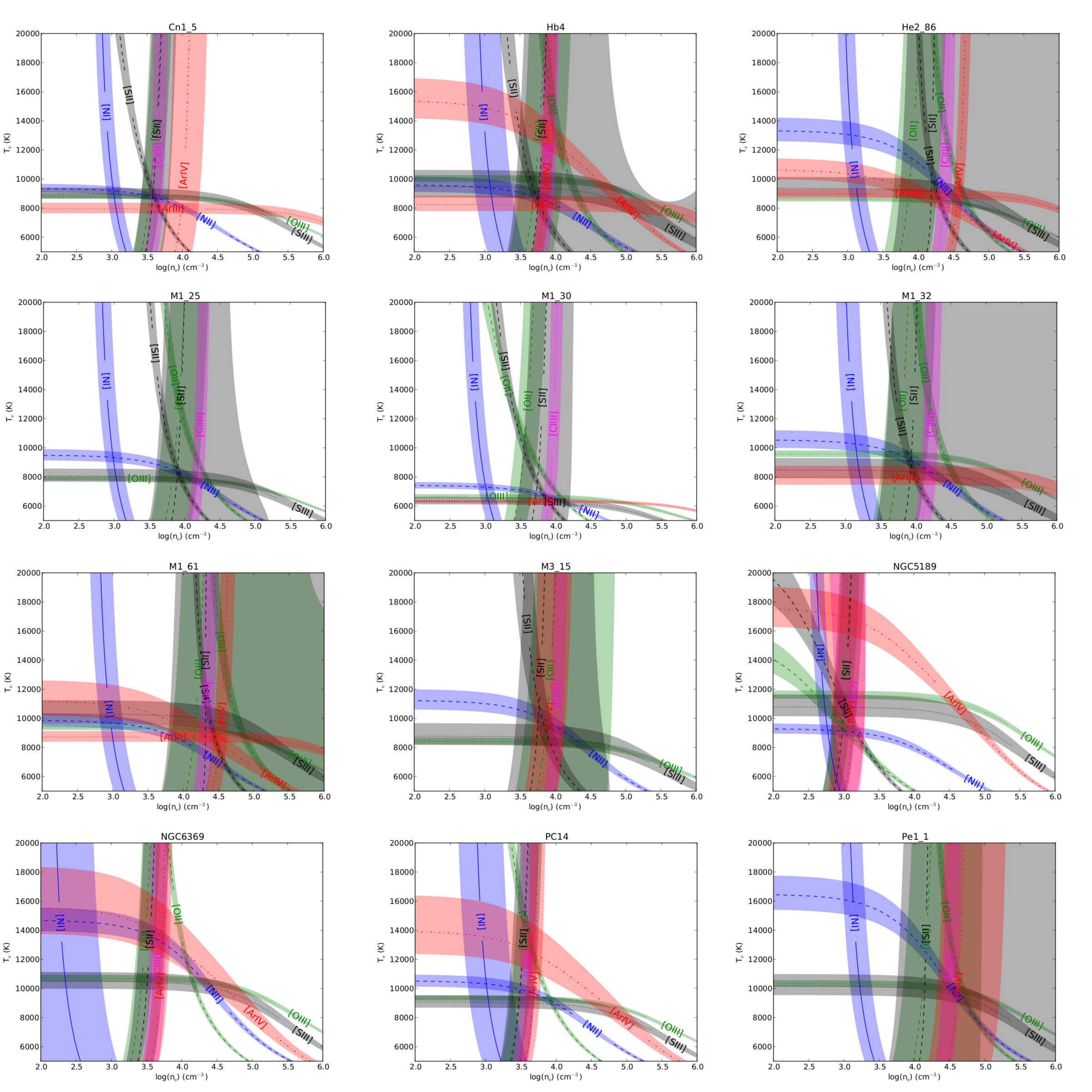}
\caption{$T_e-n_e$ diagnostic plots. Colors correspond to species: Grey for S, blue for N, green for O, magenta for Cl, and red for Ar. 
Different lines indicate ions: solid for neutral ions ({\fnitroi}), dashed for once ionized ions ({\fnii}, {\fsii} and {\fsii}), dotted 
for two ionized ions ({\foiii}, {\fsiii}, {\fariii}, and {\fcliii}), and dotted-dashed for three times ionized ions ({\fariv}).}  
\label{te_ne_diag}
\end{center}
\end{figure*}

To obtain an overall picture of the physical conditions, we have constructed $T_e-n_e$  diagnostic plots for 
all our objects, making use of the software {\sc pynebular} 
\citep{luridianaetal11}, an update of the {\sc iraf} package {\sc nebular}, rewritten in {\it Python}. These  plots 
are shown in Fig.~\ref{te_ne_diag} where all the available diagnostic ratios for {\te} and {\elecd}, in each object, 
have been included. We note in these diagrams that in general, a good average solution for $T_e-n_e$ can be found
for most of the objects. However, in several cases there are significant differences among the values of {\te} and {\elecd} derived 
from different diagnostic ions, which reflects the density and temperature structure inside the nebulae. We examine 
these differences in the following sections, to derive the most appropriate set of plasma conditions 
for the different nebular zones.

\subsection{The densities
\label{densities}}

\begin{figure}[] 
\begin{center}
\includegraphics[width=\columnwidth]{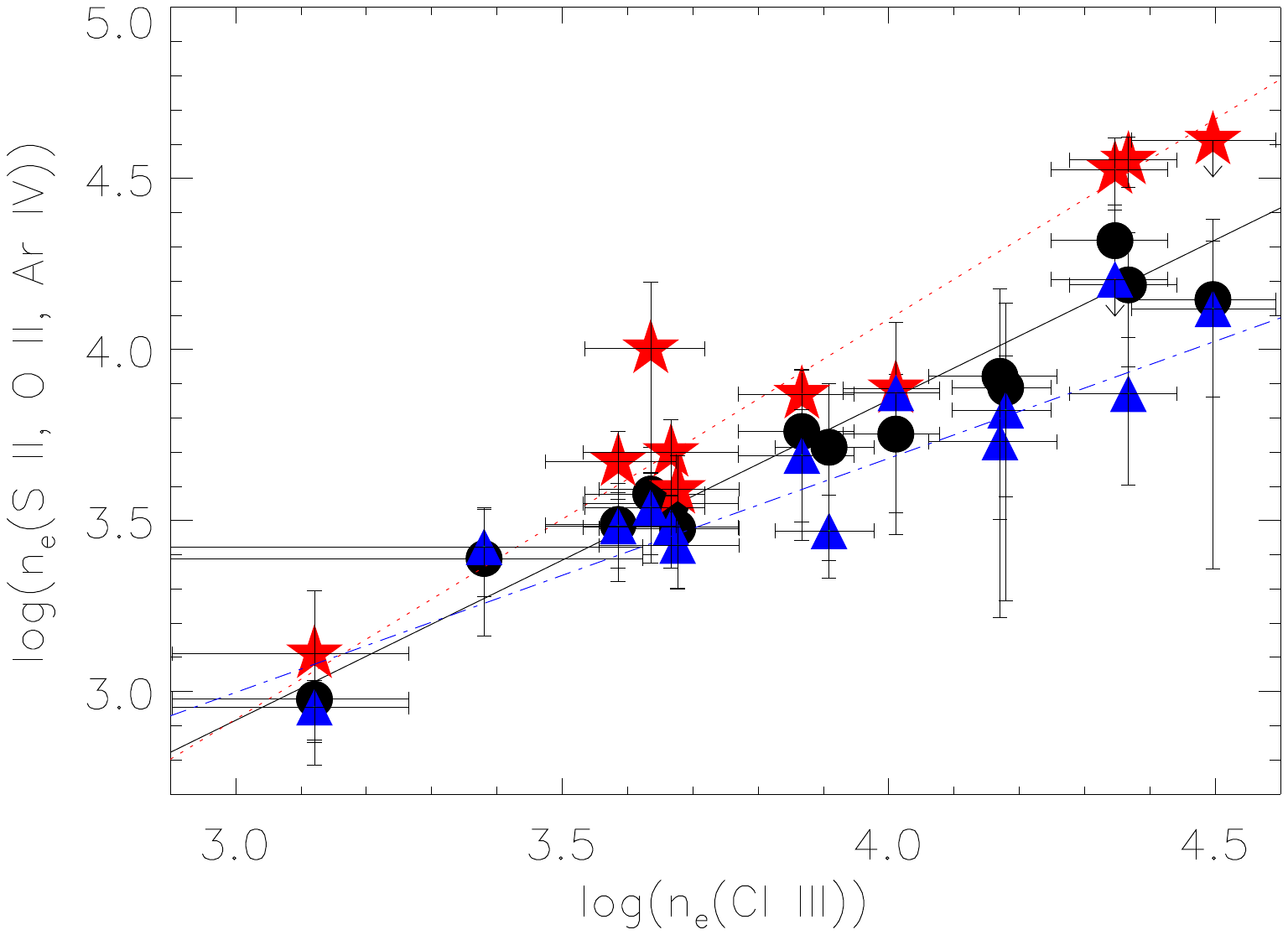}
\caption{Electron densities derived from {\fsii} $\lambda\lambda$6717/6730 (black filled circles), {\foii} 
$\lambda\lambda$3726/3729 (blue triangles), and {\fariv} $\lambda\lambda$4711/4740 (red stars)  compared to the one derived 
from {\fcliii} $\lambda\lambda$5517/5537.There is a clear correlation between all quantities. The trend 
{\elecd}({\fariv}) $>$ {\elecd}({\fcliii}) $>$ {\elecd}({\fsii}) $>$ {\elecd}({\foii}) is clear. Solid line represents 
the equality. Pointed, solid,  and dashed lines are the fits to the {\elecd}({\fariv}), {\elecd}({\fsii}), and {\elecd}({\foii}) 
$vs.$ {\elecd}({\fcliii}) data points, respectively.}
\label{dens_comp}
\end{center}
\end{figure}

We derived densities from several diagnostic ratios of {\fsii}, {\foii}, {\fcliii}, 
{\fariv}, and {\ffeiii}.
Fig.~\ref{dens_comp} shows the behaviour of {\elecd}({\foii}) (blue triangles), {\elecd}({\fsii}) (black circles), and 
{\elecd}({\fariv}) (red stars) 
$vs.$  {\elecd}({\fcliii}). A correlation is observed in all the cases. For {\elecd}({\fsii}) $vs.$ {\elecd}({\fcliii}), 
the slope (in logarithm) is 
0.93$\pm$0.05, and for {\elecd}({\foii}) the slope is about 0.69$\pm$0.05. For {\elecd}({\fariv}), the slope is 
1.17$\pm$0.06. We find, in general, that  {\elecd}({\fariv}) $>$ {\elecd}({\fcliii}) $>$ {\elecd}({\fsii}) 
$>$ {\elecd}({\foii}). However, 
the trend is not present in the objects  Cn\,1-5, Hb\,4, NGC\,5189, NGC\,6369, and PC\,14,  
for which we can adopt a density as the weighted average of all the computed density diagnostics.
For the other objects, these density differences can be explained from two points of view. On the one hand, it is 
possible that  a real and 
strong density stratification exists in these objects that has to be taken into account to properly compute the physical 
conditions and chemical abundances. On the other hand, we know that at densities higher than the 
critical density of the departure level, collisional de-excitation dominates over radiative de-excitation, suppressing 
part of the emission flux of the line; 
if this effect occurs for both lines of the density diagnostic, then the density derived would probably be lower than the real value.  
In the case of the {\foii} $\lambda$3726/$\lambda$3729 ratio, the critical density  is $\sim$3500 cm$^{-3}$ and for the 
{\fsii} $\lambda$6717/$\lambda$6731 ratio, it is about 3000 cm$^{-3}$, hence, in 
 the presence of high density clumps in the the low ionization zone, these ratios would underestimate the 
real density. However, the {\fcliii} $\lambda$5517/$\lambda$5537 and {\fariv} $\lambda$4711/$\lambda$4740 line 
ratios are free of these effects owing to the high critical densities of the departure levels for, at least, one of the 
lines ($>4\times10^4$ cm$^{-3}$).

To decide which density should be adopted, other alternatives can be used. First, 
we could use {\foii}$_{na}$ and {\fsii}$_{na}$ nebular to auroral line ratios, which in our density range are also density 
sensitive (see Fig.~\ref{te_ne_diag}). However, these ratios 
should be treated with caution because they can be affected by telluric emisson lines in the case of {\foii} $\lambda$$\lambda$ 
7320+30 lines, or might be blended with other emission lines, in the case of {\fsii} $\lambda\lambda$4068+76 lines 
(see note $^{\rm e}$ on Table~\ref{plasma}). 
In Table~\ref{plasma}, we show the densities obtained from the {\foii}$_{na}$ and {\fsii}$_{na}$ ratios, which were computed 
by assuming that {\te}({\fnii}) is representative of the low ionization zone. 
The values obtained  are much larger than those obtained from the nebular ratios and are, 
in general, consistent with the values derived from the {\fcliii} $\lambda$5517/$\lambda$5537 ratio. As said before, 
this suggests that the nebular {\foii} and {\fsii} diagnostics are insensitive to any high density clumps in the nebula.

\begin{figure}[] 
\begin{center}
\includegraphics[width=\columnwidth]{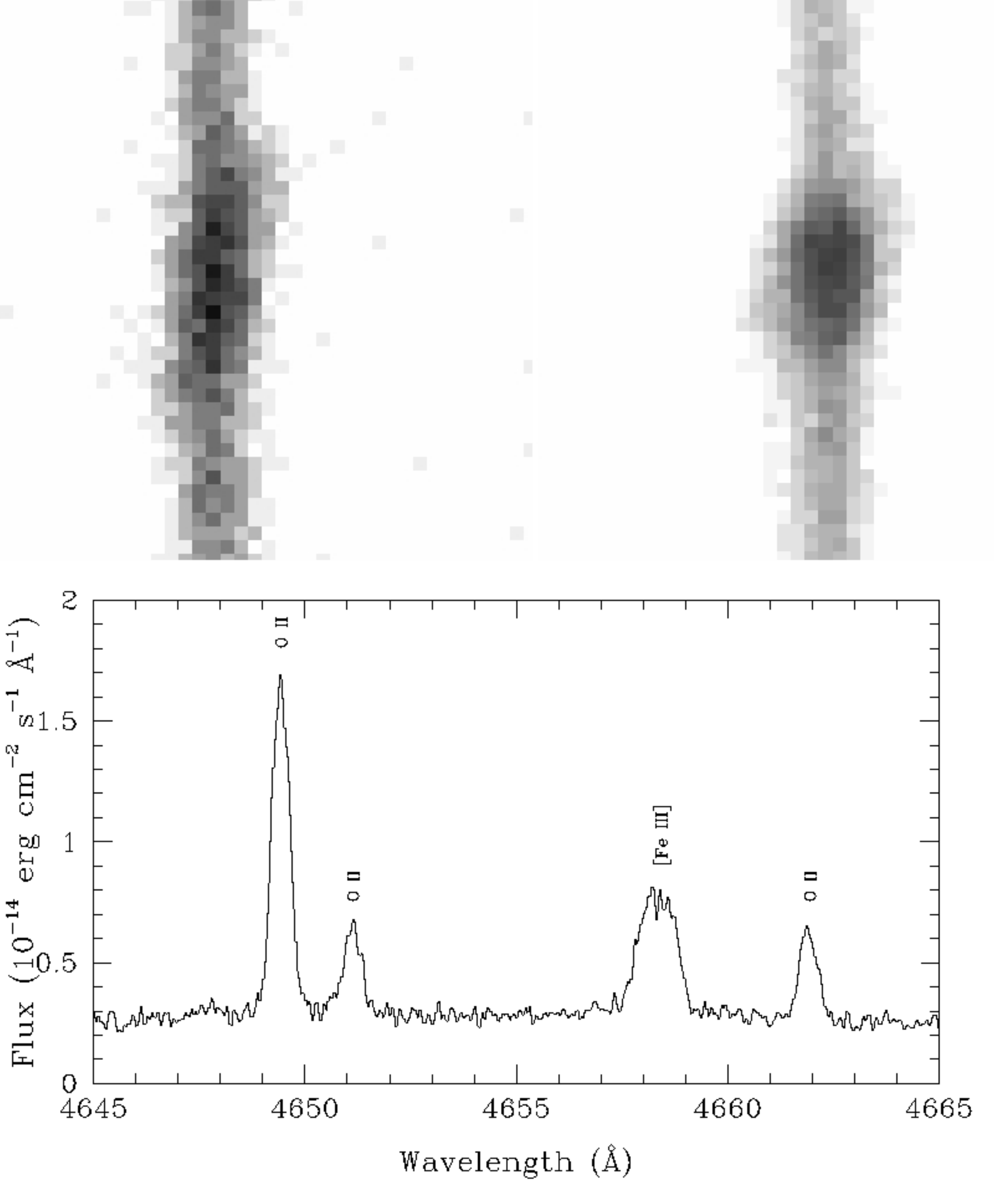}
\caption{Upper panel: Portion of the 2D echellogram showing {\ffeiii} $\lambda$5270 (left) and {\fcliii} $\lambda$5517 lines for M1-61. 
The grey scale is 
the same for both lines. It is clear that the {\ffeiii} emission is more extended that {\fcliii} emission. Lower panel: 
portion of the 1D extracted spectra showing the {\ffeiii} $\lambda$4658 line, compared with multiplet 1 {\oii} ORLs. The FWHM of the {\ffeiii} line is 
clearly larger.}
\label{feiii_lines}
\end{center}
\end{figure}

Second, we measured several {\ffeiii} lines in our spectra, which are very useful 
because their ratios provide very robust density 
diagnostics over a wider range in electron density \citep{keenanetal01}. 
We derived electron densities from the analysis of {\ffeiii} lines detected in 
several of our objects (Cn\,1-5, He\,2-86, M\,1-25, M\,1-32, M\,1-61, and PC\,14). In most of them, we 
observed at least four emission lines of the 2F and 3F multiplets, namely {\ffeiii} $\lambda\lambda$4658, 4701, 4734, and 4881. 
The density values of {\ffeiii}, {\elecd}({\ffeiii}), which are presented in Table~\ref{plasma}, were calculated by computing 
the minimum dispersion between the observed and theoretical ratios of the {\ffeiii} lines with respect to 
the bright {\ffeiii} $\lambda$4658 line, considering the observational errors. The 
theoretical emissivites were calculated by solving a 34-level model atom that uses the 
collision strengths of \citet{zhang96}, the transition probabilities of \citet{quinet96}, 
combined with the probabilities of some UV transitions estimated by 
\citet{johanssonetal00}, and adopting \te({\fnii}) as the temperature of the Fe$^{++}$ zone. 
The {\elecd}({\ffeiii}) values determined from this analysis are 
representative of the low-medium ionization zone, though their values seems to be more 
consistent with those determined from high density indicators such as {\fcliii} and 
{\fariv} line ratios. However, there is a clear exception to this rule: M\,1-61, which has an extremely 
high {\elecd}({\ffeiii}) of about 94000 cm$^{-3}$. We tested this value by computing the density after discarding various 
lines, but we reached similar values in all the cases. It is not the scope of this paper to investigate the origin of 
this discrepancy; however, after inspecting of the 2D echellograms (Fig.~\ref{feiii_lines} upper panel) and the 1D 
spectra (Fig.~\ref{feiii_lines} lower panel), one can easily see that the {\ffeiii} line structure is completely different than 
that of other CELs or ORLs, which indicates that {\ffeiii} emission originates from a different place in the PN than the other 
lines in the low and medium ionization zones. 
Hence, {\elecd}({\ffeiii}) is no longer representative of the low-medium ionization zone of this nebula.
Two {\ffeiii} lines were also detected in M\,1-30 and Pe\,1-1, but could not be used to perform this analysis.

Therefore, in agreement with the above discussion, for each object we adopted the densities that are summarized in Table~\ref{dens}.

\setcounter{table}{5}
\begin{table}[] 
\begin{center}
\caption{Adopted electron densities. } 
\label{dens}
\begin{tabular}{lcc} 
\noalign{\smallskip} \noalign{\hrule} \noalign{\smallskip}
\noalign{\smallskip} \noalign{\hrule} \noalign{\smallskip}
Object & {\elecd}(low-medium) (cm$^{-3}$) & {\elecd} (high) (cm$^{-3}$)  \\
\noalign{\smallskip} \noalign{\hrule} \noalign{\smallskip}
Cn\,1-5    & 4000$\pm$600		   & 10050$\pm$5700	   \\
Hb\,4      & 6250$\pm$1050		   & 6250$\pm$1050	   \\
He\,2-86   & 23300$\pm$4350		   & 35810$\pm$5990	   \\
M\,1-25    & 15100$\pm$2600  	   	   & 15100$\pm$2600	   \\
M\,1-30    & 8000$\pm$1000		   & 8000$\pm$1000         \\
M\,1-32	   & 15000$\pm$3300  	   	   & 15000$\pm$3300        \\
M\,1-61    & 22200$\pm$4500		   & 35590$\pm$8070	   \\
M\,3-15    & 8800$\pm$1400		   & 8800$\pm$1400	   \\
NGC\,5189  & 960$\pm$140	  	   & 960$\pm$140	   \\
NGC\,6369  & 3700$\pm$500		   & 3700$\pm$500	   \\
PC\,14	   & 3550$\pm$450	 	   & 3350$\pm$450	   \\
Pe\,1-1    & 31100$\pm$7800		   & 31100$\pm$7800		   \\
\noalign{\smallskip} \noalign{\hrule} \noalign{\smallskip}
\end{tabular}
\end{center}
\end{table}

\begin{figure}[] 
\begin{center}
\includegraphics[width=\columnwidth]{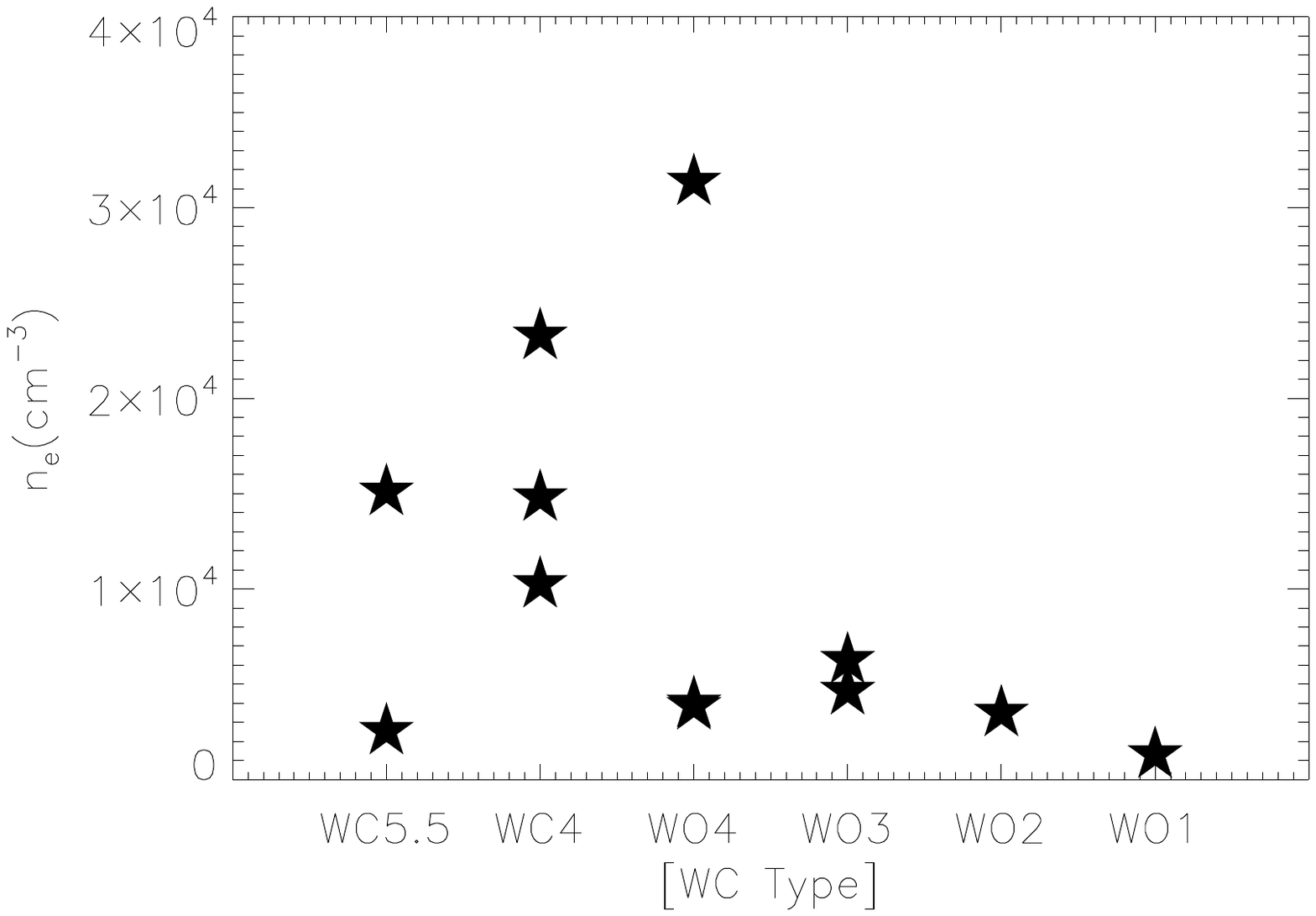}
\caption{{\elecd} as a function of the [WC] type of the central star. }
\label{ne_wrtype}
\end{center}
\end{figure}

In Fig.~\ref{ne_wrtype}, we present the behaviour of the adopted low-medium electron density $vs.$ the [WC] spectral classification 
of the central star. This diagram has been constructed several times in the literature and it shows that [WC]-early 
stars are surrounded by lower density nebulae. Most of the [WC]-late nebulae have, in contrast, a quite high density. 
It can be concluded that [WC]-early PNe are more evolved objects 
than [WC]-late PNe, which appear younger from the point of view of the nebulae and the central star as well.
This diagram and this conclusion have been used to propose that there is an evolutionary sequence from [WC]-late PNe to 
[WC]-early PNe \citep{ackerneiner03}. 
However, this sequence has been questioned from the point of view of the chemical abundances calculated for the 
central stars, for which it is 
deduced that the mean C abundance in [WC]-early stars is a factor of two lower than the C abundance in [WC]-late stars, while 
an evolutionary sequence would predict the opposite. 
\citep[][]{koesterke01,hamannetal05}. In Paper II, we will discuss this point further.

\subsection{The temperatures
\label{temperatures}}

Given that the {\foii} $\lambda\lambda$7320+30 and {\fsii} $\lambda\lambda$4068+76 lines are affected by telluric emission 
and blends with other lines, and that these ratios are also density sensitive, especially at the densities of most of the 
objects analysed here, we assumed that only {\te}({\fnii}) is representative of the low ionization zone.

\begin{figure}[] 
\begin{center}
\includegraphics[width=\columnwidth]{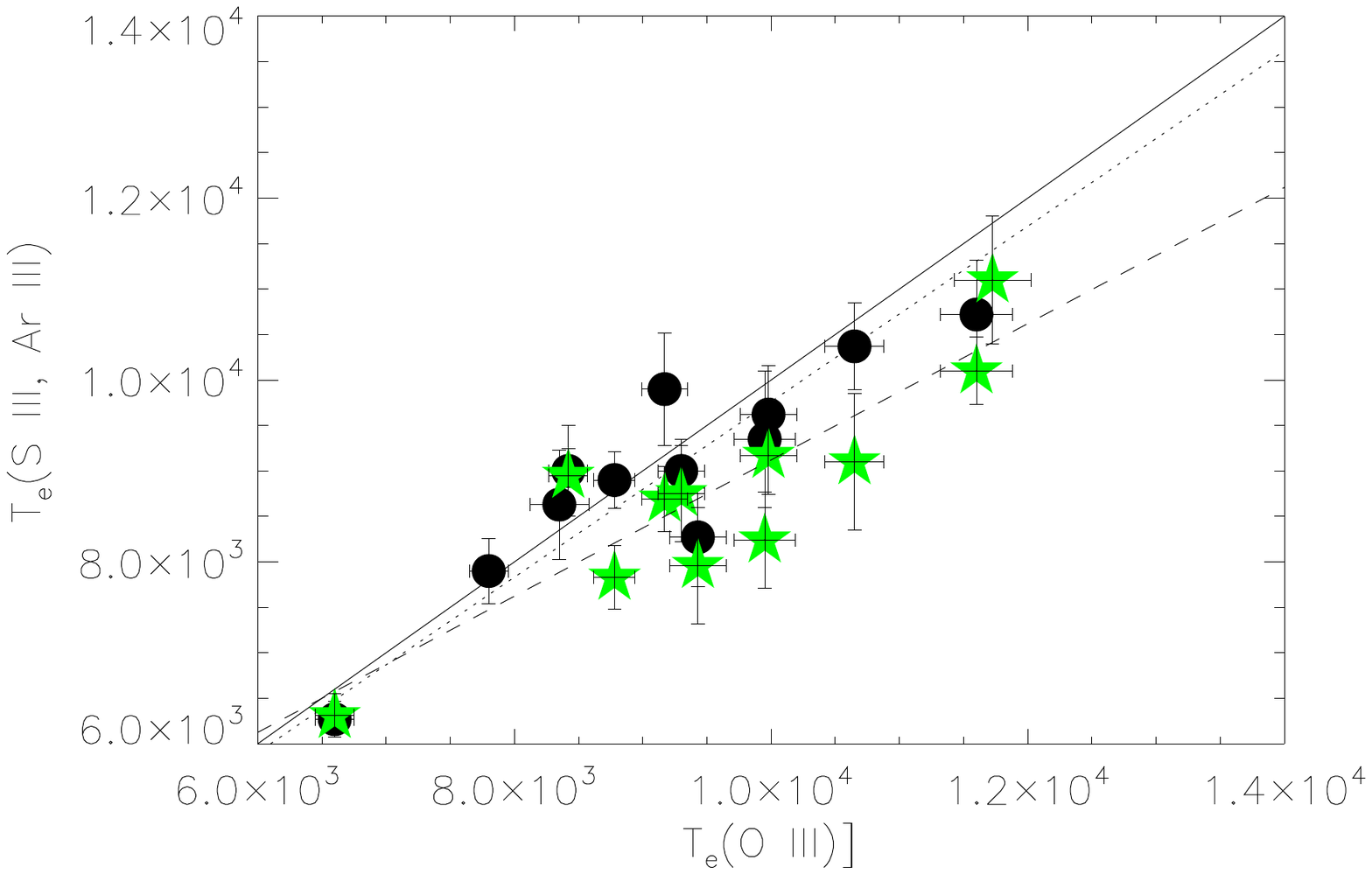}
\caption{Comparison between high ionization {\te} diagnostics. Black dots: {\te}({\fsiii}); green stars: {\te}({\fariii}). 
A very good correlation {\te}({\foiii}) vs. {\te}({\fsiii}) is found.
Solid line represent an equality.}
\label{temphigh}
\end{center}
\end{figure}

In Fig.~\ref{temphigh}, we compare the temperatures obtained from the high ionization diagnostic ratios {\te}({\foiii}), 
{\te}({\fsiii}), and {\te}({\fariii}). In this figure, the correlation between the {\foiii} and {\fsiii} temperatures has 
a slope of 0.96$\pm$0.03, which is very close the unity, and a Spearman rank correlation coefficient (SRCC) of 
r=0.84. In Fig.~\ref{temphigh}, we can also see that the {\fariii} diagnostic ratios provide somewhat lower temperatures than {\foiii} ones. 
The slope is 0.75$\pm$0.02 and the SRCC is r=0.77. Given the similarity between these temperatures, we 
assumed {\te}({\foiii}) is representative of the high ionization zone.

\begin{figure}[] 
\begin{center}
\includegraphics[width=\columnwidth]{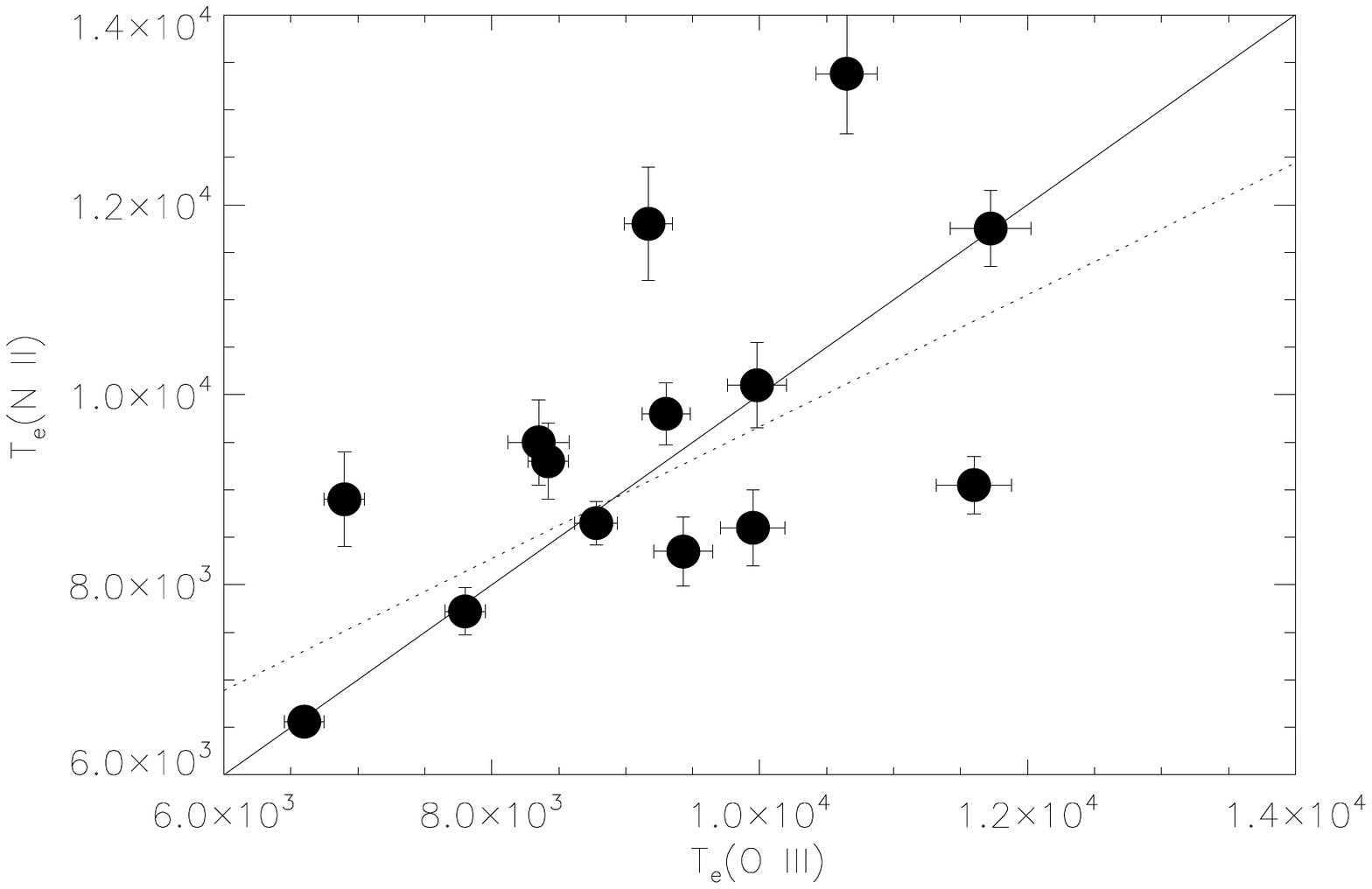}
\caption{Comparison between {\te}({\foiii}) and {\te}({\fnii}). Solid line represent equality. Pointed line is the 
fit to the data.}
\label{toiiinii}
\end{center}
\end{figure}

In general, temperatures derived from {\foiii} and {\fnii} diagnostics are similar, except in some cases where 
{\te}({\fnii}) is higher than {\te}({\foiii}) by as much as 3000 K. Only one object has a 
{\te}({\foiii}) that is higher than {\te}({\fnii}) by about 2000 K. In Fig.~\ref{toiiinii}, we compare 
both temperatures, finding that they are only weakly 
correlated, with a slope of 0.69, but a SRCC of r=0.61. This weak correlation is mainly due to the cases mentioned before.

\begin{figure}[] 
\begin{center}
\includegraphics[width=\columnwidth]{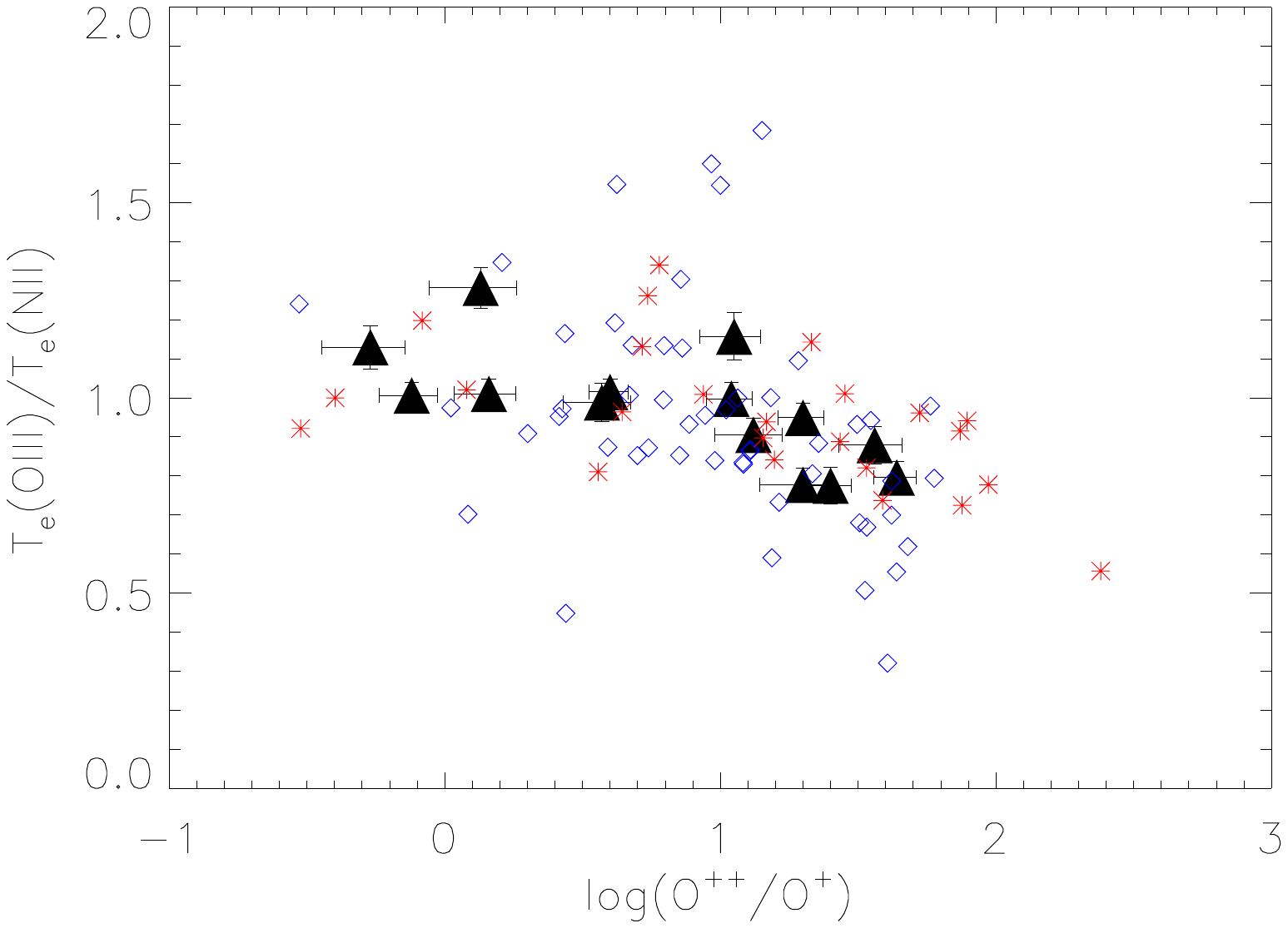}
\caption{{\te}({\foiii})/{\te}({\fnii})  vs.O$^{++}$/O$^+$. Black triangles: our data; red asterisks: \citet{penaetal01}; blue open diamonds: 
\citet{gornyetal04} data for bulge PNe}
\label{coctemp}
\end{center}
\end{figure}

To analyse in greater depth the behaviour between the temperatures in the high and low ionization zones, we present, in 
Fig.~\ref{coctemp}, the electron temperature ratio {\te}({\foiii})/{\te}({\fnii}) $vs.$ the ionization degree, 
represented by O$^{++}$/O$^+$ for our sample (filled black triangles). In general, we find two regimes in this figure: 
one zone in the lower ionization regime, where {\te}({\foiii})/{\te}({\fnii}) is about one, and a second zone where 
O$^{++}$/O$^+$ is larger than 10 and {\te}({\foiii})/{\te}({\fnii}) drops. In this zone, we find objects for which 
{\te}({\fnii}) is higher than {\te}({\foiii}). To verify that this effect \citep[already described for the 
sample analysed by][]{penaetal01}  is not due to biases introduced by the incompleteness of our particular sample of 
[WC]PNe, we overplotted  in Fig.~\ref{coctemp} the data for [WR]PNe by \citet{penaetal01}, and the data for bulge PNe 
by \citet{gornyetal04}. The trend is clear for all the samples.

This behaviour of {\te}({\fnii}) being higher than {\te}({\foiii}) is often found in photoionized nebulae, and is due to radiation 
hardening, i.~e. the lower absorption probability of  more energetic photons, which causes these photons to have longer mean free 
paths and to remain unabsorbed until the edge of the Str\"omgren sphere. \citet{penaetal01} found, from simple photoionization 
models at metallicities around solar, that the behaviour predicted by the models is mild, dependent on $T_*$, not as extreme 
as the one found here, and, in adition, not dependent on the excitation of the PNe. \citet{penaetal01} proposed as a
possible origin of this behaviour a strong inhomogeneous structure, or
additional heating by shocks and/or turbulence. Hence, to properly constrain the problem, we calculated  
a more complete grid of photoionization models.

\subsection{Photoionization modelling
\label{modelling}}

\begin{figure}[] 
\begin{center}
\includegraphics[width=\columnwidth]{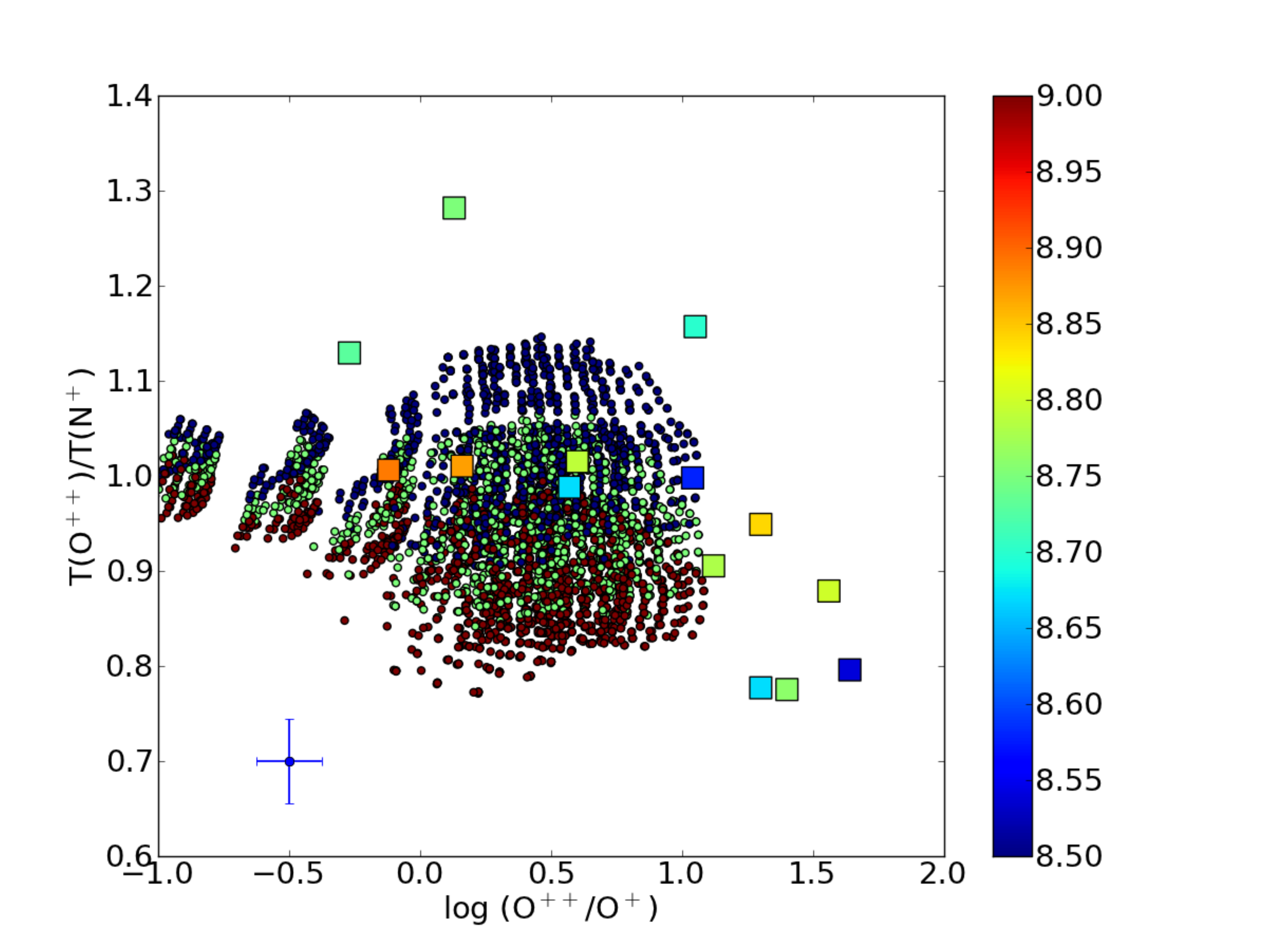}
\caption{Photoionization models computed to reproduce the observed {\te}({\foiii})/{\te}({\fnii}) $vs.$ O$^{++}$/O$^+$ behaviour. 
Models (circles) and observational data (squares) are colored according to the O/H value. At the lower left, we show the typical 
error bars for the observational data.}
\label{coctemp_mod}
\end{center}
\end{figure}

A grid of photoionization models was computed using Cloudy c10.00
code \citep{ferlandetal98}.
The grid covers the parameter space [lower value, higher
value, steps] of $T_{eff}$ (kK) = [50, 150, 10],
$log$ (O/H) = [-3.5, -3.0, 0.25], $log$ ($R_{in}$) (cm) = [15.5, 18,
0.5], $log$ (H-density) (cm-3) = [2.5, 4.5, 0.25], and $log$
($L/L_{\odot}$) = [2.5, 3.5, 0.5]. A total of about 5000 models were then
run, using  atmosphere models from \citet{rauch03} for
the ionizing spectral energy distribution, constant density shells,
and metal abundances following the O/H ratio. The models
are radiation bounded and dust-free. The mean values of the ionic
fractions and temperatures were obtained by integrating these
variables over the volume of the nebula. The results of the models are presented
in Figure~\ref{coctemp_mod},
with the observations superimposed as square symbols. The color code
is related to the O/H abundance and is the same for the models
and the observations. For the objects, the O/H abundance is that computed 
for the adopted physical conditions in each object and will be presented 
and discussed in Paper II.  

The models, which are radiation-bounded (R), do not reach values
larger than 10 for O$^{++}$/O$^+$ and do not reproduce the trend of
the observed values.

We then computed matter-bounded models (M), where the geometrical size
of the nebula is
set to 70\% of the Str\"omgren size for which, theoretically, we can
successfully
reach any high O$^{++}$/O$^+$ value, as the size of the O$^+$ region
can be reduced to values close to 0.
Real nebulae may be a combination of matter-  and radiation-bounded
components. They can thus reach higher values of O$^{++}$/O$^+$ than
the pure R nebulae shown in Fig.~\ref{coctemp_mod}. However, for any
realistic combination of M and R models, no changes in the values of
$T_e$(O$^{++}$)/$T_e$(N$^+$) are expected relative to the R models, as
the O$^{++}$ temperature is the same in both models and the
N$^{+}$ temperature is the one from the R models (N$^{+}$ is
negligible in the M models). It is interesting to notice that the smallest values for
$T_e$(O$^{++}$)/$T_e$(N$^+$) are obtained with high metallicity models
($log$ (O/H) $\sim$ 9.0 ),
but the observed nebulae are not so metal-rich. We note however, that  the models 
are relatively simple, and that, for example, for each model  the density is constant within 
the nebula and all metal abundances follow that of oxygen. 
We have to bear in mind that real objects are more complex but our simple 
approximation seems to succesfully reproduce the observed behaviour.

\section{Conclusions}

We have presented deep high-resolution spectrophotometric data
of 12 PNe with [WC] central star, obtained at LCO with the 6.5-m
Magellan telescope and the spectrograph MIKE. Data were reduced, wavelength- and
flux-calibrated and dereddened. Hundreds of lines were detected and identified for
each object, and their fluxes were measured. These [WR]PNe,
together with the three objects analysed by \citet{garciarojasetal09} 
represent the most extensive sample of this type of PNe
analysed so far, at such high resolution. The spectra were exposed
deep enough to detect, with a signal-to-noise ratio higher than $\sim$3, the weak ORLs
of {\oii}, {\cii}, and other species.

From our deep spectra, which cover a wide wavelength range (from 3350 \AA~to
9400 \AA), numerous diagnostic line ratios for {\te} and {\elecd} were
determined, from CELs and ORLs as well.  In addition, H discontinuities
(H Paschen discontinuity  in particular) were measured for this purpose.
All known recombinations effects that could perturb the CEL
diagnostic ratios were careful removed. In addition, the possible mechanisms
perturbing ORLs, such as fluorescence, departures from LTE and others, 
were considered.

Our main aim in this paper is to determine the optimal physical conditions
in the nebulae, for the accurate calculation of ionic abundances. We have performed a careful
analysis of all our available {\te} and {\elecd} and their errors to ascertain these conditions.

The conditions derived from CEL diagnostic ratios allow us to conclude the
following:

\begin{itemize}
\item The CEL diagnostic ratios usually used to determine {\elecd},
{\fsii}$\lambda$6730/$\lambda$6717, and
{\foii}$\lambda$3726/$\lambda$3729 seem to underestimate the true
nebular densities owing to their low critical densities. This occurs
particularly for the densest objects of our sample. Therefore, for these objects, we decided
to use the density derived from  the {\fcliii}$\lambda$5517/$\lambda$5537
line ratio as a representative value of the whole nebula, which agrees with density determinations using 
{\ffeiii} lines and the densities computed from the highly density-sensitive {\foii}$_{na}$ and {\fsii}$_{na}$ 
line ratios.

\item Temperature-sensitive CEL ratios were corrected for the effects of
recombination. In some cases, these effects can severely affect (up to
several thousand degrees) the {\te} determinations.
Finally, we adopted  a three-temperature scheme  for the nebulae:
{\te}({\fariv}) is used to represent the highest ionization zone, when available, 
{\te}({\foiii}) is used to represent the high ionization zone,
and {\te}({\fnii}) represent the low-ionization zone.
\end{itemize}

These physical conditions are used in Paper II to determine ionic
abundances from CEL lines.
\smallskip

A careful analysis to determine the physical conditions from ORL
diagnostic lines was also performed, with the following results:

\begin{itemize}
\item Electron temperatures were derived from  the Paschen
discontinuity relative to several Paschen recombination lines.
Although the derived values of {\te} agree with {\te} obtained from
other mechanisms, the errors are quite large because of
the stellar emission lines near the Paschen jump. We also derived {\te}
from several He lines by considering two different approaches: that of
\citet{apeimbertetal05} who consider several {\hei} lines and a 
$t^2$ parameter, and that of \citet{zhangetal05}, based on the ratio
of two {\hei} lines. In both cases, we found that  {\te}({\hei}) appears to be 
lower than {\te}({\hi}), although the effect is smaller when a $t^2$ value
is considered. However, the errors are so large (for {\hi} in
particular) that no conclusive  results can be extracted.

\item Electron temperatures were also computed from {\oii} and {\nii}
recombination lines, as proposed in several papers in the
literature \citep{wessonetal03, wessonetal05,fangliu11}.  However,
different authors have indicated that some effects, such as
fluorescence and departures from LTE, could be perturbing these lines.
We conclude that, despite our deep high-resolution data, the
uncertainties in the measurements of these faint lines dominate
over  any other effect and we cannot conclude anything about
the origin of these recombination lines.
\end{itemize}

Other phenomena such as the behaviour of density as a function of the [WC]
spectral type and the electron temperatures as a function of the
nebular ionization degree have been investigated. We have confirmed
that PNe around [WC]-early stars are evolved nebulae, while those around
[WC]-late stars are young, a result already reported in the literature.
We have analysed the behaviour of the temperatures 
found by \citet{penaetal01} of an unusually small 
{\te}({\foiii})/{\te}({\fnii}) when O$^{++}$/O$^+$ is larger than 10. An
ample  grid of photoionization models was computed with this aim. We
have found that models could reproduce
this behaviour (shown in Fig.~\ref{coctemp_mod}) if a combination of
matter-bounded and radiation-bounded models are considered, but for
the lowest $T_e$(O$^{++}$)/$T_e$(N$^+$) ratio, a too high metallicity
seem required.

The second part of this work, including ADF calculations and ionic and
total abundances for the nebulae, will be presented elsewhere
(Garc\'{\i}a-Rojas et al., Paper II, in preparation).


\begin{acknowledgements} 
This work received financial support from the  Ministerio  de Educaci\'on y Ciencia (MEC)  Espa\~nol, under project AYA2007-63030; from 
CONACYT-M\'exico under grant \#43121 and from DGAPA-UNAM, M\'exico under grants IN118405, IN112708 and IN105511. 
JGR acknowledges people and staff of Instituto de Astronom\'{\i}a at UNAM,
where part of this work were done. CM received financial support for his Sabbatical at the IAC from the Spanish MEC. 
The authors want to thank M\'onica Rodr\'{\i}guez, Grazyna Stasi\'nska, Antonio Peimbert, Manuel Peimbert 
and C\'esar Esteban for very fruitful discussions. 
\end{acknowledgements}




\Online

\onecolumn
\centering
\begin{longtable}{ccccccccc} 
\caption{Observed and reddening corrected line ratios (F(H$\beta$) = 100) and line identifications.
\label{lineid1}}\\
\hline\hline 
& & & \mc{5}{c}{Cn\,1-5} & \\
\cline{4-8} \\
$\lambda_0$ (\AA) & Ion & Mult. & $\lambda_{\rm obs}$ & $V_{\rm rad}$ (km s$^{-1}$) & $F(\lambda)/F({\rm H}\beta)$$^{\rm a}$ & $I(\lambda)/I({\rm H}\beta)$$^{\rm b}$ & Err(\%) & not \\
\hline
\endfirsthead
\caption{continued.}\\
\hline\hline 
& & & \mc{5}{c}{Cn\,1-5} & \\
\cline{4-8} \\ 
$\lambda_0$ (\AA) & Ion & Mult. & $\lambda_{\rm obs}$ & $V_{\rm rad}$ (km s$^{-1}$) & $F(\lambda)/F({\rm H}\beta)$$^{\rm a}$ & $I(\lambda)/I({\rm H}\beta)$$^{\rm b}$ & Err(\%) & not \\
\hline
\endhead
\hline
\endfoot
\endlastfoot
 3447.59 &           {\hei} &            7 & 3447.24 & -30.45 &   0.286 &   0.441 &   16 &     \\
 3478.97 &           {\hei} &           43 & 3478.76 & -18.09 &   0.169 &   0.257 &   23 &     \\
 3487.73 &           {\hei} &           42 & 3487.44 & -24.93 &   0.106 &   0.161 &   33 &     \\
 3512.51 &           {\hei} &           38 & 3512.17 & -29.03 &   0.225 &   0.338 &   18 &     \\
 3530.50 &           {\hei} &           36 & 3530.15 & -29.73 &   0.202 &   0.302 &   20 &     \\
 3554.42 &           {\hei} &           34 & 3554.05 & -31.20 &   0.232 &   0.343 &   18 &     \\
 3587.28 &           {\hei} &           32 & 3586.89 & -32.61 &   0.350 &   0.512 &   13 &     \\
 3613.64 &           {\hei} &            6 & 3613.26 & -31.52 &   0.363 &   0.525 &   13 &     \\
\end{longtable}

\clearpage

\newpage

\centering
\begin{longtable}{ccccccccc} 
\caption{Observed and reddening corrected line ratios (F(H$\beta$) = 100) and line identifications.
\label{lineid2}}\\
\hline\hline 
& & & \mc{5}{c}{Hb\,4} & \\
\cline{4-8} \\
$\lambda_0$ (\AA) & Ion & Mult. & $\lambda_{\rm obs}$ & $V_{\rm rad}$ (km s$^{-1}$) & $F(\lambda)/F({\rm H}\beta)$$^{\rm a}$ & $I(\lambda)/I({\rm H}\beta)$$^{\rm b}$ & Err(\%) & not \\
\hline
\endfirsthead
\caption{continued.}\\
\hline\hline 
& & & \mc{5}{c}{Hb\,4} & \\
\cline{4-8} \\ 
$\lambda_0$ (\AA) & Ion & Mult. & $\lambda_{\rm obs}$ & $V_{\rm rad}$ (km s$^{-1}$) & $F(\lambda)/F({\rm H}\beta)$$^{\rm a}$ & $I(\lambda)/I({\rm H}\beta)$$^{\rm b}$ & Err(\%) & not \\
\hline
\endhead
\hline
\endfoot
\endlastfoot
 3340.77 &          {\oiii} &            3 & 3340.03 & -66.42 &   0.965 &   4.541 &   14 &     \\
 3428.65 &          {\oiii} &           15 & 3427.81 & -73.45 &   0.268 &   1.115 &   19 &     \\
 3444.07 &          {\oiii} &           15 & 3443.31 & -66.17 &   2.730 &  11.128 &   12 &     \\
 3669.47 &            {\hi} &          H25 & 3668.65 & -67.01 &   0.186 &   0.578 &   22 &     \\
 3671.48 &            {\hi} &          H24 & 3670.82 & -53.89 &   0.157 &   0.488 &   24 &     \\
 3673.76 &            {\hi} &          H23 & 3673.05 & -57.95 &   0.216 &   0.667 &   20 &     \\
 3676.37 &            {\hi} &          H22 & 3675.62 & -61.17 &   0.206 &   0.634 &   21 &     \\
 3679.36 &            {\hi} &          H21 & 3678.58 & -63.57 &   0.256 &   0.786 &   18 &     \\
\end{longtable}

\clearpage

\newpage

\centering
\begin{longtable}{ccccccccc} 
\caption{Observed and reddening corrected line ratios (F(H$\beta$) = 100) and line identifications.
\label{lineid3}}\\
\hline\hline 
& & & \mc{5}{c}{He\,2-86} & \\
\cline{4-8} \\
$\lambda_0$ (\AA) & Ion & Mult. & $\lambda_{\rm obs}$ & $V_{\rm rad}$ (km s$^{-1}$) & $F(\lambda)/F({\rm H}\beta)$$^{\rm a}$ & $I(\lambda)/I({\rm H}\beta)$$^{\rm b}$ & Err(\%) & not \\
\hline
\endfirsthead
\caption{continued.}\\
\hline\hline 
& & & \mc{5}{c}{He\,2-86} & \\
\cline{4-8} \\ 
$\lambda_0$ (\AA) & Ion & Mult. & $\lambda_{\rm obs}$ & $V_{\rm rad}$ (km s$^{-1}$) & $F(\lambda)/F({\rm H}\beta)$$^{\rm a}$ & $I(\lambda)/I({\rm H}\beta)$$^{\rm b}$ & Err(\%) & not \\
\hline
\endhead
\hline
\endfoot
\endlastfoot
 3711.97 &            {\hi} &          H15 & 3711.89 &  -6.47 &   0.415 &   1.451 &    8 &     \\
 3721.83 &         {\fsiii} &           2F & 3721.76 &  -5.64 &   0.910 &   3.146 &    8 &     \\
 3721.93 &            {\hi} &          H14 &       * &       * &       * &       * &    * &     \\
 3726.03 &          {\foii} &           1F & 3725.94 &  -7.25 &   3.726 &  12.817 &    8 &     \\
 3728.82 &          {\foii} &           1F & 3728.70 &  -9.66 &   1.596 &   5.474 &    8 &     \\
 3734.37 &            {\hi} &          H13 & 3734.28 &  -7.23 &   0.662 &   2.256 &    8 &     \\
 3750.15 &            {\hi} &          H12 & 3750.07 &  -6.38 &   0.876 &   2.933 &    8 &     \\
 3770.63 &            {\hi} &          H11 & 3770.55 &  -6.35 &   1.002 &   3.277 &    8 &     \\
\end{longtable}

\clearpage

\newpage

\centering
\begin{longtable}{ccccccccc} 
\caption{Observed and reddening corrected line ratios (F(H$\beta$) = 100) and line identifications.
\label{lineid4}}\\
\hline\hline 
& & & \mc{5}{c}{M\,1-25} & \\
\cline{4-8} \\
$\lambda_0$ (\AA) & Ion & Mult. & $\lambda_{\rm obs}$ & $V_{\rm rad}$ (km s$^{-1}$) & $F(\lambda)/F({\rm H}\beta)$$^{\rm a}$ & $I(\lambda)/I({\rm H}\beta)$$^{\rm b}$ & Err(\%) & not \\
\hline
\endfirsthead
\caption{continued.}\\
\hline\hline 
& & & \mc{5}{c}{M\,1-25} & \\
\cline{4-8} \\ 
$\lambda_0$ (\AA) & Ion & Mult. & $\lambda_{\rm obs}$ & $V_{\rm rad}$ (km s$^{-1}$) & $F(\lambda)/F({\rm H}\beta)$$^{\rm a}$ & $I(\lambda)/I({\rm H}\beta)$$^{\rm b}$ & Err(\%) & not \\
\hline
\endhead
\hline
\endfoot
\endlastfoot
 3447.59 &           {\hei} &            7 & 3447.73 &  12.16 &   0.099 &   0.296 &   30 &     \\
 3478.97 &           {\hei} &           43 & 3479.27 &  25.85 &   0.065 &   0.186 &    : &     \\
 3530.50 &           {\hei} &           36 & 3530.69 &  16.13 &   0.126 &   0.345 &   25 &     \\
 3554.42 &           {\hei} &           34 & 3554.58 &  13.51 &   0.130 &   0.348 &   24 &     \\
 3587.28 &           {\hei} &           32 & 3587.42 &  11.69 &   0.163 &   0.424 &   20 &     \\
 3613.64 &           {\hei} &            6 & 3613.75 &   9.13 &   0.224 &   0.568 &   16 &     \\
 3634.25 &           {\hei} &           28 & 3634.38 &  10.71 &   0.258 &   0.643 &   14 &     \\
 3666.10 &            {\hi} &          H27 & 3666.26 &  13.08 &   0.183 &   0.444 &   18 &     \\
\end{longtable}

\clearpage

\newpage

\centering
\begin{longtable}{ccccccccc} 
\caption{Observed and reddening corrected line ratios (F(H$\beta$) = 100) and line identifications.
\label{lineid5}}\\
\hline\hline 
& & & \mc{5}{c}{M\,1-30} & \\
\cline{4-8} \\
$\lambda_0$ (\AA) & Ion & Mult. & $\lambda_{\rm obs}$ & $V_{\rm rad}$ (km s$^{-1}$) & $F(\lambda)/F({\rm H}\beta)$$^{\rm a}$ & $I(\lambda)/I({\rm H}\beta)$$^{\rm b}$ & Err(\%) & not \\
\hline
\endfirsthead
\caption{continued.}\\
\hline\hline 
& & & \mc{5}{c}{M\,1-30} & \\
\cline{4-8} \\ 
$\lambda_0$ (\AA) & Ion & Mult. & $\lambda_{\rm obs}$ & $V_{\rm rad}$ (km s$^{-1}$) & $F(\lambda)/F({\rm H}\beta)$$^{\rm a}$ & $I(\lambda)/I({\rm H}\beta)$$^{\rm b}$ & Err(\%) & not \\
\hline
\endhead
\hline
\endfoot
\endlastfoot
 3703.86 &            {\hi} &          H16 & 3702.37 &-120.65 &   0.725 &   1.329 &    7 &     \\
 3705.04 &           {\hei} &           25 & 3703.53 &-122.23 &   0.553 &   1.013 &    8 &     \\
 3711.97 &            {\hi} &          H15 & 3710.48 &-120.39 &   0.827 &   1.502 &    7 &     \\
 3712.74 &           {\oii} &            3 & 3711.46 &-103.39 &   0.066 &   0.119 &   19 &     \\
 3721.83 &         {\fsiii} &           2F & 3720.41 &-114.44 &   1.233 &   2.228 &    7 &     \\
 3721.93 &            {\hi} &          H14 &       * &       * &       * &       * &    * &     \\
 3726.03 &          {\foii} &           1F & 3724.54 &-119.93 &  21.560 &  38.868 &    7 &     \\
 3728.82 &          {\foii} &           1F & 3727.29 &-123.06 &  10.913 &  19.643 &    7 &     \\
\end{longtable}

\clearpage

\newpage

\centering
\begin{longtable}{ccccccccc} 
\caption{Observed and reddening corrected line ratios (F(H$\beta$) = 100) and line identifications.
\label{lineid6}}\\
\hline\hline 
& & & \mc{5}{c}{M\,1-32} & \\
\cline{4-8} \\
$\lambda_0$ (\AA) & Ion & Mult. & $\lambda_{\rm obs}$ & $V_{\rm rad}$ (km s$^{-1}$) & $F(\lambda)/F({\rm H}\beta)$$^{\rm a}$ & $I(\lambda)/I({\rm H}\beta)$$^{\rm b}$ & Err(\%) & not \\
\hline
\endfirsthead
\caption{continued.}\\
\hline\hline 
& & & \mc{5}{c}{M\,1-32} & \\
\cline{4-8} \\ 
$\lambda_0$ (\AA) & Ion & Mult. & $\lambda_{\rm obs}$ & $V_{\rm rad}$ (km s$^{-1}$) & $F(\lambda)/F({\rm H}\beta)$$^{\rm a}$ & $I(\lambda)/I({\rm H}\beta)$$^{\rm b}$ & Err(\%) & not \\
\hline
\endhead
\hline
\endfoot
\endlastfoot
 3703.86 &            {\hi} &          H16 & 3702.77 & -88.26 &   0.452 &   0.992 &   13 &     \\
 3705.04 &           {\hei} &           25 & 3703.88 & -93.90 &   0.379 &   0.827 &   15 &     \\
 3711.97 &            {\hi} &          H15 & 3710.89 & -87.26 &   0.624 &   1.357 &   12 &     \\
 3721.83 &         {\fsiii} &           2F & 3720.73 & -88.64 &   1.615 &   3.485 &   10 &     \\
 3721.93 &            {\hi} &          H14 &       * &       * &       * &       * &    * &     \\
 3726.03 &          {\foii} &           1F & 3724.94 & -87.73 &  34.294 &  73.758 &    9 &     \\
 3728.82 &          {\foii} &           1F & 3727.68 & -91.69 &  15.476 &  33.221 &    9 &     \\
 3734.37 &            {\hi} &          H13 & 3733.21 & -93.17 &   1.061 &   2.269 &   10 &     \\
\end{longtable}

\clearpage

\newpage

\centering
\begin{longtable}{ccccccccc} 
\caption{Observed and reddening corrected line ratios (F(H$\beta$) = 100) and line identifications.
\label{lineid7}}\\
\hline\hline 
& & & \mc{5}{c}{M\,1-61} & \\
\cline{4-8} \\
$\lambda_0$ (\AA) & Ion & Mult. & $\lambda_{\rm obs}$ & $V_{\rm rad}$ (km s$^{-1}$) & $F(\lambda)/F({\rm H}\beta)$$^{\rm a}$ & $I(\lambda)/I({\rm H}\beta)$$^{\rm b}$ & Err(\%) & not \\
\hline
\endfirsthead
\caption{continued.}\\
\hline\hline 
& & & \mc{5}{c}{M\,1-61} & \\
\cline{4-8} \\ 
$\lambda_0$ (\AA) & Ion & Mult. & $\lambda_{\rm obs}$ & $V_{\rm rad}$ (km s$^{-1}$) & $F(\lambda)/F({\rm H}\beta)$$^{\rm a}$ & $I(\lambda)/I({\rm H}\beta)$$^{\rm b}$ & Err(\%) & not \\
\hline
\endhead
\hline
\endfoot
\endlastfoot
 3512.51 &           {\hei} &           38 & 3512.73 &  18.77 &   0.089 &   0.219 &   23 &     \\
 3530.50 &           {\hei} &           36 & 3530.76 &  22.08 &   0.075 &   0.181 &   26 &     \\
 3554.42 &           {\hei} &           34 & 3554.63 &  17.71 &   0.091 &   0.217 &   22 &     \\
 3587.28 &           {\hei} &           32 & 3587.48 &  16.71 &   0.174 &   0.403 &   15 &     \\
 3613.64 &           {\hei} &            6 & 3613.86 &  18.27 &   0.222 &   0.504 &   13 &     \\
 3634.25 &           {\hei} &           28 & 3634.48 &  18.97 &   0.223 &   0.498 &   13 &     \\
 3658.64 &            {\hi} &          H34 & 3658.83 &  15.58 &   0.203 &   0.445 &   14 &     \\
 3659.42 &            {\hi} &          H33 & 3659.67 &  20.48 &   0.111 &   0.243 &   19 &     \\
\end{longtable}

\clearpage

\newpage

\centering
\begin{longtable}{ccccccccc} 
\caption{Observed and reddening corrected line ratios (F(H$\beta$) = 100) and line identifications.
\label{lineid8}}\\
\hline\hline 
& & & \mc{5}{c}{M\,3-15} & \\
\cline{4-8} \\
$\lambda_0$ (\AA) & Ion & Mult. & $\lambda_{\rm obs}$ & $V_{\rm rad}$ (km s$^{-1}$) & $F(\lambda)/F({\rm H}\beta)$$^{\rm a}$ & $I(\lambda)/I({\rm H}\beta)$$^{\rm b}$ & Err(\%) & not \\
\hline
\endfirsthead
\caption{continued.}\\
\hline\hline 
& & & \mc{5}{c}{M\,3-15} & \\
\cline{4-8} \\ 
$\lambda_0$ (\AA) & Ion & Mult. & $\lambda_{\rm obs}$ & $V_{\rm rad}$ (km s$^{-1}$) & $F(\lambda)/F({\rm H}\beta)$$^{\rm a}$ & $I(\lambda)/I({\rm H}\beta)$$^{\rm b}$ & Err(\%) & not \\
\hline
\endhead
\hline
\endfoot
\endlastfoot
 3498.64 &           {\hei} &           40 & 3499.77 &  96.81 &   0.269 &   1.253 &   32 &     \\
 3705.04 &           {\hei} &           25 & 3705.06 &   1.62 &   0.302 &   1.059 &   29 &     \\
 3713.08 &          {\neii} &            5 & 3713.27 &  15.33 &   0.464 &   1.612 &   22 &     \\
 3721.83 &         {\fsiii} &           2F & 3723.03 &  96.62 &   0.751 &   2.574 &   17 &     \\
 3721.93 &            {\hi} &          H14 &       * &       * &       * &       * &    * &     \\
 3726.03 &          {\foii} &           1F & 3727.24 &  97.32 &   3.301 &  11.265 &   10 &     \\
 3728.82 &          {\foii} &           1F & 3729.99 &  94.03 &   1.415 &   4.814 &   13 &     \\
 3734.37 &            {\hi} &          H13 & 3735.60 &  98.71 &   0.450 &   1.521 &   23 &     \\
\end{longtable}

\clearpage

\newpage

\centering
\begin{longtable}{ccccccccc} 
\caption{Observed and reddening corrected line ratios (F(H$\beta$) = 100) and line identifications.
\label{lineid9}}\\
\hline\hline 
& & & \mc{5}{c}{NGC\,5189} & \\
\cline{4-8} \\
$\lambda_0$ (\AA) & Ion & Mult. & $\lambda_{\rm obs}$ & $V_{\rm rad}$ (km s$^{-1}$) & $F(\lambda)/F({\rm H}\beta)$$^{\rm a}$ & $I(\lambda)/I({\rm H}\beta)$$^{\rm b}$ & Err(\%) & not \\
\hline
\endfirsthead
\caption{continued.}\\
\hline\hline 
& & & \mc{5}{c}{NGC\,5189} & \\
\cline{4-8} \\ 
$\lambda_0$ (\AA) & Ion & Mult. & $\lambda_{\rm obs}$ & $V_{\rm rad}$ (km s$^{-1}$) & $F(\lambda)/F({\rm H}\beta)$$^{\rm a}$ & $I(\lambda)/I({\rm H}\beta)$$^{\rm b}$ & Err(\%) & not \\
\hline
\endhead
\hline
\endfoot
\endlastfoot
 3340.77 &          {\oiii} &            3 & 3340.56 & -18.84 &   2.733 &   4.085 &    9 &     \\
 3345.83 &          {\fnev} &        3P-1D & 3345.70 & -11.66 &   1.395 &   2.081 &    9 &     \\
 3425.87 &          {\fnev} &        3P-1D & 3425.79 &  -7.01 &   3.019 &   4.374 &    8 &     \\
 3428.65 &          {\oiii} &           15 & 3428.42 & -20.11 &   0.868 &   1.256 &    9 &     \\
 3444.07 &          {\oiii} &           15 & 3443.84 & -20.02 &   6.225 &   8.963 &    8 &     \\
 3554.42 &           {\hei} &           34 & 3554.16 & -21.93 &   0.245 &   0.340 &   13 &     \\
 3568.50 &          {\neii} &            5 & 3568.28 & -18.48 &   0.416 &   0.576 &   10 &     \\
 3587.28 &           {\hei} &           32 & 3587.03 & -20.89 &   0.215 &   0.295 &   14 &     \\
\end{longtable}

\clearpage

\newpage

\centering
\begin{longtable}{ccccccccc} 
\caption{Observed and reddening corrected line ratios (F(H$\beta$) = 100) and line identifications.
\label{lineid10}}\\
\hline\hline 
& & & \mc{5}{c}{NGC\,6369} & \\
\cline{4-8} \\
$\lambda_0$ (\AA) & Ion & Mult. & $\lambda_{\rm obs}$ & $V_{\rm rad}$ (km s$^{-1}$) & $F(\lambda)/F({\rm H}\beta)$$^{\rm a}$ & $I(\lambda)/I({\rm H}\beta)$$^{\rm b}$ & Err(\%) & not \\
\hline
\endfirsthead
\caption{continued.}\\
\hline\hline 
& & & \mc{5}{c}{NGC\,6369} & \\
\cline{4-8} \\ 
$\lambda_0$ (\AA) & Ion & Mult. & $\lambda_{\rm obs}$ & $V_{\rm rad}$ (km s$^{-1}$) & $F(\lambda)/F({\rm H}\beta)$$^{\rm a}$ & $I(\lambda)/I({\rm H}\beta)$$^{\rm b}$ & Err(\%) & not \\
\hline
\endhead
\hline
\endfoot
\endlastfoot
 3703.86 &            {\hi} &          H16 & 3702.57 &-104.45 &   0.415 &   1.335 &    9 &     \\
 3705.04 &           {\hei} &           25 & 3703.72 &-106.85 &   0.289 &   0.930 &   11 &     \\
 3711.97 &            {\hi} &          H15 & 3710.68 &-104.22 &   0.470 &   1.488 &    9 &     \\
 3721.83 &         {\fsiii} &           2F & 3720.59 & -99.91 &   0.847 &   2.651 &    7 &     \\
 3721.93 &            {\hi} &          H14 &       * &       * &       * &       * &    * &     \\
 3726.03 &          {\foii} &           1F & 3724.87 & -93.35 &   8.621 &  26.875 &    6 &     \\
 3728.82 &          {\foii} &           1F & 3727.65 & -94.11 &   4.772 &  14.833 &    6 &     \\
 3734.37 &            {\hi} &          H13 & 3733.08 &-103.60 &   0.721 &   2.227 &    8 &     \\
\end{longtable}

\clearpage

\newpage

\centering
\begin{longtable}{ccccccccc} 
\caption{Observed and reddening corrected line ratios (F(H$\beta$) = 100) and line identifications.
\label{lineid11}}\\
\hline\hline 
& & & \mc{5}{c}{PC\,14} & \\
\cline{4-8} \\
$\lambda_0$ (\AA) & Ion & Mult. & $\lambda_{\rm obs}$ & $V_{\rm rad}$ (km s$^{-1}$) & $F(\lambda)/F({\rm H}\beta)$$^{\rm a}$ & $I(\lambda)/I({\rm H}\beta)$$^{\rm b}$ & Err(\%) & not \\
\hline
\endfirsthead
\caption{continued.}\\
\hline\hline 
& & & \mc{5}{c}{PC\,14} & \\
\cline{4-8} \\ 
$\lambda_0$ (\AA) & Ion & Mult. & $\lambda_{\rm obs}$ & $V_{\rm rad}$ (km s$^{-1}$) & $F(\lambda)/F({\rm H}\beta)$$^{\rm a}$ & $I(\lambda)/I({\rm H}\beta)$$^{\rm b}$ & Err(\%) & not \\
\hline
\endhead
\hline
\endfoot
\endlastfoot
 3340.77 &          {\oiii} &            3 & 3340.41 & -32.32 &   0.770 &   1.320 &    8 &     \\
 3428.65 &          {\oiii} &           15 & 3428.27 & -33.22 &   0.196 &   0.322 &   15 &     \\
 3444.07 &          {\oiii} &           15 & 3443.71 & -31.35 &   1.520 &   2.478 &    7 &     \\
 3447.59 &           {\hei} &            7 & 3447.27 & -27.83 &   0.127 &   0.207 &   21 &     \\
 3512.51 &           {\hei} &           38 & 3512.05 & -39.26 &   0.163 &   0.258 &   17 &     \\
 3530.50 &           {\hei} &           36 & 3530.13 & -31.43 &   0.126 &   0.197 &   21 &     \\
 3554.42 &           {\hei} &           34 & 3554.03 & -32.89 &   0.152 &   0.236 &   18 &     \\
 3587.28 &           {\hei} &           32 & 3586.90 & -31.77 &   0.257 &   0.394 &   13 &     \\
\end{longtable}

\clearpage

\newpage

\centering
\begin{longtable}{ccccccccc} 
\caption{Observed and reddening corrected line ratios (F(H$\beta$) = 100) and line identifications.
\label{lineid12}}\\
\hline\hline 
& & & \mc{5}{c}{Pe\,1-1} & \\
\cline{4-8} \\
$\lambda_0$ (\AA) & Ion & Mult. & $\lambda_{\rm obs}$ & $V_{\rm rad}$ (km s$^{-1}$) & $F(\lambda)/F({\rm H}\beta)$$^{\rm a}$ & $I(\lambda)/I({\rm H}\beta)$$^{\rm b}$ & Err(\%) & not \\
\hline
\endfirsthead
\caption{continued.}\\
\hline\hline 
& & & \mc{5}{c}{Pe\,1-1} & \\
\cline{4-8} \\ 
$\lambda_0$ (\AA) & Ion & Mult. & $\lambda_{\rm obs}$ & $V_{\rm rad}$ (km s$^{-1}$) & $F(\lambda)/F({\rm H}\beta)$$^{\rm a}$ & $I(\lambda)/I({\rm H}\beta)$$^{\rm b}$ & Err(\%) & not \\
\hline
\endhead
\hline
\endfoot
\endlastfoot
 3613.64 &           {\hei} &            6 & 3613.90 &  21.57 &   0.116 &   0.379 &   23 &     \\
 3634.25 &           {\hei} &           28 & 3634.52 &  22.27 &   0.173 &   0.555 &   17 &     \\
 3664.68 &            {\hi} &          H28 & 3664.84 &  13.10 &   0.121 &   0.376 &   22 &     \\
 3666.10 &            {\hi} &          H27 & 3666.31 &  17.17 &   0.126 &   0.391 &   22 &     \\
 3667.68 &            {\hi} &          H26 & 3667.96 &  22.89 &   0.133 &   0.410 &   21 &     \\
 3669.47 &            {\hi} &          H25 & 3669.72 &  20.42 &   0.159 &   0.489 &   18 &     \\
 3671.48 &            {\hi} &          H24 & 3671.76 &  22.86 &   0.198 &   0.607 &   16 &     \\
 3673.76 &            {\hi} &          H23 & 3674.05 &  23.67 &   0.227 &   0.697 &   15 &     \\
\end{longtable}

\clearpage

\end{document}